\documentclass[reprint, superscriptaddress, prd, amsmath, amssymb, nofootinbib, tightenlines, onecolumn]{revtex4-2}
\usepackage{graphicx}
\usepackage{hyperref}
\usepackage{dcolumn}
\usepackage{adjustbox}
\usepackage{bm}
\usepackage{aas_macros}
\usepackage{natbib}
\usepackage{xcolor}
\usepackage[thinc]{esdiff}
\usepackage{amsmath}
\usepackage{cprotect}
\usepackage{xspace}
\usepackage{ulem}
\usepackage{booktabs}
\usepackage{color}
\usepackage{colortbl}
\usepackage[english]{babel}

\newcommand{\planck}{\textit{Planck}\xspace}
\newcommand{\ymono}{\langle y \rangle\xspace}
\newcommand{\commander}{\texttt{Commander}\xspace}
\newcommand{\nver}{\hat{\mathbf{n}}}

\begin{document}

\defcitealias{BianchiniFabbian2022}{BF22}
\defcitealias{Abitbol2017}{A17}

\preprint{}
\title{A new constraint on the $y$-distortion with \textit{FIRAS}: robustness of component separation methods}
\author{Alina Sabyr}
\email{as6131@columbia.edu}
\affiliation{Department of Astronomy, Columbia University, New York, New York, USA 10027}
\author{Giulio Fabbian}
\email{giulio.fabbian@universite-paris-saclay.fr}
\affiliation{Université Paris-Saclay, CNRS, Institut d’Astrophysique Spatiale, 91405, Orsay, France}
\affiliation{School of Physics and Astronomy, Cardiff University, The Parade, Cardiff, CF24 3AA, UK}
\affiliation{Kavli Institute for Cosmology Cambridge, Madingley Road, Cambridge CB3 0HA, UK}
\author{J.~Colin Hill}
\affiliation{Department of Physics, Columbia University, New York, NY, USA 10027}
\author{Federico Bianchini}
\affiliation{Kavli Institute for Particle Astrophysics and Cosmology, Stanford University, 452 Lomita Mall, Stanford, CA, 94305, USA}
\affiliation{SLAC National Accelerator Laboratory, 2575 Sand Hill Road, Menlo Park, CA, 94025, USA}
\affiliation{Department of Physics, Stanford University, 382 Via Pueblo Mall, Stanford, CA, 94305, USA}

\date{\today}

\begin{abstract}
The sky-averaged Compton-$y$ distortion in the cosmic microwave background (CMB) energy spectrum, $\ymono$, provides information about energy injection in the Universe occurring at $z \lesssim 5\times10^{4}$. It is primarily sourced by the thermal Sunyaev-Zeldovich effect --- the up-scattering of CMB photons on free electrons in collapsed dark matter halos. In this work and our companion paper, Ref.~\cite{Fabbian2023}, we perform a re-analysis of the archival \textit{COBE/FIRAS} data to constrain $\ymono$. We utilize two analysis approaches: (i) fitting the sky-averaged intensity spectrum (\textit{frequency monopole}) and (ii) fitting the sky spectrum in each pixel (\textit{pixel-by-pixel}). We obtain the baseline upper limits  $\ymono < 31\times 10^{-6}$ and $\ymono < 8.3\times 10^{-6}$ (both at $95\%$ C.L.) for these two approaches, respectively. We present the first detailed comparison of these analysis methods on both mock skies and real data. Our findings suggest that accounting for the spatial variability of foregrounds via \textit{pixel-by-pixel} fitting allows for tighter constraints on $\ymono$ by a factor of $\approx 3-5$ as compared to the \textit{frequency monopole} method. We show that our \textit{frequency monopole} results agree well with predictions from Fisher forecast techniques based on the sky-averaged signal, which have been used for forecasting future spectral distortion experiments. Our results thus suggest that the scientific reach of future spectral distortion experiments can potentially be enhanced by a factor of a few via more optimal component separation methods, and we identify the \textit{pixel-by-pixel} method as one such robust way to achieve this. We discuss the implications of our improved constraints on $\ymono$ from the \textit{pixel-by-pixel} method in Ref.~\cite{Fabbian2023}.
\end{abstract}

\maketitle

\section{Introduction} \label{sec:intro}
The \textit{Far Infrared Absolute Spectrophotometer} (\textit{FIRAS}) instrument onboard the \textit{Cosmic Background Explorer} (\textit{COBE}) satellite, which operated in the 1990s, provides the best measurements of the absolutely calibrated sky emission across a range of frequencies at which the cosmic microwave background (CMB) signal is dominant.  It thus provides the best constraints to date on the energy spectrum of the CMB (e.g., \cite{Mather1990_preliminary, Fixsen1996_dist, Fixsen2009}). While the CMB spectrum is very close to that of a perfect blackbody, small shifts from a Planck distribution are expected within the standard cosmological framework (e.g., \cite{zeldovich1969, sunyaev1970relic, Illarionov1975compton, Hu1992sd}). These deviations are referred to as ``spectral distortions'' and emerge due to various processes occurring at redshifts $z \lesssim 2\times 10^{6}$, when the Universe is no longer hot and dense enough to maintain full thermal equilibrium between matter and radiation via efficient Compton, double-Compton, and Bremsstrahlung interactions (e.g.,~\cite{sunyaev1970relic, Danese1982, Burigana1991}). Between $5 \times 10^{4} \lesssim z \lesssim 2 \times 10^{6}$, the time scales of double-Compton scattering and Bremsstrahlung emission are longer than the expansion rate of the Universe, but Compton scattering remains efficient. The number of photons is preserved and any energy injection or subtraction results in a Bose-Einstein photon distribution with a non-zero chemical potential, $\mu$, or a $\mu$-type distortion.  At $z \lesssim 5 \times 10^{4}$, Compton scattering also becomes inefficient, allowing the formation of a $y$-type distortion, which will be the focus of this work \cite{zeldovich1969, sunyaev1970relic}. 

The dominant source of the Compton-$y$ distortion is the thermal Sunyaev-Zeldovich effect (tSZ) --- the inverse-Compton scattering of CMB photons off energetic free electrons primarily located in galaxy clusters and groups, with additional, smaller contributions from the intergalactic medium (IGM) and the epoch of reionization (EoR) \cite{zeldovich1969, sunyaev1970relic, Hill2015}. Some primordial contributions to the $y$-distortion are also expected, but they are predicted to be 2-3 orders of magnitude lower than the tSZ effect \cite{Chluba2012y_order}. The strength of the tSZ effect is parametrized by the Compton-$y$ parameter, which is proportional to the integrated electron pressure along the line of sight (LOS). The sky-averaged tSZ signal, i.e., its monopole, thus provides a constraint on the total thermal energy stored in electrons in the Universe. Additionally, some intracluster medium (ICM) gas is expected to reach temperatures where relativistic corrections are significant. The relativistic correction to the tSZ signal quantifies the ($y$-weighted) average temperature of the electrons, at leading order. Therefore, the $y$-distortion provides unique information about the ionized gas during structure formation and can be used to constrain baryonic feedback models \cite{Hill2015, Thiele2022}.

Direct measurements of the monopole spectral distortion signals require an absolutely calibrated CMB intensity spectrum\footnote{Indirect estimates can also be made. One could measure the cross-correlation of the spatially anisotropic component of the spectral distortions with CMB anisotropies to constrain the amplitude of the monopole signals~\cite{kite2023}. Or one could use the Compton-$y$ redshift-kernel determined from cross correlations of the CMB data with galaxy surveys, but this method requires extrapolation to high redshifts~\cite{Chiang2020}. }. The tightest upper limits on the amplitude of the spectral distortions from the \textit{FIRAS} data are the following: $|\langle\mu\rangle|< 90\times10^{-6}$ and $|\langle y\rangle|< 15\times10^{-6}$ both at 95\% confidence level (C.L.). For the $y$-distortion, the \textit{FIRAS} sensitivity was roughly one order of magnitude away from reaching the expected theoretical prediction ($\sim 2\times10^{-6}$ from halo-model calculations \cite{Hill2015}; see also Refs.~\cite{Dolag2016,Thiele2022} for estimates from hydrodynamical simulations). Recently, Ref.~\cite{BianchiniFabbian2022} (hereafter \citetalias{BianchiniFabbian2022}) re-analyzed the \textit{FIRAS} archival data and were able to tighten the limits on the $\mu$-distortion by a factor of $\approx 2$, reaching $|\langle\mu\rangle|< 47 \times10^{-6}$ (95\% C.L.). 

In this work and our companion paper, Ref.~\cite{Fabbian2023}, we perform a similar re-analysis of the \textit{FIRAS} data to place new constraints on the $y$-distortion. In this paper, we primarily focus on a re-analysis using the \textit{frequency monopole} method, which models the sky-averaged monopole signal in each observational frequency channel. Our companion paper Ref.~\cite{Fabbian2023} uses the \textit{pixel-by-pixel} method, which fits the sky spectrum in each individual pixel, as was first employed in \citetalias{BianchiniFabbian2022}. We obtain the following upper limits for the baseline set-ups: $\ymono<31\times10^{-6}$ (\textit{frequency monopole}), $\ymono<8.3\times10^{-6}$ (\textit{pixel-by-pixel}).

The goals of this paper are threefold:
\begin{itemize}
    \item \textit{Comparing analysis methods}: in principle, different data analysis techniques, if unbiased, should yield consistent results. In practice, different methods are subject to different shortcomings, mainly due to the way they deal with the complexity of the sky emission and contamination due to instrumental systematic errors. We therefore compare our \textit{frequency monopole} with the \textit{pixel-by-pixel} component separation method used in \citetalias{BianchiniFabbian2022} and Ref.~\cite{Fabbian2023}  to assess their individual strengths and weaknesses when applied to the \textit{FIRAS} data. Ref.~\cite{Abitbol2017} (hereafter \citetalias{Abitbol2017}) first speculated that using spatial information would help separate the spectral distortion monopole signal from the astrophysical foregrounds more effectively, as compared to using the sky-averaged spectrum at each frequency. In this work, we show that indeed this is the case, finding that we can achieve $\approx 4$ times tighter constraints using the \textit{pixel-by-pixel} approach as compared to the \textit{frequency monopole} approach. 
    \item \textit{Testing more complex sky models}: we experiment with fitting more complex sky models using the \textit{frequency monopole} method than those used in \citetalias{BianchiniFabbian2022}. This becomes more challenging for the \textit{pixel-by-pixel} approach as interferograms in a single pixel are significantly noisier than the sky-averaged frequency spectrum. As such, sky models requiring too many parameters are generally badly or ill constrained in this case.        
    \item \textit{Validating Fisher forecasts}: there are several satellite, balloon-borne, and ground-based experiments that have been or are currently proposed to upgrade the \textit{FIRAS} measurements of the absolutely calibrated CMB spectrum (e.g., \textit{PIXIE} \cite{pixie2011,pixie2024}, \textit{BISOU} \cite{Bisou,Bisou2022}, \textit{COSMO} \cite{Cosmo}, \textit{Voyage 2050} \cite{Voyage2050}, and new instrument concepts like \textit{SPECTER} \cite{specter}). In our work, we verify that \textit{frequency monopole}-based Fisher forecasts, which have been used to assess the performance of various future missions (such as in \citetalias{Abitbol2017}), are consistent with the level of precision that can be obtained from real data using this technique. 
\end{itemize}

Our paper is organized as follows. The \textit{FIRAS} and simulated mock data sets used in this work are presented in Sec.~\ref{sec:data}. In Sec.~\ref{sec:methods} we discuss the the details of the \textit{frequency monopole} and \textit{pixel-by-pixel} component separation methods, the adopted sky models, and inference approaches. In Sec.~\ref{sec:mock_results} we present our results on mock data from the \textit{frequency monopole} and \textit{pixel-by-pixel} methods. In Sec.~\ref{sec:data_results_monopole} and Sec.~\ref{sec:data_results_pixpix}, we discuss the results obtained from data using the \textit{frequency monopole} and the \textit{pixel-by-pixel} methods, respectively. We summarize the findings of this work and conclude in Sec.~\ref{sec:conclusion}.

\section{Data}\label{sec:data}
\subsection{COBE/FIRAS}
We use the low-spectral-resolution, absolutely calibrated \textit{FIRAS} sky maps both from the low- and high-frequency instruments, which cover the full sky in 213 frequency bands between 68 GHz and $\approx 3$ THz (with bandwidth $\Delta\nu\approx 13.6$ GHz). We adopt the official reprocessing of the data in \textsc{HEALPix} pixelization\footnote{\url{http://healpix.sourceforge.net}} at a spatial resolution of $\delta\theta \approx 3.5^\circ$ available on LAMBDA. \footnote{\url{https://lambda.gsfc.nasa.gov/product/cobe/firas_tpp_all_get.cfm}} The data provides sky emission maps in units of MJy/sr, which we use throughout this work. In our baseline analysis we use only frequencies $\nu<800$~GHz, as modeling the astrophysical emission at higher frequencies is more complex. In Appendix \ref{app:high_frequencies}, we explore extending our analysis to $\approx 1.9$ THz. We do not utilize frequencies higher than $\approx 1.9$ THz due to the strong level of noise in the data. Although the CMB signal becomes almost entirely negligible in the channels of the high-frequency instrument ($\nu\gtrsim 612$ GHz) and the noise significantly increases, we investigate whether using the higher frequency channels can help better constrain the foregrounds and thus tighten the error bars on $\ymono$.

In this paper, we present results based on two different data analysis methods that require a different manipulation and contraction of the full general \textit{FIRAS} data covariance. The covariance matrix of the \textit{FIRAS} data has both frequency ($\nu$) and pixel ($p$) dependence. Following the notation in \citetalias{BianchiniFabbian2022} and in the \textit{FIRAS} explanatory supplement \footnote{\url{https://lambda.gsfc.nasa.gov/product/cobe/firas_exsupv4.cfm}} \cite{firas_supp}, we can write the covariance of two elements of the \textit{FIRAS} data cube $\hat{I}^{\rm FIRAS}_{\nu}(\nver_p)\equiv \hat{I}^{\rm FIRAS}_{\nu p}$ in the $\nver_p$ pixel direction as 
\begin{eqnarray}
    \mathbb{C}_{\nu p\nu'p'} &=& {\rm Cov}(\hat{I}^{\rm FIRAS}_{\nu p},\hat{I}^{\rm FIRAS}_{\nu'p'})\nonumber \nonumber \\ 
    &=& C_{\nu \nu^{\prime}}\left(\delta_{p p^{\prime}} / N_{p}+\beta_{p}^{k} \beta_{p^{\prime} k}+0.04^{2}\right) \nonumber \\ 
    &+& S_{p \nu} S_{p^{\prime} \nu^{\prime}}\left(J_{\nu} J_{\nu^{\prime}}+G_{\nu} G_{\nu} \delta_{\nu \nu^{\prime}}\right) \nonumber \\ 
    &+& P_{\nu} P_{\nu^{\prime}}\left(U^{2} \delta_{p p^{\prime}} / N_{p}+T^{2}\right).
\label{eq:covariance}
\end{eqnarray}

\noindent where $N_{p}$ is the pixel weight accounting for the inhomogeneity of the survey and hence the noise properties. The other terms describe the other main sources of uncertainty:
\begin{enumerate}
    \item Detector noise is described by the $C_{\nu\nu'}$ matrix that accounts for the frequency-frequency noise correlation due to instrumental effects.
    \item Destriper errors are captured by $\beta^{p}_{k}$. As suggested in, e.g., Ref.~\cite{Odegard2019} we use the so-called orthogonal stripes available on LAMBDA and we sum over all $k$-th stripes to construct $\beta$.\footnote{The \textit{FIRAS} explanatory supplement suggests adding a factor of $0.04^{2}$ to account for the uncertainty in the stripes.}
    \item Bolometer and emissivity gain uncertainties are captured by $J_{\nu}J_{\nu'}$ (JCJ) and $G_{\nu}G_{\nu}$ (PEP), respectively.  The gain uncertainties only have a frequency dependence; the $S_{p\nu}$ vectors correspond to the absolute sky brightness after CMB monopole subtraction at a reference $T_{0}=2.7255$ K.
    \item Internal calibrator and absolute temperature errors are captured by $U^{2}$ (PUP) and $T^{2}$ (PTP), respectively.  Here, $P_{\nu}=\partial B_{\nu}(2.728\,\mathrm{K})/\partial T$ is assumed to be a CMB dipole signal at a given temperature, $U=180\, \mu$K is the uncertainty in calibrator temperature as suggested in Sec.~7.4.5 of the \textit{FIRAS} explanatory supplement, and $T=0.002$ K based on the publicly available calibration uncertainty file available on LAMBDA.
\end{enumerate}

\begin{figure}
    \centering
    \includegraphics[width=0.5\columnwidth]{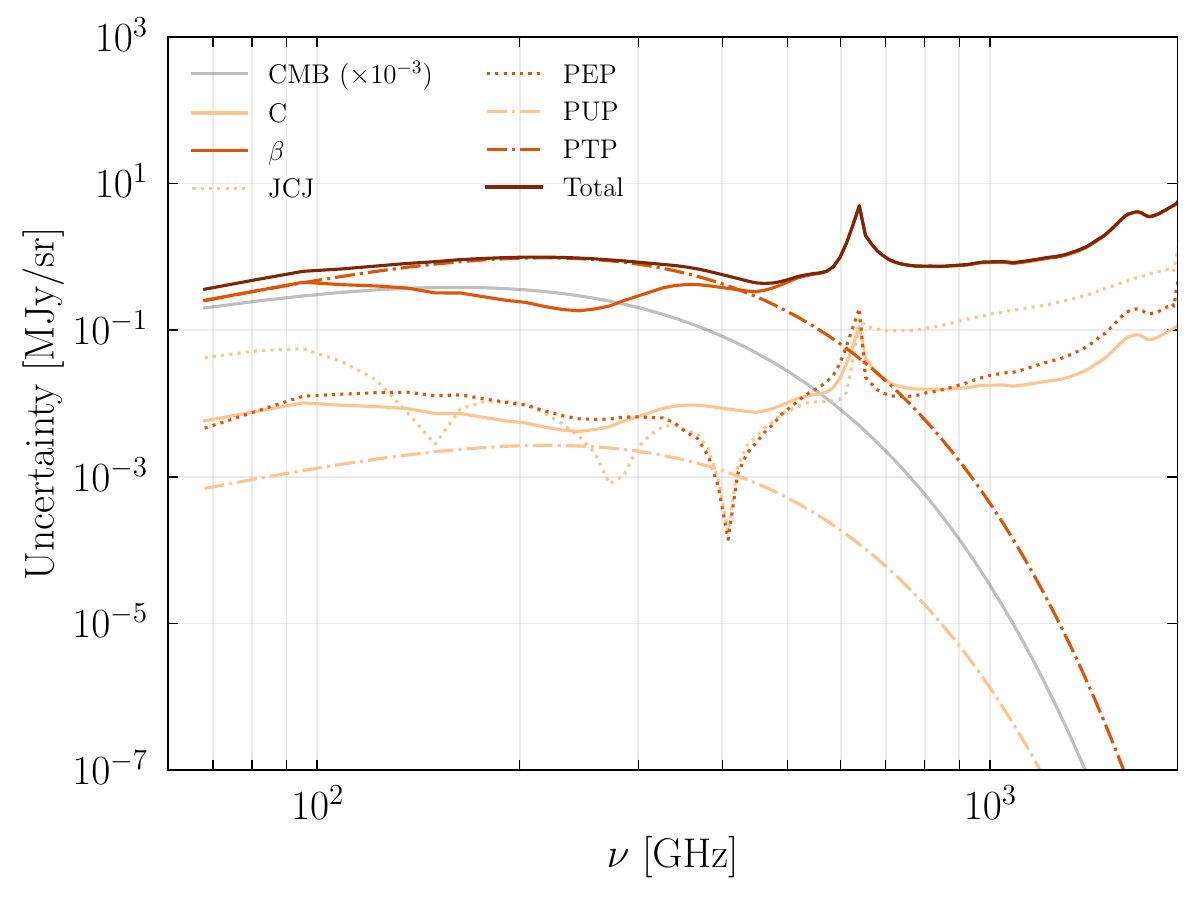}
    \caption{Different components of the statistical and systematic uncertainties of the \textit{FIRAS} measurements for both the low- and high-frequency instruments. The PTP and PUP terms describe systematic calibration uncertainties, the PEP and JCJ terms describe statistical calibration errors, $C$ represents instrumental noise, and $\beta$ captures residual errors from the destriping procedure. The light grey curve shows the CMB monopole signal divided by 1000, for comparison.}
    \label{fig:noise}
\end{figure}

Fig.~\ref{fig:noise} shows the contribution of each of these terms to the total uncertainty of the sky-averaged signal in each frequency band. The dominant sources of uncertainty for the \textit{frequency monopole} are the absolute temperature errors at low frequencies ($<400$ GHz) and instrumental noise at high frequencies. Note that the noise contributions differ for the \textit{pixel-by-pixel} method, where it is dominated by the instrumental noise across all frequencies (see Fig.~2 in \citetalias{BianchiniFabbian2022}).

\subsection{Simulated data sets}\label{sec:simulated_data_sets}
\begin{figure*}[!htbp]
\includegraphics[width=\textwidth]{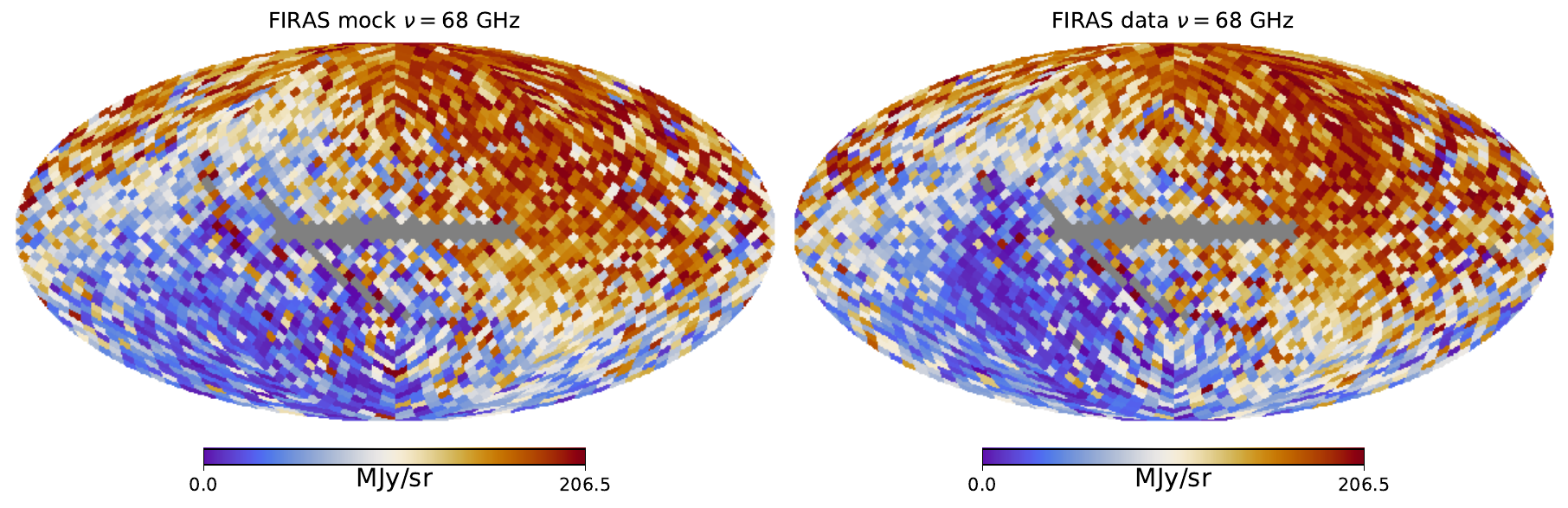}\\
\includegraphics[width=\textwidth]{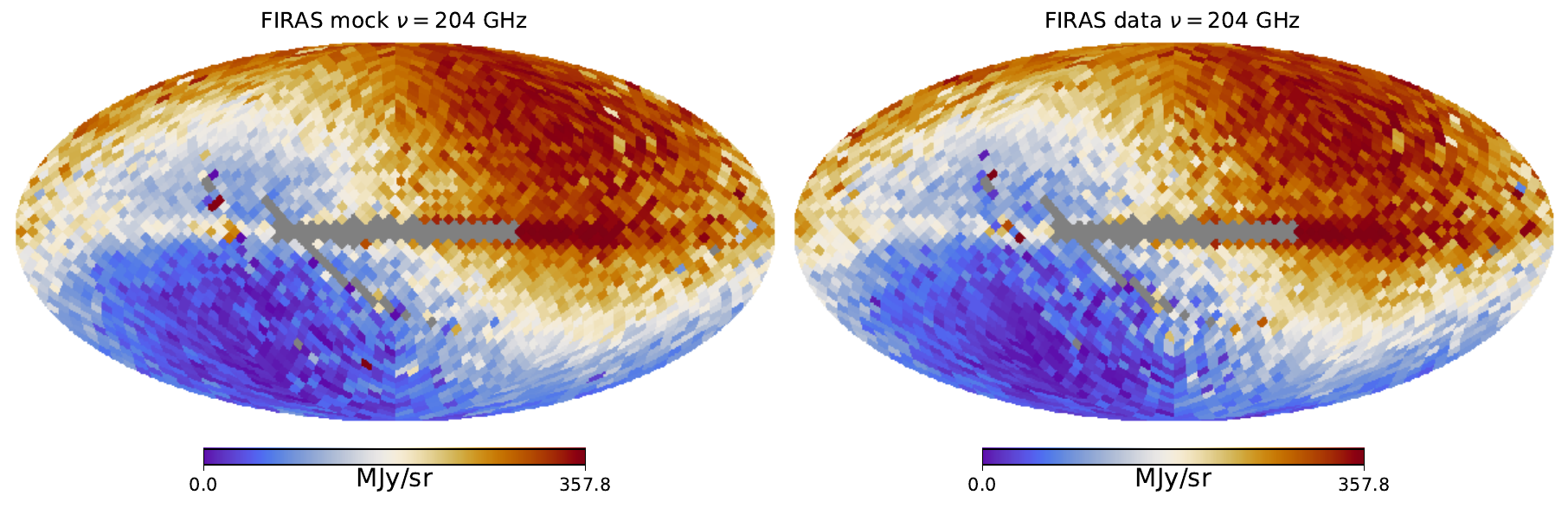}\\
\includegraphics[width=\textwidth]{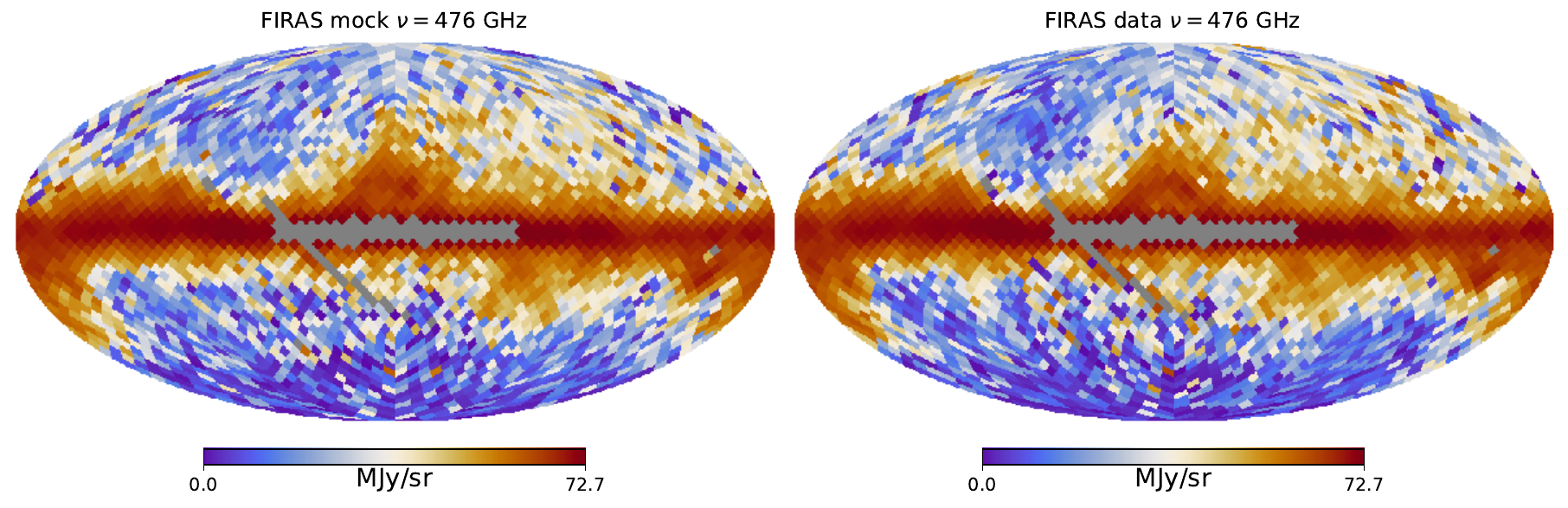}
\caption{Simulated \textit{FIRAS} data set (left) including spatially varying Galactic dust, synchrotron, CO lines, AME, free-free, and CIB compared to the \textit{FIRAS} data (right) at various observed frequencies (top to bottom) of the low-frequency instrument. Pixels removed by the \textit{FIRAS} destriper mask are shown in grey.}
\label{fig:mocks-lowf}
\end{figure*}

\begin{figure*}[!htbp]
\includegraphics[width=\textwidth]{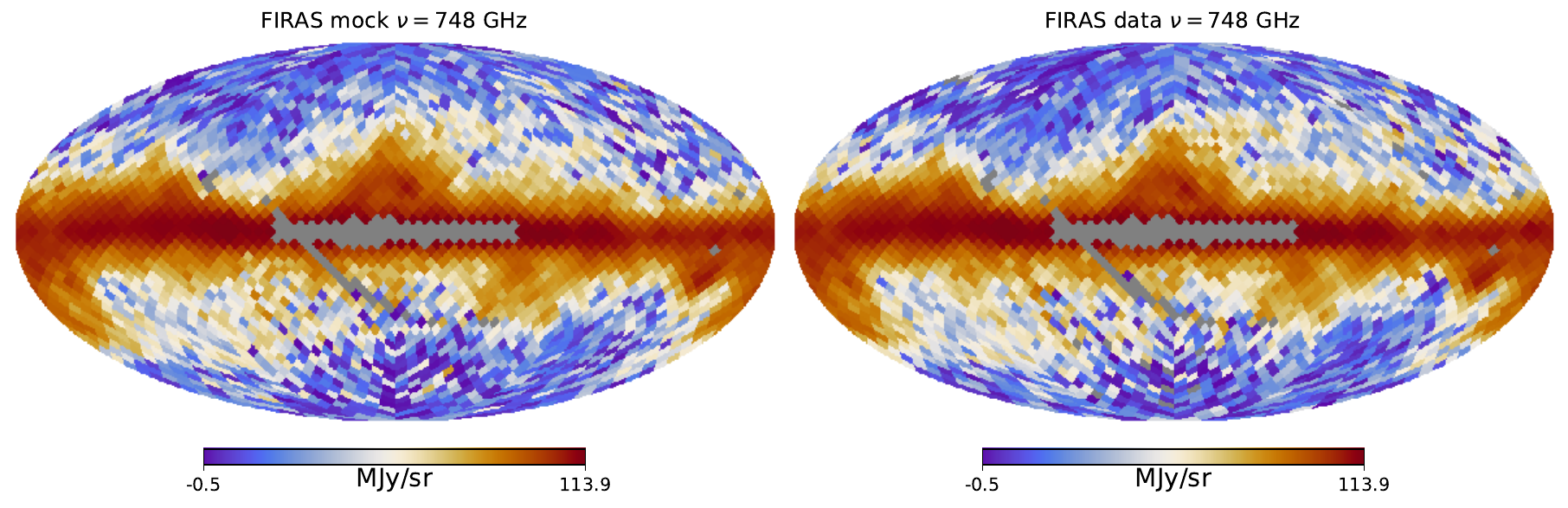}\\
\includegraphics[width=\textwidth]{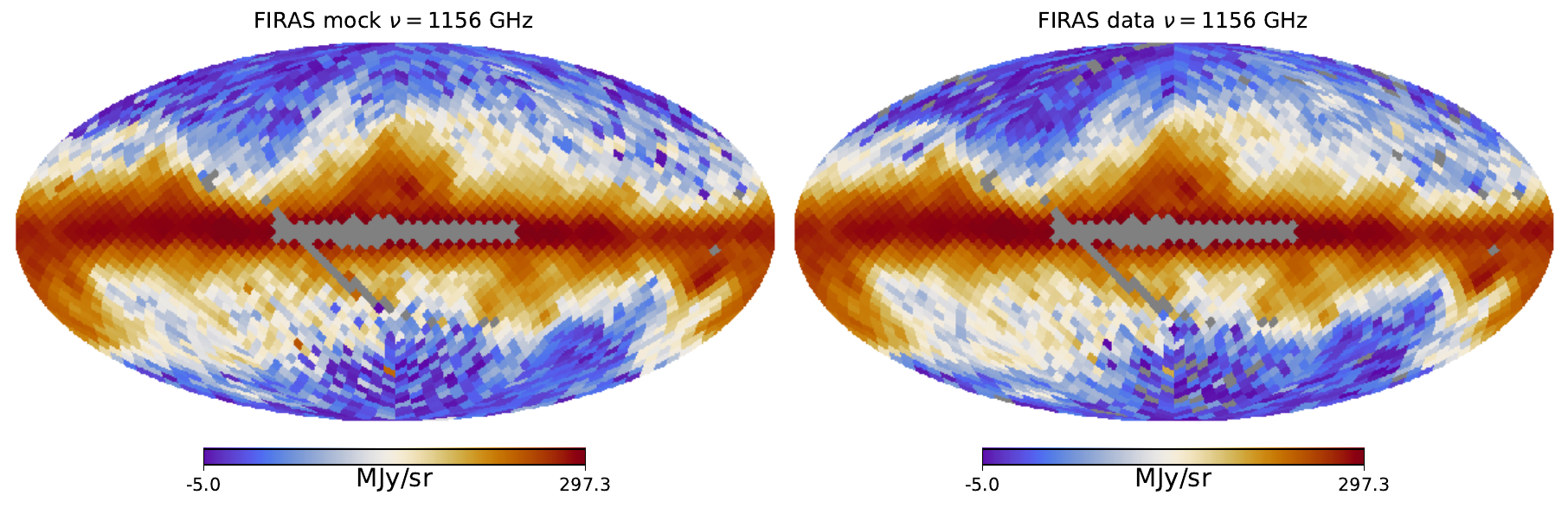}\\
\includegraphics[width=\textwidth]{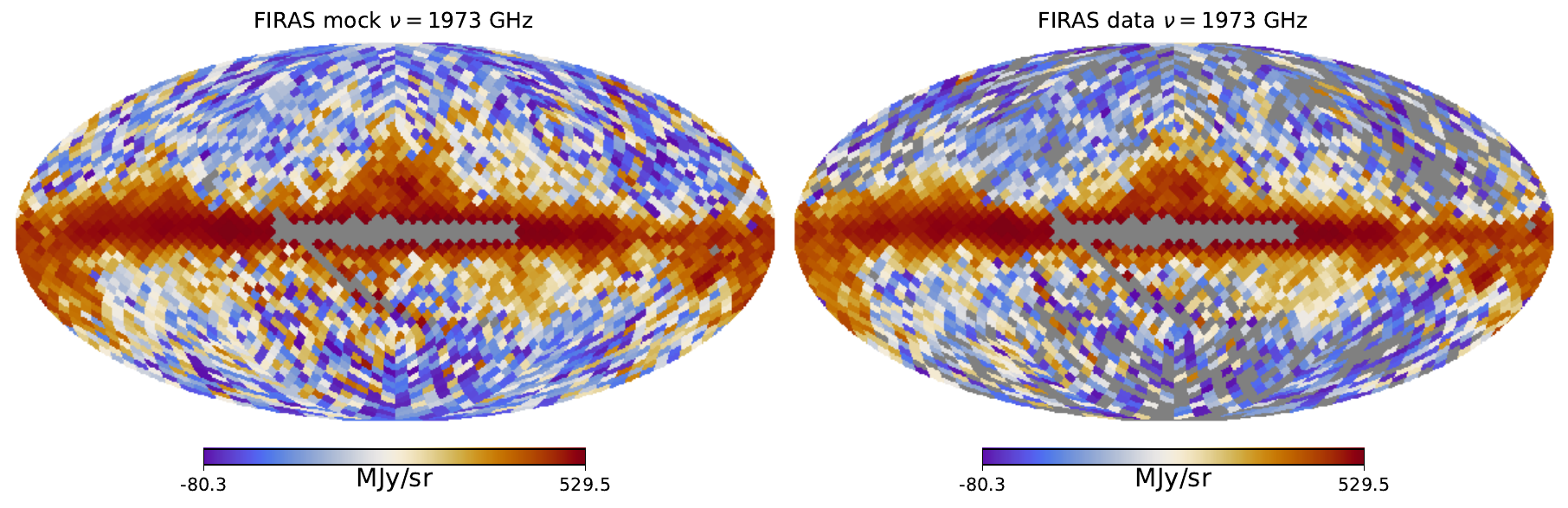}
\caption{Same as Fig.~\ref{fig:mocks-lowf}, but showing various frequency channels of the \textit{FIRAS} high-frequency instrument.}
\label{fig:mocks-high}
\end{figure*}

To verify that we can recover $\ymono$ in the presence of astrophysical foregrounds both with the \textit{frequency monopole} approach of this work and the \textit{pixel-by-pixel} method used in \citetalias{BianchiniFabbian2022} and Ref.~\cite{Fabbian2023}, we construct simulations of the \textit{FIRAS} data cube. These mock data sets include the \textit{FIRAS} instrumental noise and the various Galactic and extragalactic sky emission components across all \textit{FIRAS} frequencies, as described below. Some examples of these mock sky maps are shown in Figs.~\ref{fig:mocks-lowf}  and~\ref{fig:mocks-high}. 

\subsubsection{Noise and systematic uncertainties} 
As previously discussed, the \textit{FIRAS} data are affected by several instrumental systematics. However, in the absence of an end-to-end model to go from the raw data to the sky maps for the \textit{FIRAS} data set, we do not add any instrumental systematics to our simulated maps. A possible work-around could be incorporating systematic errors via random noise realizations based on the full \textit{FIRAS} covariance (i.e., including all the systematic uncertainty terms). However, the maps constructed with this approach do not match the data with respect to the overall variance and deliver unrealistically noisy data sets.  We find the off-diagonal destriper errors $\beta_{p}^{k} \beta_{p^{\prime} k}$ to be mainly responsible for this mismatch, suggesting an inaccurate modeling of such components in the original \textit{FIRAS} covariance model.  An additional simple indication of such inaccuracies from these terms is that the random realizations of the \textit{FIRAS} noise accounting for the pixel-pixel correlation (i.e., including all the terms of the first line of Eq.~\ref{eq:covariance}) are large enough to hide the expected CMB dipolar anisotropies induced by the motion of our local frame at frequencies where the data clearly display a dominant dipolar pattern. As such, we drop all the terms in the covariance except the $C$ terms and the weights describing the inhomogeneity of the noise across the sky ($N_{p}$) in the construction of the sky mocks, effectively assuming that the noise is uncorrelated between pixels. We refer the reader to the description of the \textit{frequency monopole} method in Sec.~\ref{sec:monopole} and tests performed on mocks in Sec.~\ref{sec:mock_results} for more details.
\subsubsection{Galactic foregrounds}
To model the Galactic foreground signal,
we use the \commander low-resolution maps available in the second release of the \planck data (dust, synchrotron, anomalous microwave emission (AME), and free-free) as templates \cite{Planck2016FG}. We also use the Type 1 CO emission line maps produced for the first \planck release in 2013 \cite{Planck2013CO}. These are generally noisier than other CO maps derived from \planck, but are more robust to dust contamination in the Galactic plane. They were in fact built from a modified internal linear combination algorithm using maps derived with single-detector data within a given frequency channel \cite{Planck2013CO}. The algorithm relies on the differences between the bandpasses of single detectors to separate CO lines and dust and, as such, it is less affected by the complexity of components with smooth SEDs in the Galactic plane. We neglect any additional emission lines as we do not have reliable low-noise and high-resolution observations of these components and the only ones that are available are derived from the \textit{FIRAS} data itself. We use the emission models of \commander detailed in Ref.~\cite{Planck2016FG} to rescale the data from their reference frequency to all the other \textit{FIRAS} frequencies, accounting for the spatial dependence of all the model parameters fit from the data in each pixel. After rescaling the emission to a given frequency, we  convolve the resulting map with the \textit{FIRAS} scanning beam that includes the instantaneous $7^\circ$ top-hat optical response and its variation due to the satellite motion. Whenever it is required to convert the map units from thermodynamic temperature to flux units, we use the \planck official values of Ref.~\cite{Planck2016FG}. 

Finally, we note that \planck is not an absolutely calibrated instrument and its frequency bands are not sensitive to the global monopoles of the data. These are degenerate with the offset induced by the $1/f$ noise during the spinning of the satellite and are removed by the destriping approach adopted in the mapmaking step of the data analysis. There is, therefore, no guarantee that the overall monopole of each of the sources of emission matches that observed by \textit{FIRAS}. We note, however, that many of the offsets in the \planck frequency maps have been set to values derived from \textit{FIRAS} or astrophysical models. Therefore, we expect the final \commander maps to have offsets that are in the ballpark of those that \textit{FIRAS} observed, which suffices for our purposes here.  \citetalias{BianchiniFabbian2022} verified that this was the case for e.g., their \textit{FIRAS} component-separated dust amplitude map, which agreed with the values found from the \commander map.

To test the robustness of our foreground modeling against the complexity of the sky emission, we construct a set of simple simulated data sets, which contain only Galactic dust with the amplitude of the dust SED varied across pixels, while its temperature is fixed at $T_{\rm d} = 20.88$ K and its spectral index at $\beta_{\rm d} = 1.51$, corresponding to the median values of these parameters computed from the \commander maps. We refer to these maps as {\it constant dust} mocks.

\subsubsection{Extragalactic components}
We add the CMB monopole assuming a blackbody at $T_0=2.7255$ K with a dipolar modulation with an amplitude ($v/c\approx 0.001$) and a direction ($(l,b) = (264.021^\circ, - 48.253^\circ)$)
 matching the \planck observations. Additionally, we include a random realization of the CMB temperature anisotropies based on the the \planck 2018 best-fit $\Lambda$CDM cosmology \cite{Planck2018cosmo}. Finally, we also include a CIB monopole using the fiducial modified blackbody (MBB) model from \citetalias{Abitbol2017}, to which we add anisotropies based on a power spectrum calculated with the halo model from Ref.~\cite{Maniyar2021}. We assume as baseline $\ymono=0$ in order to assess the amplitude of foreground-induced bias in the recovery of $\ymono$ and neglect any anisotropy induced by galaxy clusters. For some of the mock tests, we consider cases with non-zero $\ymono=20\times10^{-6}$. We refer to the mocks that contain both Galactic and extragalactic components as the \textit{all foregrounds} mocks.


\subsection{Sky masks}
\begin{figure}[h]
\includegraphics[width=0.5\columnwidth]{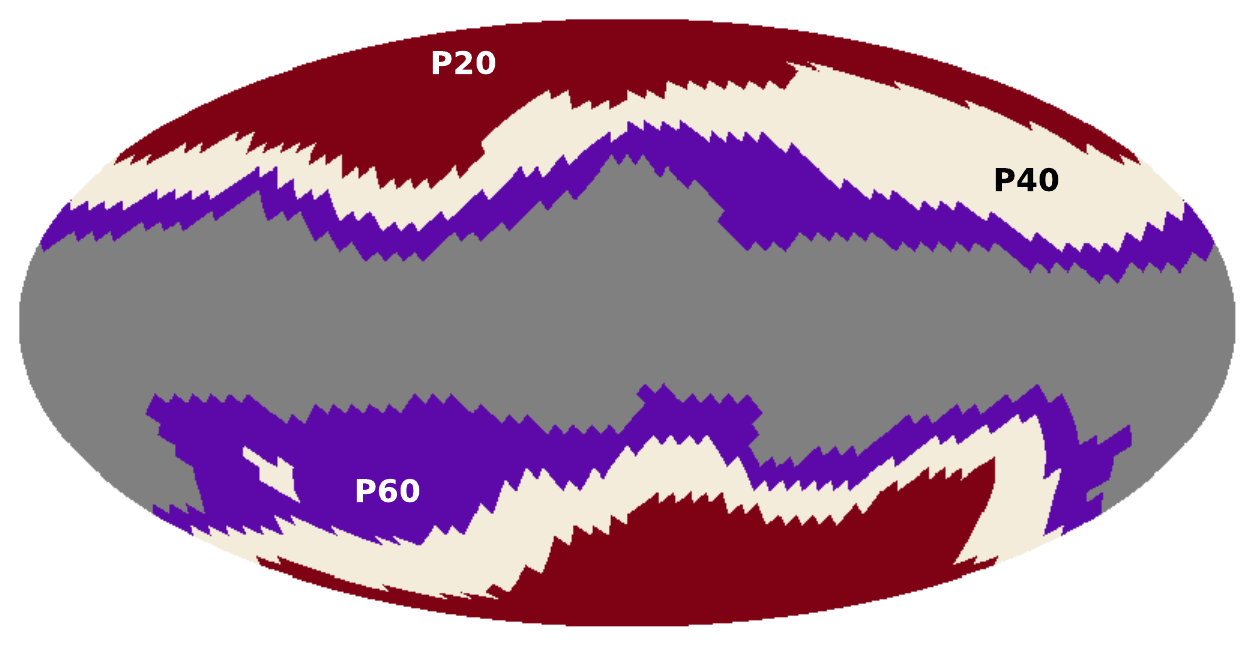}
\caption{\planck masks used in this work (in addition to the \textit{FIRAS} destriper mask). P20, P40, and P60 refer to masks that preserve $20\%, 40\%$, and $60\%$ of the sky, respectively.}
\label{fig:masks}
\end{figure}

To remove excess foreground emission near the Galactic plane, we use both the destriper mask of the \textit{FIRAS} data release and the public binary \planck Galactic masks following the analysis in \citetalias{BianchiniFabbian2022}. 
The destriper mask removes pixels not observed by \textit{FIRAS} and those not included in the destriping procedure. The \planck masks were built by thresholding the 353 GHz map after subtracting the CMB signal. We consider only the \planck  masks that retain $20\%$, $40\%$, and $60\%$ of the sky, which we will refer as P20, P40, and P60, respectively. We show these masks in Fig.~\ref{fig:masks}.  We do not apodize the masks, as we do not estimate angular power spectra in this work.

\section{Methods}\label{sec:methods}
\subsection{Analysis approaches}
We follow two approaches to analyze the \textit{FIRAS} data. The \textit{pixel-by-pixel} method leverages the statistical anisotropy of the foreground emission and the broad frequency coverage of \textit{FIRAS} to remove foregrounds with simple but spatially adaptive models, while the \textit{frequency monopole} exploits the large sky coverage to reduce the noise in the measurements. We outline the two methods and their main differences in the following subsections. 

\subsubsection{Frequency Monopole}\label{sec:monopole}
\begin{figure}
    \centering
    \includegraphics[width=0.7\columnwidth]{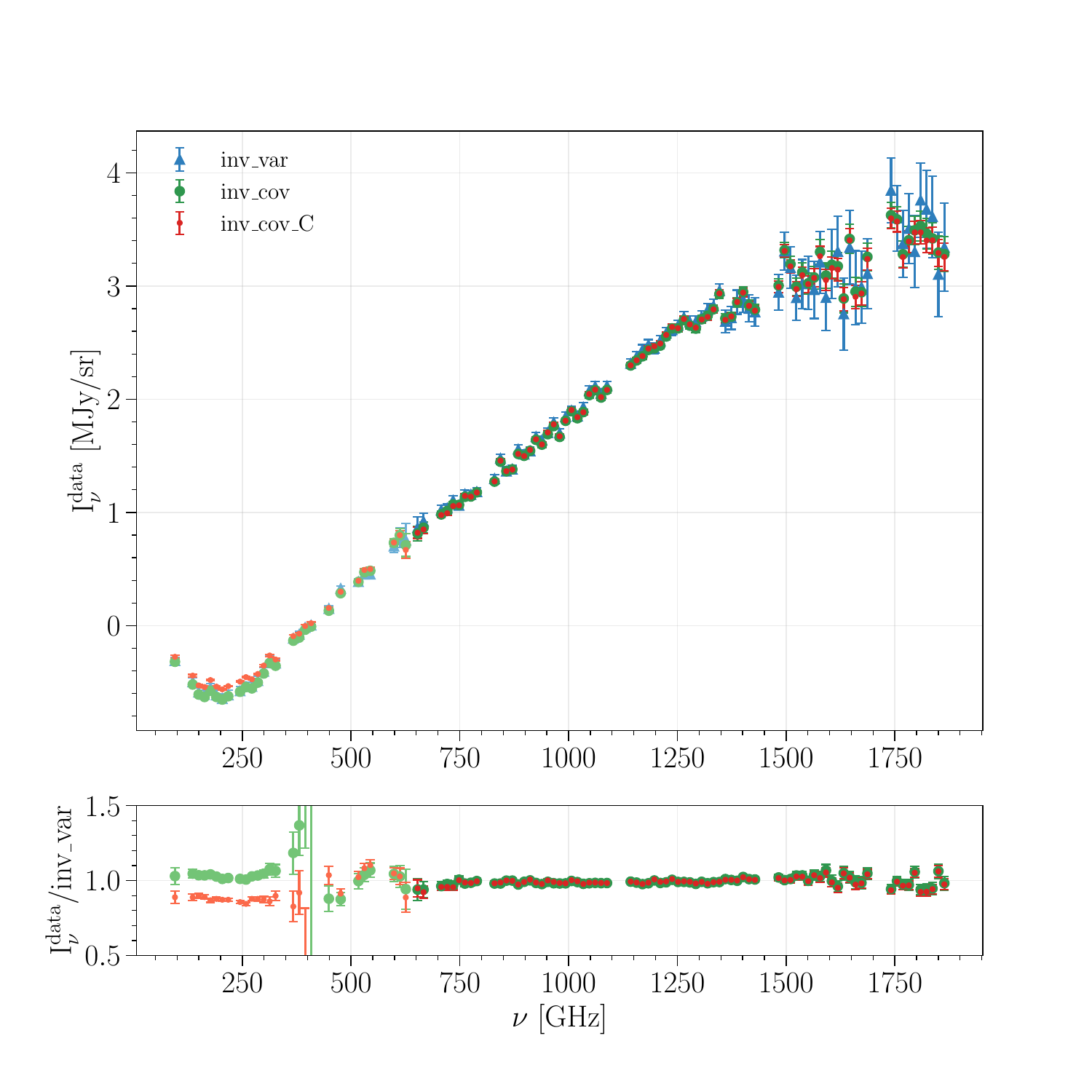}
    \cprotect\caption{\textit{Upper panel:} \textit{FIRAS} frequency monopoles computed using three different weighting approaches in the pixel averaging: \verb|inv_cov|, \verb|inv_var|, and \verb|inv_cov_C| (see Sec.~\ref{sec:monopole}) with a blackbody spectrum at $T_{0}=2.7255$ K subtracted. The plot shows all frequencies up to $\approx 1.9$ THz, although our fiducial analysis is restricted to $\nu<800$~GHz. The transition between the low- and high-frequency instruments occurs at $\approx600$ GHz. Some of the discontinuities observed in the data are due to the removal of frequency channels that may be contaminated by Galactic emission lines, as described in Sec.~\ref{sec:freq_ranges} and Appendix~\ref{app:emission}.
    \textit{Lower panel:} The ratio between our fiducial averaging approach to obtain the frequency monopoles (\verb|inv_var|) and the other two methods.}
    \label{fig:monopoles}
\end{figure}

Using the pixelized \textit{FIRAS} maps, we consider three approaches for computing the sky-averaged monopole of each frequency band in the data:
\begin{itemize}
    \item Inverse-covariance weighting (\verb|inv_cov|), in which we weight the pixels using the full pixel-pixel covariance of each frequency channel (i.e., $\mathbb{C}_{\nu p\nu p'}$ in Eq.~\eqref{eq:covariance}) and mask the unused pixels after matrix inversion.
    \item Inverse-variance weighting (\verb|inv_var|), in which we use only the diagonal part of the full covariance in the pixel weighting and handle the masked pixels as in the \verb|inv_cov| case. 
    \item Inverse-covariance-C weighting (\verb|inv_cov_C|), in which we use only the instrumental noise component of the covariance (i.e., the $C_{\nu \nu^\prime}\delta_{p p^{\prime}} / N_{p}$ term  in Eq.~\eqref{eq:covariance}) in the pixel weighting and handle the masked pixels as in the \verb|inv_cov| case.  
\end{itemize}
The \verb|inv_cov| method accounts for all the error terms in the averaging,  \verb|inv_var| excludes the pixel-pixel correlations (i.e., it assumes $p=p'$ in Eq.~\eqref{eq:covariance}), while \verb|inv_cov_C| only includes the effect of the noise inhomogeneity (described by the $C$ term and by the pixel weight in Eq.~\eqref{eq:covariance}). We show in Fig.~\ref{fig:monopoles} the values of the monopoles for each frequency obtained with these different methods. In particular, we show the residual emission after we have subtracted a blackbody spectrum at $T_{0}=2.7255$ K and we include only the channels that we use in the \textit{frequency monopole} analysis as outlined further in Sec.~\ref{sec:freq_ranges}. These three different methods allow us to test the impact and reliability of each uncertainty component of the full \textit{FIRAS} covariance. 

\subsubsection{Pixel-by-pixel}\label{sec:pix-by-pix}
The \textit{pixel-by-pixel} approach follows closely the analyses in \citetalias{BianchiniFabbian2022} and our companion paper, Ref.~\cite{Fabbian2023}. We refer the reader to those papers for more extensive discussions on the method. In the \textit{pixel-by-pixel} approach, we fit a sky model to the frequency spectrum in each pixel independently. We include all the systematic uncertainty terms of the \textit{FIRAS} covariance, i.e., taking the $p=p^\prime$ limit in Eq.~\eqref{eq:covariance}. The fitting procedure outputs a map for each free parameter of the model (e.g., the value of $y$ in each pixel) and then we simply compute the average across the map. We remove pixels for which the reduced $\chi^2$ value of the fit gives a PTE $<10^{-4}$ from the averaging operation.  However, we note that this cut has no impact on our final results for $\ymono$. This analysis approach is sub-optimal as it does not fit all the pixels at the same time, but it is computationally more tractable as it avoids the inversion of the full covariance matrix, which contains $\mathcal{O}(10^7)$ elements. 

Since the \textit{FIRAS} measurements are spatially correlated, as is evident from Eq.~\eqref{eq:covariance}, we need to account for the correlation between the estimates obtained in each sky pixel when computing the monopole of the map. The baseline model that we fit to the data is non-linear. Therefore, we estimate the pixel-pixel correlation from the covariance of the chains used in the MCMC fits and then use this to compute an inverse-covariance-weighted average for a given sky component. This approach differs from \citetalias{BianchiniFabbian2022}, which assumed the native \textit{FIRAS} pixel-pixel correlation $P_{pp^\prime}$ when computing the monopole of the $\mu$-distortion map obtained after component separation. The majority of the foreground models investigated in that work are effectively linear, while here we consider more complex models involving more foreground parameters describing non-linear degrees of freedom. We find that this choice does not impact the published results in \citetalias{BianchiniFabbian2022}; further validation tests that we perform using the mock data generated in this work are discussed in Appendix~\ref{sec:mu}. 

We believe our estimate of the pixel-pixel covariance from the MCMC chains provides a better representation of the correlation between the measurements in the presence of a non-linear sky model as it is derived self-consistently from the data. Because of the limited signal-to-noise in each pixel, if the estimates of the $y$-distortion parameter are correlated by the common underlying systematics between the neighboring or distant pixels, this approach should be able to recover the correlations without making any specific assumptions about the systematics. 

As we discuss further below, some of the results from the \textit{frequency monopole} method seem to suggest that the pixel-pixel correlation in the \textit{FIRAS} model is not properly modeled or it does not accurately match the correlation in the data. Therefore, for the sake of completeness we compare the results obtained for the monopole constraints with our baseline approach and also with the results obtained assuming different models of correlations between pixels in the $y$-parameter map obtained from the pixel-by-pixel method. Such models either assume the same pixel-pixel correlation as the original \textit{FIRAS} data or no correlation at all.

\subsubsection{Frequency Ranges}\label{sec:freq_ranges}
In our main analysis, we consider both the low- and high-frequency \textit{FIRAS} data. In the \textit{frequency monopole} method, we remove some frequency channels as described below, while in the \textit{pixel-by-pixel} fitting, we keep most of the frequency channels, since removing them has little effect on the results. We only drop the lowest two frequencies for the \textit{pixel-by-pixel} method to be consistent with choices made for the \textit{frequency monopole} analysis. The frequency bands above 640 GHz are included in the high-frequency instrument of \textit{FIRAS} but the channels between $600\,{\rm GHz}<\nu\leq 640\,{\rm GHz}$ are present in both the high- and low-frequency instruments. For the \textit{pixel-by-pixel} method, we decide to keep the channels having the lower noise level. For the \textit{frequency monopole} method, we discard the overlapping channels as described below.

We remove the highest frequency channel in the low-frequency instrument data and the lowest three frequency channels in the high-frequency instrument data due to their overlap and the systematics introduced by the presence of a beam-splitter in the optical chain of \textit{FIRAS} \cite{firas_supp}. 
Additionally, we remove the two lowest frequency channels in the low-frequency instrument data due to possible systematic errors in the lowest bands observed by \textit{FIRAS}, a Fourier transform spectrometer (FTS), which operated close to its band-limit (\cite{privcomm}, \citetalias{Abitbol2017}). Finally, we remove the channels that may be contaminated by Galactic plane emission lines based on the lines detected in Ref.~\cite{Fixsen1999_lines}\footnote{\url{https://lambda.gsfc.nasa.gov/product/cobe/firas_linemaps.html}} (see Table \ref{tab:emission_lines}). We do so by discarding frequencies that fall into the same frequency bin as an emission line, conservatively assuming that it could be broadened by $\pm1\%$. This leaves us with 27 low- and 9 high-frequency channels in our baseline analysis, which considers either only the low-frequency data ($\nu_{600}$) or the data up to $\approx$ 800 GHz ($\nu_{800}$):

\begin{itemize}
    \item low-frequency instrument data only: 27 channels, $\approx 95-626$ GHz ($\nu_{600}$)
    \item low- and high-frequency instrument data up to 800 GHz: 36 channels, $\approx 95-626$ GHz and $\approx 653-789$ GHz ($\nu_{800}$)
\end{itemize}

We include a plot showing the frequencies used in the \textit{frequency monopole} analysis and those that were removed in Appendix \ref{app:emission}.

\subsection{Sky model}\label{sec:sky_models}
We adopt the sky model used in \citetalias{Abitbol2017}\footnote{\url{https://github.com/asabyr/sd_foregrounds/tree/firas} (modified version of \url{https://github.com/mabitbol/sd_foregrounds})}, who used this model to forecast spectral distortion constraints from upcoming satellite experiments via a Fisher matrix approach.  We model the total sky emission after subtracting a reference blackbody spectrum at $T_{0}=2.7255$ K \cite{Fixsen2009}, $I^{\rm sky}_{\nu}$, as the sum of a blackbody temperature deviation ($\Delta B_{\nu}$), a Compton-$y$ distortion ($I^{y}_{\nu}$), and foreground emission ($I_{\nu}^{\rm fg}$):
\begin{equation}
\label{eq:totsky}
    I_{\nu}^{\rm sky}=\Delta B_{\nu} + I^{y}_{\nu} + I_{\nu}^{\rm fg}, 
\end{equation}
\noindent where $\Delta B_{\nu}$ represents the deviation of the CMB blackbody spectrum at $T_{\rm CMB}$ from the assumed blackbody at the reference temperature, $T_{0}$:
\begin{equation}
\label{eq:Bnu}
    \Delta B_{\nu}=\frac{\partial B_{\nu}}{\partial T} \bigg\rvert_{T=T_0} \Delta T = \frac{I_0}{T_0}\frac{x^{4}e^{x}}{(e^{x}-1)^{2}}\Delta T.
\end{equation}
\noindent Here, $I_0=(2h/c^{2})(k_{\rm B}T_{0}/h)^{3}\approx270$ MJy/sr, $x=(h\nu)/(k_{\rm B}T_{0})$, $\Delta T=T_{\rm CMB}-T_{0}$ is a free parameter to be fit from the data, $h$ is Planck's constant, $c$ is the speed of light, and $k_{\rm B}$ is Boltzmann's constant. This term captures any deviation in the true CMB temperature from that of the reference blackbody, as well as the signal from the CMB dipole, which can leak into our estimate of the monopole due to the non-trivial mask applied to the maps.  Fig.~\ref{fig:CMB_dT} in Appendix~\ref{app:dT} shows that the inferred value of $\Delta T$ varies for the different sky masks considered in this work, which is expected because the CMB dipole contribution is different in each case.

In Eq.~\eqref{eq:totsky}, $I^{y}_{\nu}$ is the sky-averaged $y$-distortion, which has the following spectral dependence \cite{zeldovich1969, sunyaev1970relic}:
\begin{equation}
\label{eq:ydist}
I^{y}_{\nu}=I_{0}\frac{x^{4}e^{x}}{(e^{x}-1)^{2}}\left[x\coth\left(\frac{x}{2}\right)-4\right] \ymono,
\end{equation}
where $\ymono$ is the free parameter characterizing the amplitude of the distortion, to be inferred from the data.  The value of $\ymono$ is expected to be on the order of $10^{-6}$ based on halo-model predictions of the dominant contribution from the ICM \cite{Hill2015,Thiele2022}.  It is proportional to the integrated electron pressure along the line-of-sight and, thus, quantifies the total thermal energy stored in the electrons, mainly in galaxy groups and clusters. 

In our baseline analysis, we consider dust as the primary foreground component, since at high frequencies ($\gtrsim 100$ GHz), the sky signal is dominated by the thermal dust emission both of Galactic and extragalactic origin. Dust grains within our Galaxy absorb UV radiation from stars and re-emit it in the infrared. Similarly, the cumulative dust emission from star-forming galaxies in the Universe gives rise to the cosmic infrared background (CIB). Dust emission can be well described by an MBB spectrum up to $\approx 900$ GHz \cite{Planck2014FG,Planck2016FG}. Since we do not extend our main analysis to frequencies higher than $\sim800$ GHz, we adopt the simple MBB model with three free parameters (amplitude $A_{\rm d}$, dust temperature $T_{\rm d}$, and spectral index $\beta_{\rm d}$)\footnote{Note that here we use a fixed value for $\nu_{0}$ compared to the fiducial \citetalias{Abitbol2017} set-up.}:
\begin{equation}
\begin{split}
    I^{\rm dust}_{\nu}=A_{\rm d}\left(\frac{\nu}{\nu_{0}}\right)^{\beta_{\rm d}+3}\frac{1}{e^{x_{\rm d}}-1}\\
    \mathrm{where} \quad x_{\rm d}=\frac{h\nu}{k_{\rm B}T_{\rm d}}\quad\mathrm{and}\quad \nu_{0}=353\, \mathrm{GHz}. 
\end{split}
\end{equation}
We do not make any assumptions on the origin of the dust emission in our baseline model fits. Therefore, this simple MBB SED is expected to capture both the CIB and the residual Galactic dust emission that remains in the maps after applying our sky masks.  We experiment with fitting two MBBs of this form in Sec.~\ref{sec:data_results_monopole} and Appendix~\ref{sec:more_fgs}, where one MBB is intended to fit the Galactic dust and another MBB has a fixed SED ($\beta_{\rm CIB}=1.59, T_{\rm CIB}=11.95$~K) based on the CIB constraints from Ref.~\cite{McCarthyHill2023}.  

In addition to the CMB, the $y$-distortion, and dust, we test fitting additional foreground components in Sec.~\ref{sec:data_results_monopole} and Appendix~\ref{sec:more_fgs}. At the intermediate frequencies used in this analysis, extragalactic CO\footnote{We do not fit for the Galactic CO since for the \textit{frequency monopole} method, we remove the frequency channels near the emission lines.} and Galactic free-free (FF) emission are expected to contribute to the sky brightness (see Fig.~2 in \citetalias{Abitbol2017}), although note that the free-free contribution is expected to be heavily suppressed by masking the Galactic plane, where the emission is mostly concentrated.  Nevertheless, we explore its inclusion in our sky model as a test. Extragalactic CO refers to the integrated emission of rotational carbon monoxide lines from star-forming galaxies in the Universe \cite{Righi2008}. 
Since there is no well-defined SED for this emission, we follow \citetalias{Abitbol2017} and use a CO model template based on spectra from Ref.~\cite{Mashian2016}, with one free overall amplitude parameter that rescales the template, $A_{\rm CO}$:
\begin{equation}
    I_{\nu}^{\rm CO}=A_{\rm CO}\Theta_{\rm CO}(\nu), 
\end{equation}
where $\Theta_{\rm CO}$ is the CO emission template. Thermal free-free (Bremsstrahlung) emission is generated by electron-ion scattering within the Galaxy. Following \citetalias{Abitbol2017}, we use the SED from Ref.~\cite{Draine2011}, which can be written as:
\begin{equation}
\begin{split}
    I^{\rm FF}_{\nu}=A_{\rm FF}\left(1+\ln\left[1+\left(\frac{\nu_{\rm ff}}{\nu}\right)^{\sqrt{3}/\pi}\right]\right)\\ \mathrm{with}\quad
    \nu_{\rm ff}=\nu_{\rm FF}(T_{\rm e}/10^{3} \, \rm K)^{3/2} \\ \mathrm{and}\quad T_{\rm e}=7000 \,\mathrm{K}, \quad \nu_{\rm FF}=255.33 \,\mathrm{GHz}.
\end{split}
\end{equation}
This model includes one free parameter (the amplitude $A_{\rm FF}$), since at the \textit{FIRAS} frequencies the spectrum does not strongly depend on the electron temperature $T_{e}$. Finally, at low frequencies, we also consider the inclusion of synchrotron radiation emitted by cosmic-ray electrons accelerated by Galactic magnetic fields, since in this regime this component is expected to dominate. At the \textit{FIRAS} frequencies ($\gtrsim 60$ GHz), the synchrotron spectrum is well approximated by a power-law, since flattening of the spectrum has only been shown at lower frequencies \cite{Planck2016FG}. Therefore, we model the synchrotron emission as a power-law with the amplitude, $A_{\rm s}$, and the spectral index, $\beta_{\rm s}$, as free parameters:
\begin{equation}
    I^{\rm synch}_{\nu}=A_{\rm s}\left(\frac{\nu}{\nu_{0}}\right)^{\beta_{\rm s}}
\end{equation}

\noindent where $\nu_{0}=100$ GHz and the spectral index is expected to be $\beta_{\rm s}\approx -1$ in intensity units. 

\subsection{Inference}\label{sec:inference}

Both of the analysis approaches described above fit, in some way, a parametric sky model to a multi-frequency data set (the sky spectra for the \textit{pixel-by-pixel} and the sky monopoles in different frequencies for the \textit{frequency monopole} method). We infer the parameters $\vec{\theta}$ of our sky model using a standard Bayesian approach, sampling a Gaussian likelihood defined as 
\begin{equation}
    \ln{\mathcal{L}}(I_{\nu}^{\rm FIRAS}|\vec{\theta})=-\frac{1}{2}\Delta_{\nu}^{T}(\vec{\theta})\hat{C}_{\nu\nu'}^{-1}\Delta_{\nu'}(\vec{\theta})
\end{equation}
\noindent with a Markov Chain Monte Carlo (MCMC) sampler. Here, $\Delta_{\nu}(\vec{\theta})$ is the difference between the observed data, $I_{\nu}^{\rm FIRAS}$ (sky averaged or individual pixel frequency spectrum), and the sky model, $I_{\nu}^{\rm sky}(\vec{\theta})$; $\hat{C}_{\nu\nu'}$ is the covariance matrix of the data. We use the hat notation to denote that this covariance is different depending on the adopted method.

For the \textit{frequency monopole}, we use the frequency-frequency correlation described by the instrumental noise covariance (C-term) and the errors given by the averaging methods (Sec.~\ref{sec:methods}) to turn this correlation into a covariance matrix. We follow  \citetalias{BianchiniFabbian2022}  in our choice to only use the correlation from the C-term as we observe regions of the sky far from the Galactic plane where correlated calibration errors are subdominant. For the \textit{pixel-by-pixel} case, $\hat{C}_{\nu\nu'} = \mathbb{C}_{\nu p\nu' p'}$ and we perform the fit in every pixel including all the terms of Eq.~\eqref{eq:covariance}.

\begin{figure}[htbp!]
    \centering
    \includegraphics[width=0.5\columnwidth]{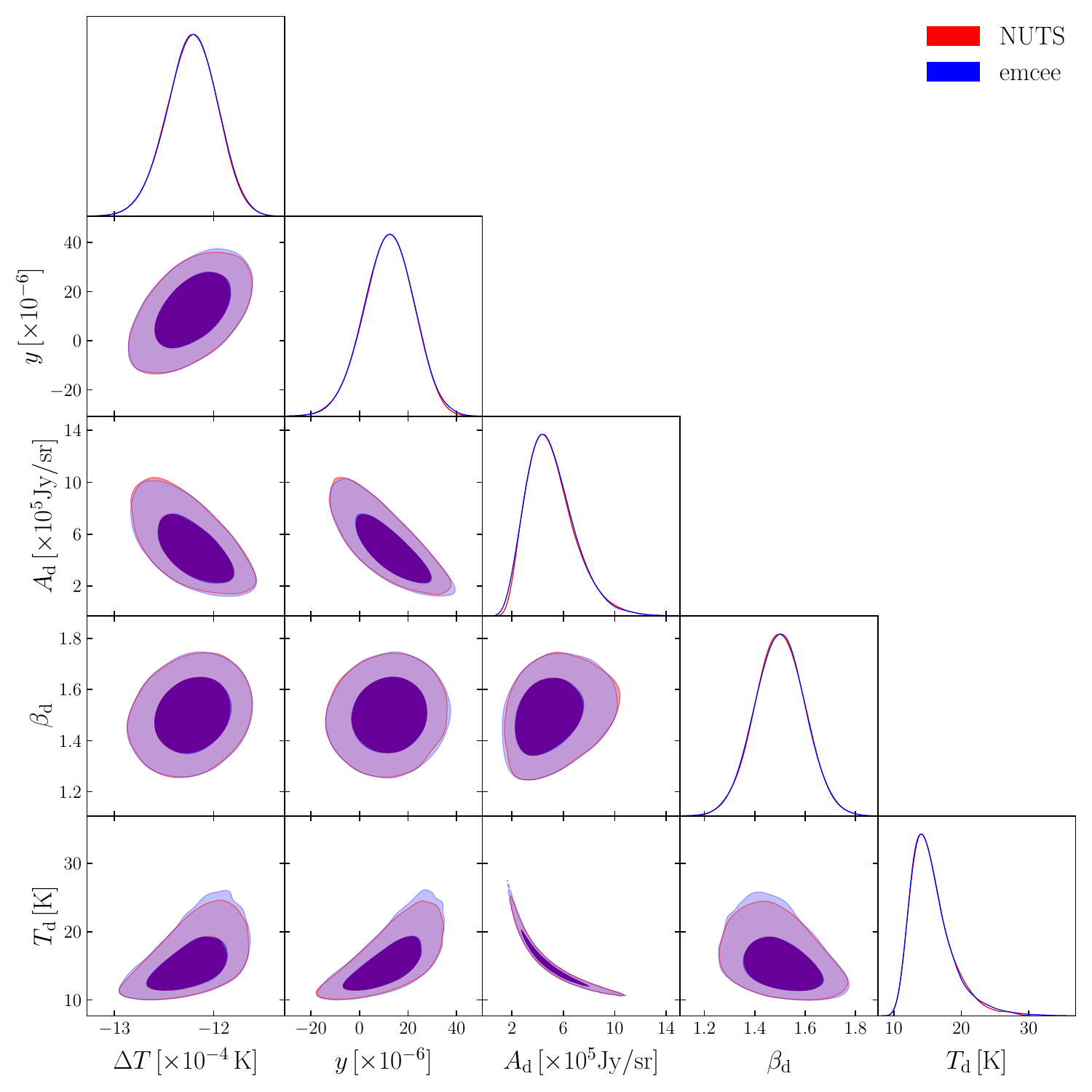}
    \cprotect\caption{Comparison of posteriors between the \verb|NUTS| and \verb|emcee| samplers for our baseline analysis set-up: $\nu_{600}$ and P40, fitting a model that includes dust as the only foreground. The two samplers give consistent results.}
    \label{fig:emcee_vs_NUTs}
\end{figure}

In the \textit{frequency monopole} case, we use two MCMC samplers: the No-U-Turn Sampler (\verb|NUTS|) implemented in \verb|numpyro| \cite{numpyro2019composable,numpyro2019pyro} and the affine-invariant ensemble sampler implemented in \verb|emcee|~\cite{emcee2013}.  For the \verb|NUTS| sampler, we run 10 chains for 10,000 steps (+10,000 steps for burn-in) at a tree depth of 10-25 depending on the data and model. We use the effective sample size (at least 100 independent samples per chain) and rank normalized split $\hat{R}$ \cite{Rhat2021} statistic ($< 1.01)$ via the \verb|ArviZ| \cite{arviz_2019} library to check that our runs have converged. For the \verb|NUTS| sampler, we are required to sample each parameter from a probability distribution. Therefore, we additionally impose wide uniform priors on $\Delta T \in [-10^{8},10^{8})/a\approx[-0.37,0.37)$, $y \in [-10^{6}, 10^{6})/a\approx[-0.0037,0.0037)$, $A_{\rm CO}\in[-1000,1000]$, and $A_{\rm s}, A_{\rm FF}, A_{\rm CIB}$, $A_{\rm d} \in [-10^{8},10^{8})$, where $a\approx2.7\times10^{8}$.\footnote{We use a scaled version of $\Delta T$ and $\ymono$ in the sampling to reduce dynamic range.} For \verb|emcee| runs, we use the integrated auto-correlation time of the chains as our convergence diagnostic (total number of samples $> 50\times$maximum auto-correlation length). We use 100 walkers and 100,000 or 500,000 steps. For both samplers, we discard the first 30$\%$ of the samples as burn-in.\cprotect\footnote{For the \verb|NUTS| sampler, we discard $30\%$ from post-burn-in samples.} We use \verb|emcee| to run some tests on mocks and check the stability of our models, but since we find that convergence requires longer computation time, we use \verb|NUTS| for the flat prior runs on mocks, results run on data, and all the final data posteriors shown in this paper. We check that the two samplers give consistent results (e.g., see Fig.~\ref{fig:emcee_vs_NUTs}). In the \textit{pixel-by-pixel} analysis we only use \verb|emcee| with 25 walkers and 300,000 samples per walker, and assess the convergence of the chains in the same way.

\subsection{Sky Mask Consistency Criterion}
To assess the robustness of our component separation approaches as a function of the sky fraction, we test the consistency of the $\ymono$ estimates across the sky masks. If the foregrounds are removed successfully and the differences in the estimates of $\ymono$ from two different Galactic masks are only due to the noise and sample variance fluctuations, the covariance of such differences is simply the difference of the covariances of the estimates. This follows from the fact that all the estimates are effectively obtained from nested data sets  \cite{nested-masks}, e.g., pixels included in the estimates using the P40 mask are a subset of the pixels included in the P60 mask, and so forth. As such, we can set a consistency criterion, which we call $C_{\rm mask}$, between the results obtained from two different nested masks: 
\begin{equation}
C_{\rm mask}=\frac{\left|\ymono^{\rm mask\,1}-\ymono^{\rm mask\,2}\right|}{\sqrt{\left|\sigma_{\ymono}^{2, \rm mask\,1}-\sigma_{\ymono}^{2, \rm mask\,2}\right|}}\lesssim 2.
\label{eq:consistency-test}
\end{equation}

\section{Results from Mocks}\label{sec:mock_results}
\subsection{\textit{Frequency Monopole}}

\begin{figure}
    \centering
    \includegraphics[width=0.75\columnwidth]{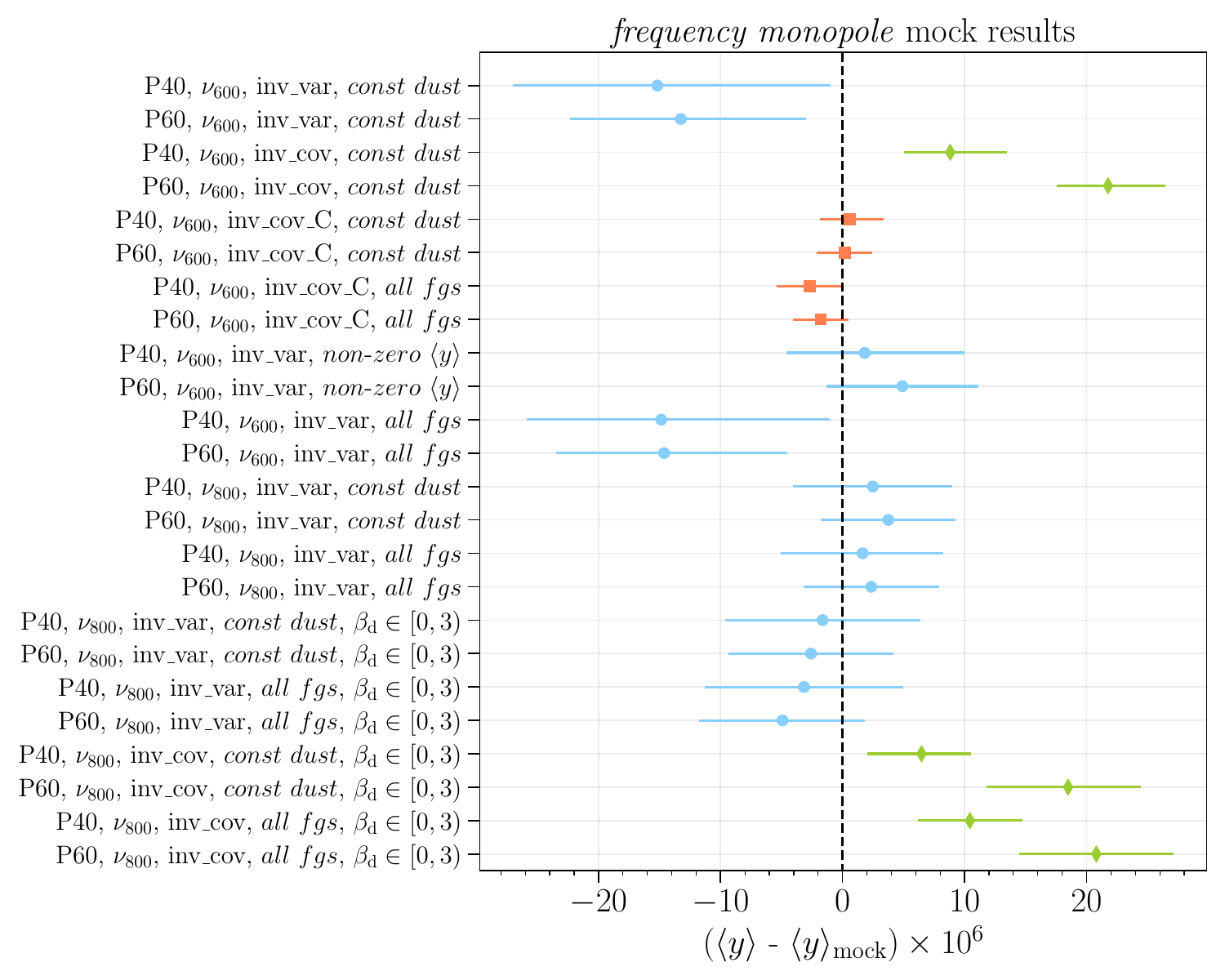}
    \cprotect \caption{Summary of the results on mocks from the \textit{frequency monopole}. We show the error bars obtained from \verb|inv_var| (blue), \verb|inv_cov| (green), and \verb|inv_cov_C| (orange) on mocks with the dust SED parameters fixed across pixels (\textit{const.~dust}), mocks that contain dust with fixed SED parameters across pixels and a non-zero $\ymono$ component (\textit{non-zero $\ymono$}), and mocks that include several Galactic and extragalactic foregrounds (\textit{all fgs}) (details of the simulated data are described in Sec.~\ref{sec:simulated_data_sets}). For the dust model, we use a flat prior on $T_{\rm d}$ and a Gaussian prior on $\beta_{\rm d}$ for the first 16 cases and a flat prior on $\beta_{\rm d}$ in the last 8 cases shown.}
    \label{fig:summary_mock_monopole}
\end{figure}

In order to assess different methods for averaging over pixels to generate the \textit{frequency monopole} spectrum (Sec.~\ref{sec:monopole}), we perform tests on mock data (Sec.~\ref{sec:simulated_data_sets}). Unless otherwise stated, we fit a sky model with a blackbody temperature deviation, $y$-distortion, and dust with a flat prior on $T_{\rm d}\in [0,100)$~K or $T_{\rm d}\in[0,100]$~K \cprotect \footnote{We note that for the \verb|emcee| sampler, we use priors with both lower and upper interval inclusive, whereas \verb|NUTS| samples from an interval that excludes the upper bound. Since the two samplers are consistent as shown in Fig.~\ref{fig:emcee_vs_NUTs}, we do not expect this choice to have any significant effects.} and a Gaussian prior on $\beta_{\rm d}\in \mathcal{N} (1.51, 0.1)$ in the tests described below (this is the baseline set-up we apply to data --- see rows 1-2 in Table \ref{tab:models}). The results from the tests on mocks using the \textit{frequency monopole} spectrum are summarized in Fig.~\ref{fig:summary_mock_monopole} and Table~\ref{tab:mock_results_monopole}.

We first perform tests on simple {\it constant dust} mocks that contain CMB signal and a dust component whose SED amplitude varies across pixels, but the $T_{\rm d}$ and $\beta_{\rm d}$ values are fixed across pixels. Since these maps do not contain any $y$-distortion and include a very simple dust signal, we expect to recover a constraint on $\ymono$ that is consistent with zero. The first four points in Fig.~\ref{fig:summary_mock_monopole} show that the \verb|inv_var| (blue circle) method gives stable results for different sky fractions, while the \verb|inv_cov| (green diamond) results in a $\sim5.1\sigma$ bias using the P60 mask.

The mocks do not contain any systematic effects that correlate pixels.  Thus, any weighting of pixels should lead to unbiased results but with different levels of optimality (i.e., different error bars). If the full covariance correctly models the systematics, the \verb|inv_cov| weighting should minimize their effects. The majority of the pixel-pixel correlation is introduced by the part of the covariance matrix related to the destriper errors (the $\beta$ vectors in Sec.~\ref{eq:covariance}). The other covariance components that include correlation between different pixels are the JCJ and PEP instrument gain terms that are subdominant and only become more important in the Galactic plane pixels, which are masked out in the computation of $\ymono$. The correlated destriper errors are modeled from just a handful of templates estimated from the data itself. Therefore, we speculate that the matrix does not precisely describe this correlation for pixels with low signal-to-noise and the matrix itself might not be well conditioned. This could lead to the bias we see with the \verb|inv_cov| method, even in the simplest case of a \textit{constant dust} mock.

To test our hypothesis that the systematic errors in the covariance are not well modeled, we consider the \verb|inv_cov_C| method, which averages pixels accounting only for the different instrumental noise in each pixel and no systematic effects. Indeed, similar to the case of the \verb|inv_var| averaging, we find that this gives stable results both for the simplest \textit{constant dust} case and the most complex \textit{all foregrounds} mock (dust, synchrotron, CIB, CO, AME, FF foregrounds), as shown in Fig.~\ref{fig:summary_mock_monopole} (orange square). In our analysis, we choose to adopt the \verb|inv_var| averaging method for the \textit{frequency monopole}, since it includes all systematic errors in a single pixel as described in Sec.~\ref{sec:methods}. Systematics, in fact, represent a significant fraction of the uncertainties, as can be seen in the significantly smaller error bars using \verb|inv_cov_C| (orange square) in comparison with the larger error bars from \verb|inv_var| (blue circle).

In the tests on mocks, we do not include results for the P20 mask because at this sky fraction, the chains do not fully converge with the \verb|inv_var| method according to our metrics (this is likely due to the low signal on such a small sky fraction). Relaxing our convergence criteria and considering the results of these chains even if they are not perfectly converged, we obtain a constraint on $\ymono$ from P20 that is still consistent with zero within $1\sigma$. Using low frequencies alone, we find that the dust amplitude is consistent with zero within $\sim1-2\sigma$ even for P40/P60. We note that in the \verb|inv_var| runs, we find a preferred T$_{\rm d}$ that is lower than the input in the the \textit{constant dust} mock (e.g., the posterior mean T$_{\rm d} = 11.4^{+0.6}_{-4.2}$~K for P40). This suggests that using low frequencies alone does not provide enough sensitivity to constrain the spectral shape of the dust component, even with a Gaussian prior on one of the dust SED parameters.

We further test the \verb|inv_var| method on mocks with a non-zero $y$-distortion ($\ymono=20\times10^{-6}$) and with additional sky components.  We recover $\ymono$ consistent with the input value in the simulated data (see error bars in points 9-12 in Fig.~\ref{fig:summary_mock_monopole}). 

Similarly, we test the \verb|inv_var| method using frequencies up to 800 GHz. Points 13-16 in Fig.~\ref{fig:summary_mock_monopole} show the results for the simplest and the most complex mock cases using frequencies up to 800 GHz and fitting the same sky model, which includes dust as the only foreground component. We find that we can recover $\ymono$ consistent with zero and that including higher frequencies in our MCMC runs leads to a higher mean value of $T_{\rm d}$ from the \textit{constant dust} mocks, which is consistent with the input simulation value of 20.88~K within $1\sigma$. 

\begin{table}[h]
\begin{tabular}{|lc|}
\colrule
Analysis set-up & $\ymono$ $\times 10^{6}$ \\
\bottomrule
 \rowcolor{gray!20}\multicolumn{2}{|c|}{{\it Constant dust}}                    \\ 
\toprule
$^{\dagger}$ P40, $\nu_{600}$, \verb|inv_var| & $-15.2^{+14.2}_{-11.8}$ \\ 
P60, $\nu_{600}$, \verb|inv_var|  & $-13.3^{+10.3}_{-9.1}$ \\ 
P40, $\nu_{600}$, \verb|inv_cov|  & $8.8^{+4.6}_{-3.8}$ \\ 
P60, $\nu_{600}$, \verb|inv_cov|  & $21.8^{+4.7}_{-4.2}$ \\ 
P40, $\nu_{600}$, \verb|inv_cov_C|  & $0.6^{+2.8}_{-2.4}$ \\ 
P60, $\nu_{600}$, \verb|inv_cov_C|  & $0.2\pm2.3$\\ 
\colrule
$^{\dagger}$ P40, $\nu_{600}$, \verb|inv_var|, $\ymono\times 10^6=20$  & $21.8^{+8.2}_{-6.4}$\\ 
P60, $\nu_{600}$, \verb|inv_var|, $\ymono\times 10^6=20$  & $24.9\pm6.2$\\ 
\colrule
P40, $\nu_{\rm 800}$, \verb|inv_var| & $2.5\pm6.5$ \\ 
P60, $\nu_{\rm 800}$, \verb|inv_var|& $3.8\pm5.5$ \\
\colrule
$^{*}$ P40, $\nu_{\rm 600}$, \verb|inv_var|, $\beta_{\rm d}\in[0, 3)$& $-19.6_{-12.0}^{+14.1}$ \\ 
$^{*}$ P60, $\nu_{\rm 600}$, \verb|inv_var|, $\beta_{\rm d}\in[0, 3)$& $-18.6_{-11.3}^{+10.6}$\\ 
\colrule
$^{*}$ P40, $\nu_{\rm 600}$, \verb|inv_cov|, $\beta_{\rm d}\in[0, 3)$&  $7.1_{-5.0}^{+5.7}$\\ 
$^{*}$ P60, $\nu_{\rm 600}$, \verb|inv_cov|, $\beta_{\rm d}\in[0, 3)$&  $22.4^{+6.0}_{-5.1}$\\ 
\colrule
P40, $\nu_{\rm 800}$, \verb|inv_var|, $\beta_{\rm d}\in[0, 3)$& $-1.6\pm 8.0$ \\ 
P60, $\nu_{\rm 800}$, \verb|inv_var|, $\beta_{\rm d}\in[0, 3)$& $-2.6\pm 6.7$\\ 
\colrule
P40, $\nu_{\rm 800}$, \verb|inv_cov|, $\beta_{\rm d}\in[0, 3)$&  $6.5_{-4.4}^{+4.0}$\\ 
P60, $\nu_{\rm 800}$, \verb|inv_cov|, $\beta_{\rm d}\in[0, 3)$&  $18.5^{+5.9}_{-6.7}$\\ 
\bottomrule
 \rowcolor{gray!20}\multicolumn{2}{|c|}{{\it All foregrounds}}                    \\ 
\colrule
$^{\dagger}$ P40, $\nu_{600}$, \verb|inv_var|  & $-14.9^{+13.8}_{-11.0}$\\ 
P60, $\nu_{600}$, \verb|inv_var|  & $-14.6^{+10.1}_{-8.8}$\\ 
P40, $\nu_{600}$, \verb|inv_cov_C|   & $-2.7\pm2.7$\\ 
P60, $\nu_{600}$, \verb|inv_cov_C|   & $-1.8\pm2.3$\\
\colrule
P40, $\nu_{\rm 800}$, \verb|inv_var|& $1.6^{+6.6}_{-6.7}$ \\ 
P60, $\nu_{\rm 800}$, \verb|inv_var|& $2.4^{+5.6}_{-5.5}$ \\ 
\colrule
$^{*}$ P40, $\nu_{\rm 600}$, \verb|inv_var|, $\beta_{\rm d}\in[0, 3)$&  $-18.4_{-12.4}^{+13.7}$\\ 
$^{*}$ P60, $\nu_{\rm 600}$, \verb|inv_var|, $\beta_{\rm d}\in[0, 3)$& $-19.0_{-11.0}^{+11.1}$\\ 
\colrule
$^{*}$ P40, $\nu_{\rm 600}$, \verb|inv_cov|, $\beta_{\rm d}\in[0, 3)$& $12.1_{-4.2}^{+5.7}$\\ 
P60, $\nu_{\rm 600}$, \verb|inv_cov|, $\beta_{\rm d}\in[0, 3)$& $24.4_{-4.8}^{+5.7}$ \\ 
\colrule
P40, $\nu_{\rm 800}$, \verb|inv_var|, $\beta_{\rm d}\in[0, 3)$&  $-3.2_{-8.2}^{+8.1}$\\ 
P60, $\nu_{\rm 800}$, \verb|inv_var|, $\beta_{\rm d}\in[0, 3)$& $-4.9\pm 6.8$\\ 
\colrule
P40, $\nu_{\rm 800}$, \verb|inv_cov|, $\beta_{\rm d}\in[0, 3)$& $10.4\pm4.3$ \\ 
P60, $\nu_{\rm 800}$, \verb|inv_cov|, $\beta_{\rm d}\in[0, 3)$& $20.8\pm 6.3$ \\ 
\colrule
\end{tabular}
\cprotect\caption{\label{tab:mock_results_monopole} Results from mocks using the \textit{frequency monopole} (also shown in Fig.~\ref{fig:summary_mock_monopole}). We use a flat prior for $T_{\rm d}$ and a Gaussian prior for $\beta_{\rm d}$ in our baseline model. We also show the results when using a flat prior for $\beta_{\rm d}$. Set-ups denoted with $^{*}$ are not fully converged according to the requirements set in Sec.~\ref{sec:inference} due to lack of sensitivity to both $T_{\rm d}$ and $\beta_{\rm d}$, when only flat priors are used. The results are still unbiased when using \verb|inv_var| and have $\hat{R} \leq1.1$. The fiducial analysis set-up that we adopt for the \textit{frequency monopole} is denoted with $^{\dagger}$.} 
\end{table}

Finally, we test if the bias in the \verb|inv_cov| results is due to prior effects in addition to or instead of the pixel correlation effects. We do this by checking our results on mocks using flat priors on both $\beta_{\rm d}\in[0,3)$ and $T_{\rm d}\in[0,100)$. In the case of the simplest \textit{constant dust} mocks, the results do not converge well for neither the \verb|inv_var| nor \verb|inv_cov| methods using just the low frequencies. This is not surprising given the expected shape and the frequency extent of the dust SED. Using $\nu_{800}$, the posteriors converge both for P40 and P60 using \verb|inv_var| and \verb|inv_cov|. We find $\ymono$ is consistent with zero for \verb|inv_var| for both sky masks, while for \verb|inv_cov|, $\ymono$ is biased at $\approx 2.8\sigma$ for P60. For the \textit{all foregrounds} mock, we find that for \verb|inv_var| the results again do not converge for low frequencies, but are consistent with zero for $\nu_{\rm 800}$ for P40 and P60. For \verb|inv_cov|, we find a $\sim5.1\sigma$ bias using $\nu_{600}$ for P60 and $\sim2.4/3.3\sigma$ biases for P40/P60 using $\nu_{\rm 800}$. For completeness, we include all the results with flat priors in Table \ref{tab:mock_results_monopole}, but only include results from $\nu_{800}$ in Fig.~\ref{fig:summary_mock_monopole}. For most of the set-ups using $\nu_{600}$, the MCMC runs do not fully converge as indicated by the lower effective sample size for some of the parameters. However, since $\hat{R} \leq 1.1$, we still include these results here. 

We compute the consistency criterion, $C^{\rm mock}_{\rm mask}$ (Eq.~\ref{eq:consistency-test}), for the results on mocks in Table~\ref{tab:mock_results_monopole}. We find that \verb|inv_var| results are always consistent. The \verb|inv_cov_C| results are also all consistent. The \verb|inv_cov| results are not consistent across the masks, with $C^{\rm mock}_{\rm mask}\approx2-10$.

The results using flat priors suggest that in addition to the possible pixel correlation effects that we described earlier, we are also seeing some modeling and prior effects since the biases are alleviated with flat priors on both dust parameters in the case of the \textit{constant dust} mock. In these tests, dust is barely detected and the amplitude is consistent with zero in most cases, so the modeling is unlikely to be improved by simply adding more components. We choose to use the \verb|inv_var| averaging method for the \textit{frequency monopole} data analysis to remain conservative, since the results are stable with this method, independent of priors and across all mocks. In particular, we adopt \verb|inv_var| with a Gaussian prior on $\beta_{\rm d}$, $\nu_{600}$, and the P40 sky mask as our fiducial set-up, but we also report constraints from the set-ups using other sky masks and using $\nu_{800}$ with either a Gaussian prior on $\beta_{\rm d}$ or a flat prior on $\beta_{\rm d}\in[0,3)$. Although \verb|inv_cov_C| also gives stable results, we do not use this averaging method because it does not incorporate systematic errors and, as such, gives artificially smaller error bars (see Fig.~\ref{fig:summary_mock_monopole} and Table~\ref{tab:mock_results_monopole}).

\subsection{Pixel-by-pixel}\label{sec:mock-resuls-pixpix}
Similar to the results presented in the previous subsection, we run tests on mock data for the \textit{pixel-by-pixel} method to assess the trade-offs involved in this approach and the similarities and differences with the \textit{frequency monopole} method. Following the mock results described earlier, we restrict these tests to fitting the simplest sky model that only includes $\Delta T$, $y$, and the dust component (i.e., $A_{\rm d}, \beta_{\rm d}, T_{\rm d}$) as free parameters. Moreover, \citetalias{BianchiniFabbian2022} showed that the low-frequency foregrounds such as synchrotron or free-free emission were measured with less than $2\sigma$ significance using the \textit{pixel-by-pixel} approach, and as such we decide not to investigate them further here. This would in fact add too many degrees of freedom that become quickly degenerate and these components would not be constrained by the data outside of the Galactic plane, where these components have a small amplitude and the signal-to-noise in each pixel is therefore low. Therefore, we limit the comparison and robustness tests to the following:
\begin{itemize}
\item We compare the baseline set-up using $\nu_{600}$ and $\nu_{800}$.  
\item For both of the frequency ranges, we compare the results obtained using different priors on the foreground parameters: assuming flat priors $\beta_{\rm d} \in\left[0,3\right]$ and $T_{\rm d}\in\left(0,100\right]$, or assuming a Gaussian prior on $\beta_{\rm d}\in \mathcal{N}(1.51,0.1)$, consistent with the one adopted in the \textit{frequency monopole} method.
\item We compare the results obtained on simple \textit{constant dust} mocks, as well as results from the \textit{all foregrounds} mocks, for all the variations described in the two bullet points above.
\end{itemize}

\begin{figure}[t]
\centering
\includegraphics[width=0.49\columnwidth]{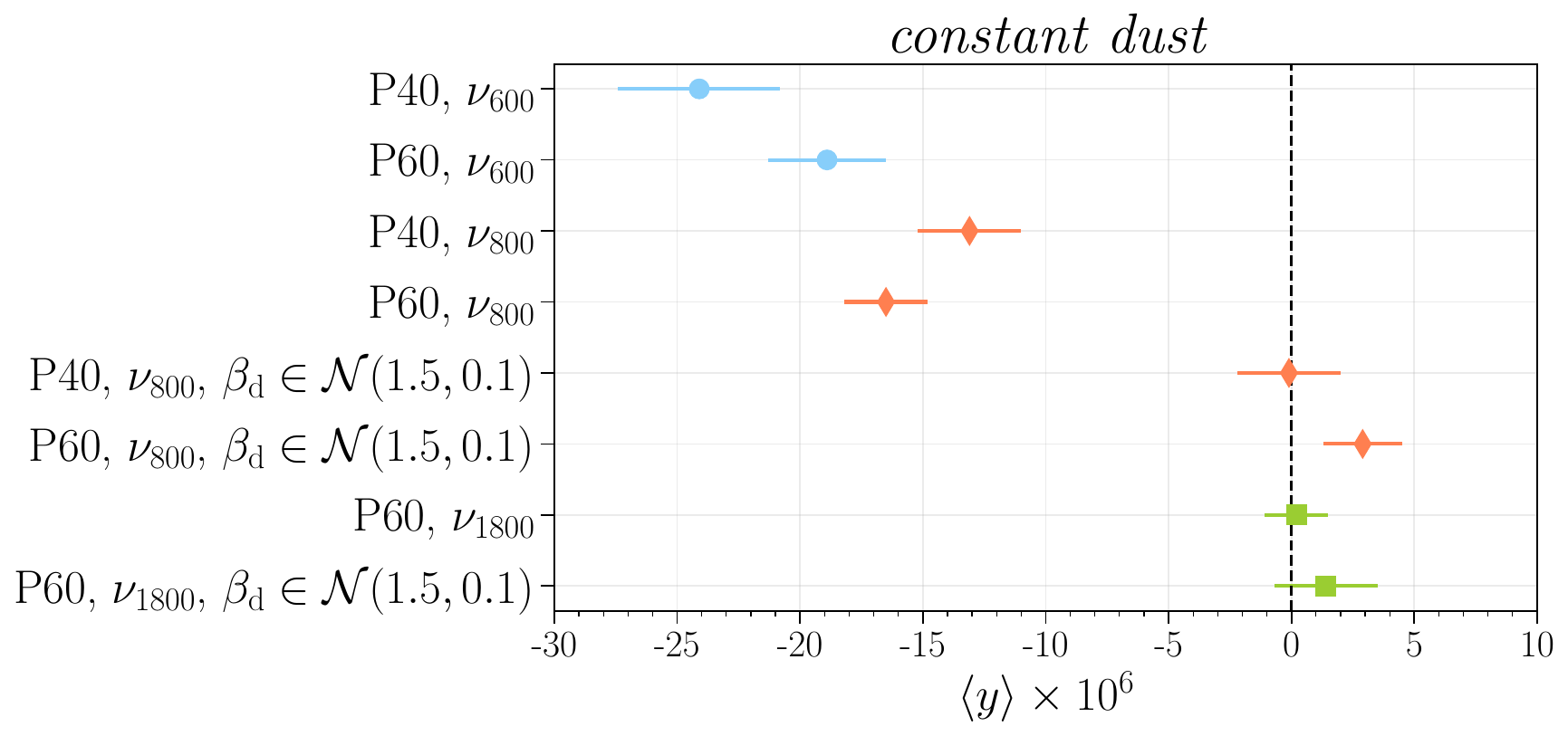}
\includegraphics[width=0.49\columnwidth]{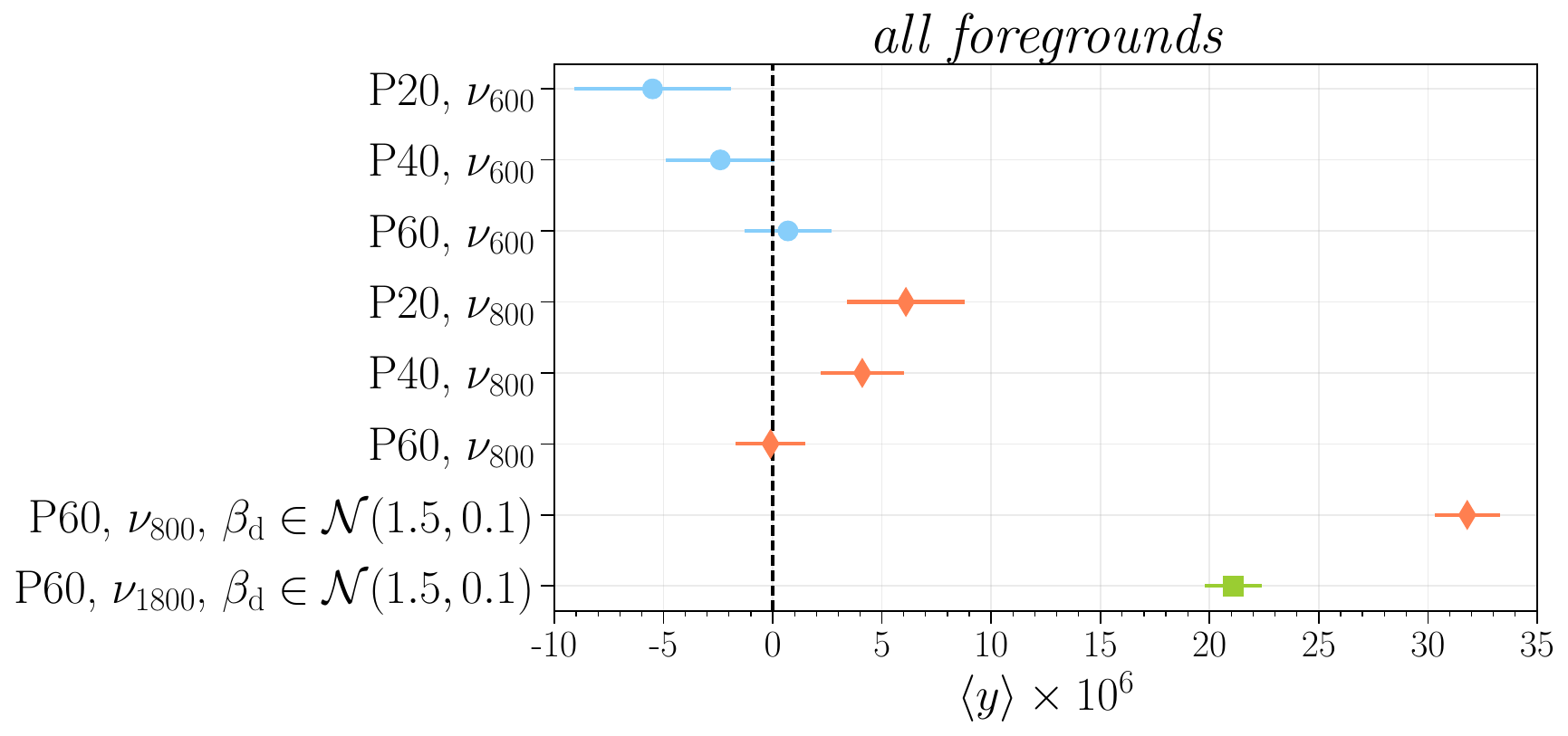}
\caption{Summary of the $\ymono$ results of the \textit{pixel-by-pixel} method for different analysis set-ups and mocks including different types of foregrounds. Different choices for $\nu_{\rm max}$ are shown in different colors and markers. The left panel shows results for mock data including only Galactic dust with constant shape of the SED, while the right panel shows results for mocks that include spatially varying SEDs for Galactic dust, CO, CIB, and low-frequency foregrounds.}
\label{fig:ymono_mock_pixel_recap}
\end{figure}

For our final analysis on the data presented in Ref.~\cite{Fabbian2023}, we investigated additional foreground models, specifically targeting the robustness to the presence of the CIB, but we do not present them here to limit our results to the models directly comparable with the \textit{frequency monopole} method. We present results for mocks having $\ymono=0$ for simplicity but confirm that these results also hold true for non-zero values of $\ymono$. As previously stated, unlike in the \textit{frequency monopole} method, we do not explicitly remove frequencies contaminated by emission lines, as we find this to have no impact on our results, given the fact that they are mainly connected with the Galactic emission and their additional contribution is easily absorbed in the foreground model. For the few cases mentioned in this section where we include data at $\nu>1$ THz, we remove frequencies that are affected by the presence of molecular lines. In particular, the [C II] line is detected with high significance by \textit{FIRAS} and has a non-negligible extragalactic component, being a tracer of star formation (see, e.g., \cite{anderson2022} and discussion therein). The results of this analysis, discussed in the following, are summarized in Table \ref{tab:y_results_mocks_pixpix} and Fig.~\ref{fig:ymono_mock_pixel_recap}.

\subsubsection{Constant dust mocks}
For the \textit{constant dust} mocks, we find that the wide priors for $\beta_{\rm d}$ and $T_{\rm d}$ often drag $\ymono$ to unphysical, negative values. For example, using the P60 mask, we obtain $\ymono = (-18.9 \pm 2.4) \times 10^{-6}$, and we find even larger negative values using the cleanest patches of the sky ($\ymono = (-24.1 \pm 3.3) \times 10^{-6}$ using the P40 sky mask). This is most likely an artifact due to an interplay of several factors, including the discrepancy between the simple sky model of the mock data and the foreground model used in the fitting, the degeneracy between $\beta_{\rm d}$ and $T_{\rm d}$ when using frequencies only up to $\nu_{600}$, and the wide priors. Adding high-frequency information up to $\nu_{\rm 800}$ improves the results somewhat and yields $\ymono = (-13.1 \pm 2.1) \times10^{-6}$ for P40, reducing the bias on $\ymono$ from $\sim 7\sigma$ to $\sim 6\sigma$. Additionally, if we impose a Gaussian prior on $\beta_{\rm d}$ as in the \textit{frequency monopole} case, we retrieve unbiased results. For the P60 (P40) mask, we obtain $\ymono = (2.9 \pm 1.6) \times 10^{-6}$ ($\ymono = (-0.1\pm 2.1) \times10^{-6}$). The results agree with the results of the \textit{frequency monopole} method shown in Fig.~\ref{fig:summary_mock_monopole}, although with smaller error bars. 

As additional consistency checks, we compute $\ymono$ for the following set-ups: (1) assuming that the pixels of the map retain the correlation as originally modeled by the \textit{FIRAS} data through the $\beta$ vectors or (2) that they are completely uncorrelated and, thus, $\ymono$ is computed as a simple inverse-variance-weighted average of the pixel values of the $y$ map. These two approaches can be considered analogous to \verb|inv_cov| and \verb|inv_var|, respectively, in the \textit{frequency monopole} method. While the size of the error bars on $\ymono$ when adopting \verb|inv_cov| is similar, the mean value is biased: $\ymono = (6.3 \pm 1.9) \times 10^{-6}$, despite being obtained from a clean region of the sky using P40.\footnote{The destriping templates and, thus, the destriper-induced error correlations are slightly different in the low- and high-frequency channels of \textit{FIRAS}, as the stripes are not the same between the two data sets. We assume the correlation is dominated by the correlation of the low-frequency instrument as the majority of the channels used in the fit come from that instrument. However, to properly account for the errors in each frequency band, it is necessary to use the exact stripes of the high-frequency instrument when including some of its frequencies in the analysis.} This agrees with the results of the \verb|inv_cov| approach for the \textit{frequency monopole} case and further suggests that the pixel-pixel correlation of \textit{FIRAS} is problematic when non-linear foreground models are fitted. We therefore will not further discuss this additional approach. On the other hand, assuming there is no correlation between pixels allows us to recover unbiased but sub-optimal results, with $\ymono = (2.2 \pm 2.6) \times 10^{-6}$. We note that the error and the central value agree well with the results of the \textit{frequency monopole} method obtained on the same mock and mask as shown in Table \ref{tab:mock_results_monopole}. These findings suggest that the \textit{pixel-by-pixel} method is able to retrieve $\ymono$ estimates that minimize additional variance due to foreground residuals and systematics and, as such, have lower overall uncertainties. We test extending the frequency range to $\nu_{\rm max}=1.8$ THz (denoted $\nu_{1800}$), which yields $\ymono = (0.2 \pm 1.3) \times 10^{-6}$ ($\ymono = (1.4 \pm 2.1) \times 10^{-6}$) with the baseline (Gaussian) prior on P60. Moreover, the use of high frequencies allows us to retrieve unbiased results even without assuming a tight prior on $\beta_{\rm d}$, as the dust model is well constrained.

\subsubsection{All foregrounds mocks}
\begin{figure*}[!htbp]
\centering
\includegraphics[width=\textwidth]{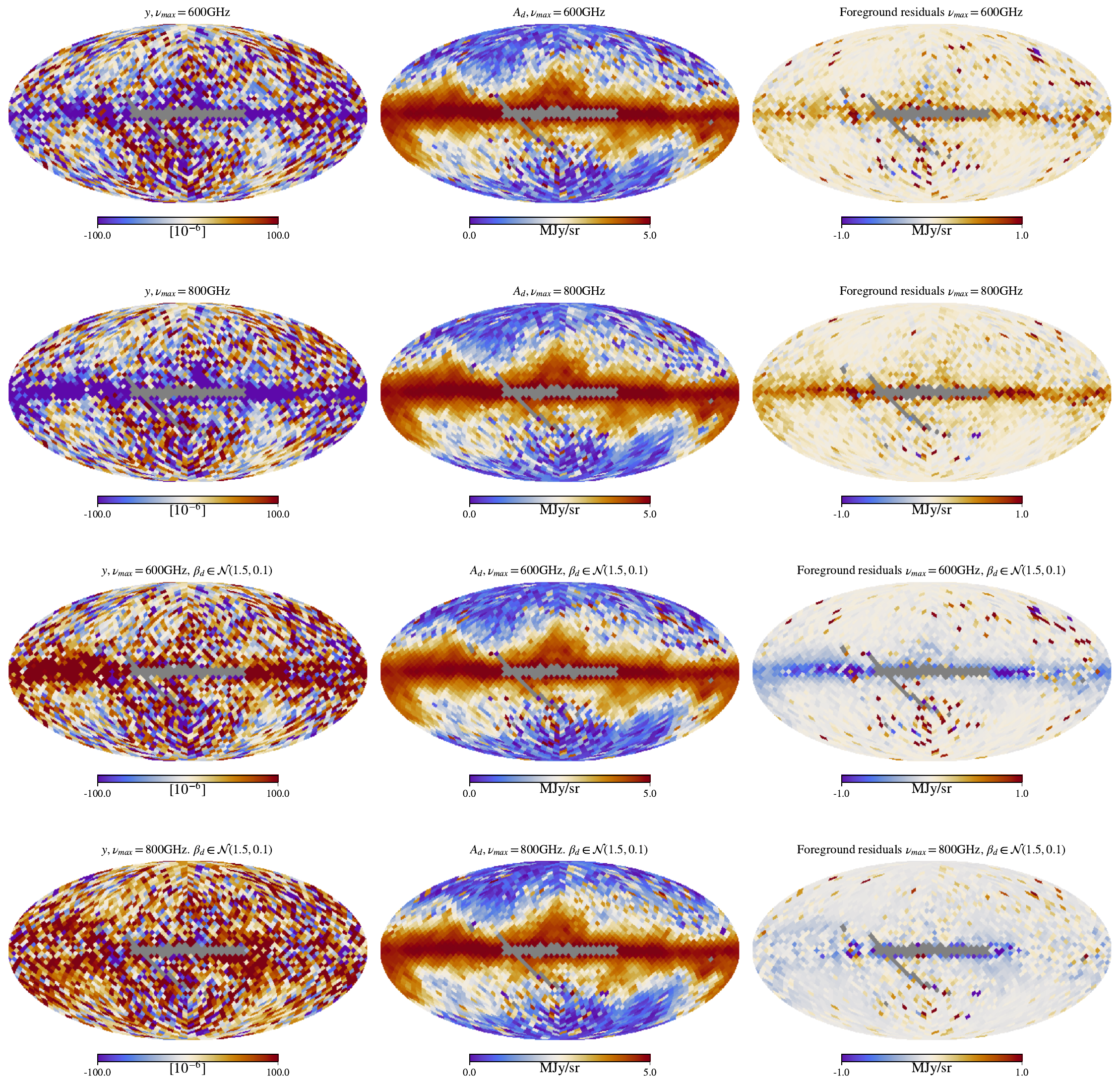}\\
\includegraphics[width=\textwidth]{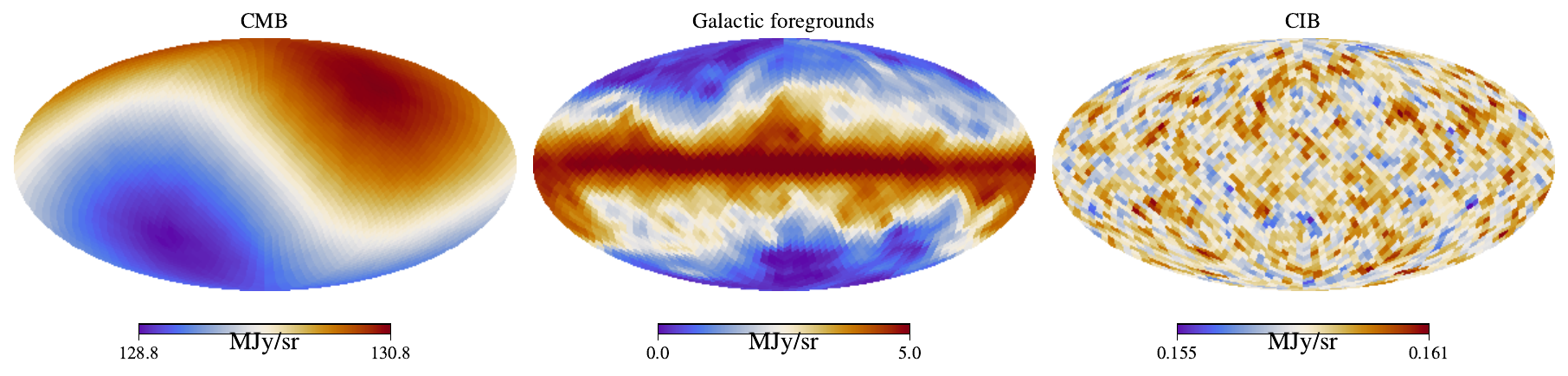}\\
\caption{Results of the \textit{pixel-by-pixel} fitting for different analysis set-ups applied to the mock data including all Galactic and extragalactic foregrounds (top to bottom, first four rows). From left to right, we show the $y$-distortion maps, the amplitude of the fitted foreground model in each pixel, and the difference between this and the sum of the input foregrounds. The bottom panel shows the input sky components at the reference frequency of $353$ GHz.}
\label{fig:pixpix-mocks-allfg}
\end{figure*}

The trends that we have observed with the \textit{constant dust} mock tests allow us to build intuition on the trade-offs in the \textit{pixel-by-pixel} method.  However, in order to validate the analysis performed on the real data, we need to perform a similar study on the more realistic mocks that include additional Galactic and extragalactic foregrounds. In Fig.~\ref{fig:pixpix-mocks-allfg}, we show the results of this analysis: maps of the dust amplitude and the recovered $y$ values in each pixel.

\begin{figure}[!htbp]
\includegraphics[width=0.6\columnwidth]{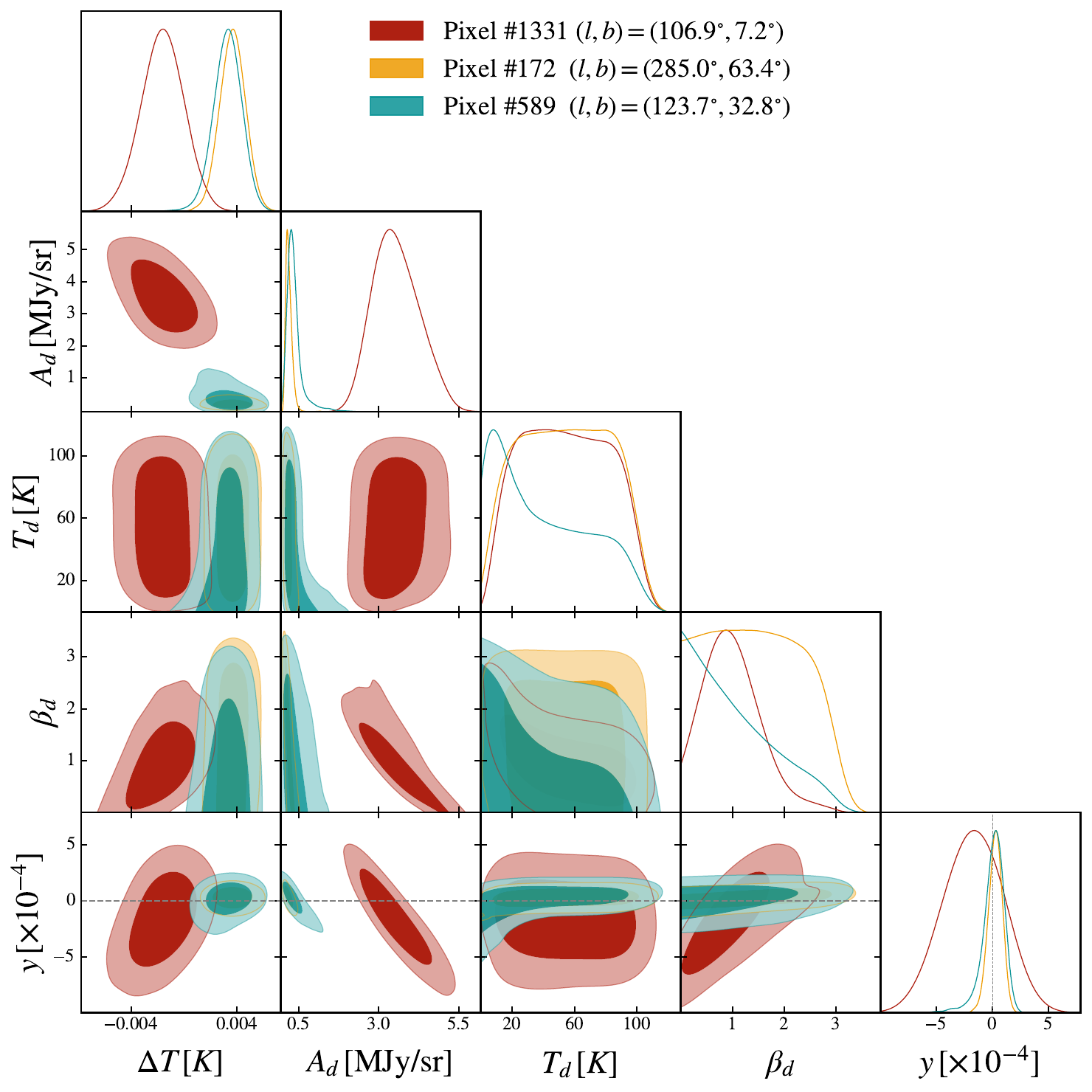}
\caption{Example posteriors of the sky model parameters derived from mock data including Galactic and extragalactic foregrounds for different sky pixels. For pixels in the Galactic plane (red), at least one of the foreground SED parameters, $\beta_{\rm d}$, can be moderately constrained from the data. Pixels at high Galactic latitude (yellow) have low foreground amplitudes and are dominated by the CIB. As such, only the amplitude of the foreground model can be constrained, while the spectral parameters are unconstrained and dominated by the priors. Pixels at intermediate latitude (green) display a behavior in between these two.}
\label{fig:allfgmocks-pixchains-summary}
\end{figure}

When all Galactic and extragalactic foreground components are included, the parameters of our reference foreground model capture the superposition of the two modified blackbody components (dust and CIB) as well as other foreground SEDs (e.g., residual sub-dominant synchrotron) and, as such, do not agree necessarily with the expected $\beta_{\rm d}$ and $T_{\rm d}$ values, either for the Galactic dust or the CIB. For $\nu_{600}$, with flat priors on both $\beta_{\rm d}$ and $T_{\rm d}$, we recover unbiased estimates of $\ymono$ across all three masks (P20, P40, P60). Typical behavior of the parameter chains is shown in Fig.~\ref{fig:allfgmocks-pixchains-summary}.  With this set-up, we also find all the results obtained on the P60, P40, and P20 masks to be consistent between each other following the consistency criterion defined in Eq.~\ref{eq:consistency-test}, $C_{\rm mask}$.

Extending our frequency coverage to $\nu_{800}$, we retrieve unbiased results at the $2\sigma$ level for P40 (P60), with $\ymono= (4.1 \pm 1.9) \times 10^{-6}$ ($\ymono = (-0.1 \pm 1.6) \times 10^{-6}$). If we consider the P20 mask, which should include regions of the sky mainly contaminated by the CIB rather than Galactic dust, we obtain results that are still roughly consistent with the expectations at the $\sim 2.3\sigma$ level ($\ymono = (6.1 \pm 2.7) \times 10^{-6}$), but the results are inconsistent between masks at the $3-4\sigma$ level based on the $C_{\rm mask}$ metric. As such, care should be employed when interpreting the results including high-frequency data from \textit{FIRAS}, and additional consistency and stability tests should be carried out. We also stress that these results do not account for any instrumental systematics that affect the data. The calibration errors are particularly prominent in the high-frequency data and further complications may arise when using these in the analysis of real data. 

Enforcing a prior on $\beta_{\rm d}$ based on the Galactic dust value worsens the overall fit and leads to biased results for P40 and P60.  When applying such a prior, we find that $\ymono$ is only unbiased on the P20 mask and for $\nu_{600}$, in which case presumably only one component, the CIB, dominates and thus any mismodeling of $\beta_{\rm d}$ is re-absorbed in $A_{\rm d}$ and $T_{\rm d}$ sufficiently well.  Using $\nu_{\rm 800}$ and a Gaussian prior on $\beta_{\rm d}$ leads to biased results on $\ymono$ for all masks, as the frequency range allows one to partially break the degeneracy between $\beta_{\rm d}$ and $T_{\rm d}$ and the combination of dust and CIB cannot be precisely accommodated if one of the parameters is tightly constrained.
 
As an additional stress test, we extend the frequency range to $\nu_{1800}$, but unlike the case of the simple \textit{constant dust} mocks, we are not able to recover unbiased results with any of the set-ups or sky masks. A more complex model that would perhaps be capable of dealing with the sky complexity needs to be used as the extended frequency range makes the component separation more prone to errors due to incorrect assumptions of the sky complexity. We do not pursue further this direction, even if in principle we could improve $\sigma_{\ymono}$ by over $60\%$.  We choose to restrict our analysis to the regime where a simple foreground model seems to constrain $\ymono$ measurements in a more robust way using fewer assumptions. Similarly to what was done for the \textit{frequency monopole} case and discussed in Appendix~\ref{app:high_frequencies}, we try implementing a component separation approach based on the moment expansion method to be employed with this extended frequency range, but we find that, on a single sky pixel level and despite the higher maximum frequency, the parameters of the model are largely degenerate or unconstrained. From these tests and further checks described in the following sections, we adopt the set-up with flat priors on both dust parameters using $\nu_{600}$, the P60 mask, and pixel-pixel correlation computed from the MCMC runs as our fiducial analysis set-up for the \verb|pixel-by-pixel| method. For completeness and direct comparison to the \textit{frequency monopole} method, we also quote results for the set-ups that use $\nu_{800}$, other sky masks, and a Gaussian prior on $\beta_{\rm d}$.

\begin{table}
\begin{tabular}{|lc|}
\colrule
Analysis set-up & $\ymono \times 10^{6}$ \\
\colrule
 \rowcolor{gray!20}\multicolumn{2}{|c|}{{\it Constant dust}}\\ 
\colrule
P40, $\nu_{600}$& $-24.1\pm 3.3$ \\ 
$^{\dagger}$ P60, $\nu_{600}$& $-18.9 \pm 2.4$ \\ 
\colrule
P40, $\nu_{800}$ & $-13.1 \pm 2.1$ \\ 
P40, $\nu_{800}$, $\beta_{\rm d}\in\mathcal{N}(1.5,0.1)$& $-0.1\pm 2.1$ \\ 
P40, $\nu_{800}$, $\beta_{\rm d}\in\mathcal{N}(1.5,0.1)$, \verb|inv_cov|& $6.3\pm 1.9$ \\ 
P40, $\nu_{800}$, $\beta_{\rm d}\in\mathcal{N}(1.5,0.1)$, \verb|inv_var|& $2.2\pm 2.6$ \\ 
P60, $\nu_{800}$& $-16.5\pm 1.7$ \\ 
P60, $\nu_{800}$, $\beta_{\rm d}\in\mathcal{N}(1.5,0.1)$& $2.9\pm 1.6$ \\ 
\colrule
P60, $\nu_{1800}$& $0.2\pm 1.3$ \\ 
P60, $\nu_{1800}$, $\beta_{\rm d}\in\mathcal{N}(1.5,0.1)$& $1.4\pm 2.1$ \\ 
\colrule
 \rowcolor{gray!20}\multicolumn{2}{|c|}{{\it All foregrounds}}                    \\ 
\colrule
P20, $\nu_{600}$& $-5.5 \pm 3.6$ \\  
P40, $\nu_{600}$& $-2.4 \pm 2.5$ \\  
$^{\dagger}$ P60, $\nu_{600}$& $0.7\pm 2.0$ \\ 
P20, $\nu_{600}$, $\beta_{\rm d}\in\mathcal{N}(1.5,0.1)$& $2.4 \pm 3.9$ \\  
P40, $\nu_{600}$, $\beta_{\rm d}\in\mathcal{N}(1.5,0.1)$& $6.8 \pm 2.5$ \\  
P60, $\nu_{600}$, $\beta_{\rm d}\in\mathcal{N}(1.5,0.1)$& $11.9 \pm 2.1$ \\  
\colrule
P20, $\nu_{800}$& $6.1 \pm 2.7$ \\  
P40, $\nu_{800}$& $4.1 \pm 1.9$ \\  
P60, $\nu_{800}$& $-0.1\pm 1.6$ \\ 
P20, $\nu_{800}$, $\beta_{\rm d}\in\mathcal{N}(1.5,0.1)$& $29.0 \pm 2.5$ \\  
P40, $\nu_{800}$, $\beta_{\rm d}\in\mathcal{N}(1.5,0.1)$& $31.6 \pm 1.8$ \\  
P60, $\nu_{800}$, $\beta_{\rm d}\in\mathcal{N}(1.5,0.1)$& $31.8 \pm 1.5$ \\  
\colrule
P60, $\nu_{1800}$& $-23.2\pm 1.3$ \\ 
P60, $\nu_{1800}$, $\beta_{\rm d}\in\mathcal{N}(1.5,0.1)$& $21.1 \pm 1.3$ \\
\colrule
\end{tabular}
\caption{\label{tab:y_results_mocks_pixpix} Summary of results from mock data for the \textit{pixel-by-pixel} method for different analysis set-ups: Galactic mask, frequency range, prior on $\beta_{\rm d}$, and the signal-averaging method. If not specified, the $\ymono$ computation accounts for pixel-pixel correlations estimated from the MCMC chains and uniform priors $\beta_{\rm d}\in\left[0,3\right]$ and $T_{\rm d}\in\left(0,100\right]$~K in the foreground model. The fiducial analysis set-up that we adopt for the \textit{pixel-by-pixel} method is denoted with $^{\dagger}$.} 
\end{table}

\subsubsection{Robustness and consistency tests}
\begin{figure}[!htbp]
\centering
\includegraphics[width=0.7\columnwidth]{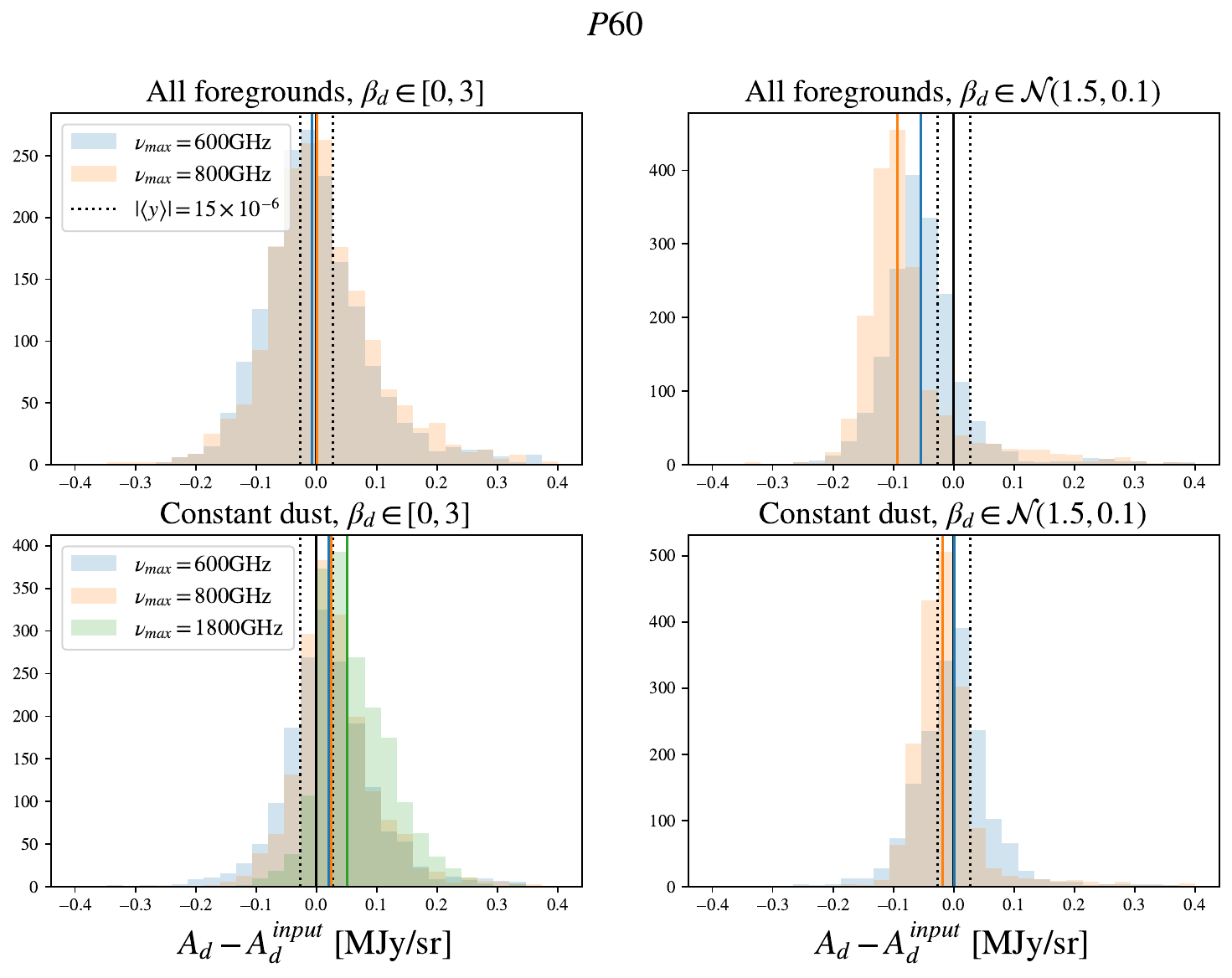}\\
\caption{Statistics of the foreground residuals in the $y$ maps for different mock data sets, using the P60 mask. Residuals are computed as the difference between the foreground model evaluated for the best-fit parameters in each pixel and the input sky at the reference frequency of $353$~GHz. The top (bottom) two rows show the statistics computed on the \textit{all foregrounds} (\textit{constant dust}) mocks. The medians of the histograms are shown as solid lines. The black solid lines correspond to a null residual and the dotted black to the brightness of $|\ymono |=15\times 10^{-6}$, corresponding to the value of the original \textit{FIRAS} upper limit.}
\label{fig:fg_mock_pixel_recap}
\end{figure}

To further validate the robustness of our component separation approach, we study in detail the foreground residuals in the component-separated $y$ map. We compute the foreground residuals as the difference in the recovered foreground amplitude at our reference frequency with respect to the sky model described in the previous section. In Fig.~\ref{fig:fg_mock_pixel_recap}, we show the distribution of the foreground residuals in each pixel on the P60 mask for different component separation set-ups discussed in the previous section. The median of the distribution of the residuals is very close to zero for our set-up with flat priors for $\beta_{\rm d}$ and including all foregrounds. Different frequency ranges retrieve similar level of residuals consistent with zero. Imposing a prior on $\beta_{\rm d}$ clearly shifts the median of the distribution away from null by $1\sigma$ or more, thus exhibiting a mismodeling of the foreground amplitude that gets worse as we increase the maximum frequency.

Imposing a Gaussian prior on $\beta_{\rm d}$ allows us to retrieve residuals consistent with null in the case of mock data that only includes dust. Relaxing the prior introduces degeneracies in the parameters, leading to biased estimates of the foreground amplitude that are only marginally mitigated by the increase in the frequency range. These results are consistent with the trends reported in the $\ymono$ estimates of Table~\ref{tab:y_results_mocks_pixpix}, reinforcing our interpretation that the set-up that avoids tight priors on $\beta_{\rm d}$ improves our ability to absorb the CIB and minimizes the foreground residuals. Further examples of consistency between mock and real data are discussed in Appendix~\ref{sec:mock-robustness}.

As a further consistency check of the foreground separation fidelity, we estimate the monopole of the dust intensity $\langle A_{\rm d}\rangle$ as the mean of the $A_{\rm d}$ map for the different choices of $\nu_{\rm max}$ and priors on $\beta_{\rm d}$. We then compare it with the monopole expected from our reference foreground model. Inside the Galactic plane (defined as the inverse of the official \textit{Planck} mask retaining 70\% of the sky), we find consistent values between $\langle A_{\rm d}\rangle$ retrieved from mock data and the value expected from our foreground model. For regions at high Galactic latitude, i.e., on the P20 footprint, we find inconsistent values at the $\sim2.5\sigma$ level when applying a Gaussian prior on $\beta_{\rm d}$. We also recover a value of $\langle T_0 \rangle = T_{\rm CMB}+\langle \Delta T\rangle$ consistent with the input $T_{\rm CMB}$ used for the mocks.

Finally, we use the mock data to derive an estimate of the foreground residual in our final $\ymono$ measurement for a given experimental set-up. As such, we define the residual as the difference between the amplitude of the total foreground emission as given by our fiducial sky model and the amplitude of the foreground inferred fixing the parameters of the foreground model adopted for component separation to their best-fit values retrieved from the mock data. We compute maps of the residuals $R^{fg}_\nu$ for all the frequencies used in a given analysis set-up.  For each of these residual maps, we then compute their monopole $\langle R^{fg}_\nu\rangle_p$ as the weighted average across pixels where the weights per pixel are the inverse variance of the corresponding pixel in the retrieved $y$ map, similarly to what is done on the data when computing $\ymono$. We choose to adopt the P60 mask as baseline to carry out this analysis and adopt the error on the weighted average as error on the foreground residuals' monopole, $\sigma({\langle R^{fg}_\nu\rangle}_p)$.

\begin{figure}[!htbp]
\centering
\includegraphics[width=.5\columnwidth]{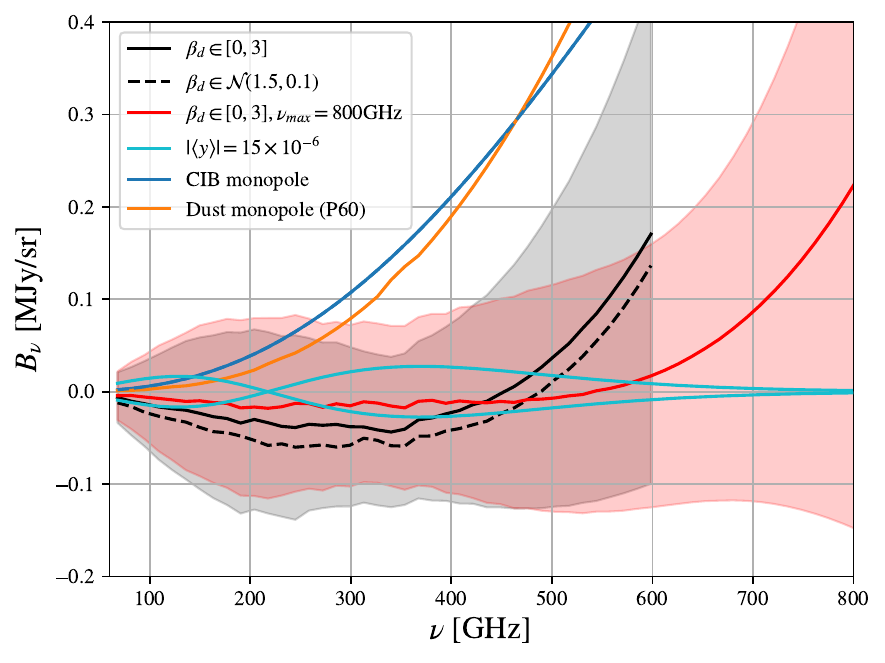}
\caption{Brightness of the foreground residual extracted from mock data as the difference between the sky model evaluated at the best fit point obtained from mock data and the input mock data themselves. The solid lines represent the monopole $\langle R^{fg}_\nu\rangle_p$ of the residual per frequency map computed on the P60 Galactic mask, and its uncertainty is shown as shaded area. The main analysis setups discussed in Sec.~\ref{sec:mock-resuls-pixpix} are shown in black and red, while monopoles of other sky components are shown for reference in other colors.}
\label{fig:fg_res_sed}
\end{figure}

The spectral shape of $\langle R^{fg}_\nu\rangle_p$ cannot be fit by an equivalent amplitude of a $y$-like distortion as shown in Fig.~\ref{fig:fg_res_sed}.  We therefore compute the inverse-covariance-weighted average across frequencies of $\langle R^{fg}_\nu\rangle_p$ and use this as a proxy to quantify the total foreground residual brightness, $\langle\langle R^{fg}_\nu\rangle_{p}\rangle_\nu$. For this purpose we use the \textit{FIRAS} frequency-frequency covariance,\footnote{As this covariance varies pixel by pixel, in particular for the component due to the destriping error, we contract Eq.~\eqref{eq:covariance} as $ \sum_p\sum_p^\prime \mathbb{C}_{\nu\nu^\prime pp^\prime}/N_{pix}^2$, where $N_{pix}$ is the number of pixels in the P60 mask.} to which we add $\sigma({\langle R^{fg}_\nu\rangle}_p)$ as an additional source of error. For $\nu_{600}$ ($\nu_{800}$), we obtain a residual foreground brightness $\langle\langle R^{fg}_\nu\rangle_{p}\rangle_\nu$ consistent with null. The set-up with $\nu_{800}$ displays the lowest total foreground residual brightness but with the highest error, while $\nu_{600}$ displays the lowest mean foreground brightness error. Such a trade-off holds when changing the Galactic mask. However, as $\nu_{600}$ shows the lowest variation in the value of the mean residual, we identify this as the most robust set-up in the absence of instrumental systematics. As an additional metric, we compute the inner product of the $\langle R^{fg}_\nu\rangle_p$ vector with a $y$-distortion SED to assess how these two emissions project onto one another. We obtain a value $\lesssim 0.05$, giving a good indication of minimal overall foreground contamination in our final measurement.

\section{Results from Data: \textit{frequency monopole} method}\label{sec:data_results_monopole}
In Table~\ref{tab:models}, we summarize the models that we fit in our analysis. We find the best-fit parameters and run MCMCs for some of the set-ups. The best-fit results are summarized in Table~\ref{tab:models}, and the MCMC results in Table~\ref{tab:y_results} and Table~\ref{tab:y_results_other_FGs}. To find the best-fit results for each model, we minimize $-\ln P(\theta|I_{\nu}^{\rm FIRAS})$, where $P(\theta|I_{\nu}^{\rm FIRAS})$ is the posterior probability distribution, using \verb|scipy.optimize.minimize| (i.e., $ P(\theta|I_{\nu}^{\rm FIRAS})=\mathcal{L} (I_{\nu}^{\rm FIRAS}|\theta)\pi(\theta)$, where $\pi$ are the priors). When we use flat priors, we assume that the log-prior for that parameter is simply zero and use them to specify the inclusive bounds for the minimizer. We perform the minimization iteratively, re-initializing at the best-fit parameter values until $-\ln(P)$ is found to have the same value ten times consecutively. For the baseline fits (set-ups 1, 2), we initialize at the MCMC posterior mean values. For the rest of the set-ups, 3-14, we initialize with best-fit parameter values found for the baseline set-ups 1-2, while setting any additional parameters to zeros. We run the minimization for ten different starting points (randomly perturbed values within $0.5-1.5\times$ of the fiducial initial values). We list the best-fit parameter values in Table~\ref{tab:models_bestfit_values}. We note that since we do not run MCMCs for all set-ups, we cannot quantify the errors on the parameters. Thus, many of these best-fit values should be interpreted with caution but we include them here for completeness. We can then compute the corresponding best-fit $\chi^{2}$ for each set-up ($\chi^{2}=-2\ln\mathcal{L}$). In Table~\ref{tab:models}, we list the $\chi^{2}$ for the baseline set-ups and $\Delta\chi^{2}$ for the rest (i.e., $\Delta\chi^{2}$ is computed with reference to the baseline model 1 or 2 depending on the frequency range used). We also include the probability to exceed (PTE) and the number of degrees of freedom (DOF) (DOF $=$ number of data points $-$ number of free parameters). 

We find that fits using the \textit{frequency monopole} computed from P20 and P40 have acceptable PTE values, while fits using P60 are marginally acceptable using $\nu_{600}$ and generally too low using $\nu_{800}$. For the latter, while adding additional components to the foreground model improves the fits in most cases, the improvement is still not sufficient. The simple baseline dust models provide a reasonable fit to P20/P40 so we restrict our main analysis to these results, which we have also tested on the mocks. In Appendix~\ref{sec:more_fgs}, we discuss the results for $\ymono$ when fitting additional foregrounds.

\begin{table*}[t]
\centering
\begin{tabular}{|cl|l|l|l|}
\colrule
&Foregrounds& Free parameters&$\chi^{2}$ or $\Delta\chi^{2}$ [DOF]& PTE\\
&(frequency range)& + extra priors &P20/P40/P60&P20/P40/P60\\
\colrule
1. &Dust ($\nu_{600}$) & $\Delta T$, $y$, $A_{\rm d}$, $T_{\rm d}$, $\beta_{\rm d}$ & 21.2/28.2/44.2 [22]&0.510/0.168/0.003\\
2. &Dust ($\nu_{\rm 800}$) & $\Delta T$, $y$, $A_{\rm d}$, $T_{\rm d}$, $\beta_{\rm d}$ & 30.2/37.2/64.9 [31]& 0.505/0.204/$3.48\times10^{-4}$\\
\colrule
3. & Dust + CO ($\nu_{600}$)& $\Delta T$, $y$, $A_{\rm d}$, $T_{\rm d}$, $\beta_{\rm d}$, $A_{\rm CO}$ &-0.3/-1.9/-4.1 [21]&0.466/0.192/0.007\\
4. & Dust + CO ($\nu_{\rm 800}$) & $\Delta T$, $y$, $A_{\rm d}$, $T_{\rm d}$, $\beta_{\rm d}$, $A_{\rm CO}$ & 0.0/-0.7/-2.7 [30]& 0.453/0.193/0.001\\
\colrule
5. & Dust + FF ($\nu_{600}$) & $\Delta T$, $y$, $A_{\rm d}$, $T_{\rm d}$, $\beta_{\rm d}$, $A_{\rm FF}$ &-0.2/-1.0/-3.5 [21]&0.461/0.164/0.006\\
6. & Dust + FF ($\nu_{800}$) & $\Delta T$, $y$, $A_{\rm d}$, $T_{\rm d}$, $\beta_{\rm d}$, $A_{\rm FF}$&0.0/-0.6/-3.0 [30]& 0.454/0.188/0.001\\
\colrule
7. & Dust + CO + FF ($\nu_{600}$) & $\Delta T$, $y$, $A_{\rm d}$, $T_{\rm d}$, $\beta_{\rm d}$, $A_{\rm CO}$, $A_{\rm FF}$ & -0.4/-2.2/-5.8 [20]&0.410/0.166/0.008\\
8. & Dust + CO + FF ($\nu_{\rm 800}$) & $\Delta T$, $y$, $A_{\rm d}$, $T_{\rm d}$, $\beta_{\rm d}$, $A_{\rm CO}$, $A_{\rm FF}$ & 0.0/-0.9/-3.6 [29]& 0.403/0.163/$4.36\times10^{-4}$\\
\colrule
9. & Dust + CIB ($\nu_{600}$) & $\Delta T$, $y$, $A_{\rm d}$, $T_{\rm d}$, $\beta_{\rm d}$, $A_{\rm CIB}$ &  -1.3/-2.9/-6.1 [21]&0.531/0.234/0.013\\
10. & Dust + CIB ($\nu_{\rm 800}$) & $\Delta T$, $y$, $A_{\rm d}$, $T_{\rm d}$, $\beta_{\rm d}$, $A_{\rm CIB}$ &0.0/-0.1/-0.7 [30]& 0.453/0.173/$2.82\times10^{-4}$\\
\colrule
11. &Dust + Sync ($\nu_{600}$) & $\Delta T$, $y$, $A_{\rm d}$, $T_{\rm d}$, $\beta_{\rm d}$, $A_{\rm s}$, $\beta_{\rm s} \in \mathcal{N}(-1, 0.5)$&  0.0/-0.8/-3.0 [20]&0.387/0.123/0.004\\
12. &Dust + Sync ($\nu_{800}$) & $\Delta T$, $y$, $A_{\rm d}$, $T_{\rm d}$, $\beta_{\rm d}$, $A_{\rm s}$, $\beta_{\rm s} \in \mathcal{N}(-1, 0.5)$ & 0.0/-0.6/-2.9 [29]& $0.402/0.155/3.47\times 10^{-4}$\\
\colrule
13. &Dust ($\nu_{600}$) & $\Delta T$, $y$, $A_{\rm d}$, $T_{\rm d}$, $\beta_{\rm d}\in[0,3)$& -0.3/-0.9/-1.8 [22]&0.532/0.198/0.006\\
14. &Dust ($\nu_{\rm 800}$) & $\Delta T$, $y$, $A_{\rm d}$, $T_{\rm d}$, $\beta_{\rm d}\in[0,3)$ & -0.3/0.0/-0.3 [31]& 0.520/0.204/$3.84\times 10^{-4}$\\
\colrule
\end{tabular}
\caption{\label{tab:models} Foreground models fit using the \textit{frequency monopole}, list of free parameters and the corresponding priors, best-fit $\chi^2$ for the baseline set-ups (1 or 2) or $\Delta\chi^2$ for the set-ups 3-14 computed with respect to 1 or 2, number of degrees of freedom (DOF $=$ number of data points $-$ free parameters), and probability to exceed (PTE) values. We use priors $T_{\rm d}\in[0,100)$, $\beta_{\rm d} \in N(1.51, 0.1)$ (unless specified otherwise in column 2). We use a Gaussian prior on $\beta_{\rm d}$ based on the mean posterior results from \planck \cite{Planck2016FG}. We list the best-fit parameter values in Table~\ref{tab:models_bestfit_values}.} 
\end{table*}

\begin{table*}[h]
\centering
\begin{tabular}{|cl|l|l|}
\colrule
&Foregrounds & Best-fit parameter values\\
&(frequency range)&P20/P40/P60\\
\colrule
1.&Dust ($\nu_{600}$) & $\Delta T \times10^{3}=-1.38/-1.22/-1.48$ K, $\ymono\times10^{6}=17.3/12.1/9.71$,\\
&& $A_{\rm d}\times10^{-5}=4.20/4.89/5.87$ Jy/sr, $T_{\rm d}=14.2/14.9/15.0$ K, $\beta_{\rm d}=1.51/1.51/1.50$\\
2.&Dust ($\nu_{\rm 800}$) & $\Delta T \times 10^{3} = -1.38/-1.22/-1.47$ K, $\ymono\times10^{6}=15.8/11.9/12.1$,\\
&&$A_{\rm d}\times10^{-5}=4.06/4.83/5.29$ Jy/sr, $T_{\rm d}=14.5/15.0/15.7$ K, $\beta_{\rm d}=1.51/1.51/1.50$\\
\colrule
3. & Dust + CO ($\nu_{600}$) & $\Delta T\times10^{3}=-1.44/-1.32/-1.60$ K, $\ymono\times10^{6}=40.2/51.1/58.3$,\\
&& $A_{\rm d} \times10^{-5}=2.57/2.46/2.87$ Jy/sr, $T_{\rm d}=17.8/20.8/21.2$ K\\
& &$\beta_{\rm d}=1.51/1.51/1.51$, $A_{\rm CO}=24.6/42.6/53.0$ Jy/sr\\
4. & Dust + CO ($\nu_{\rm 800}$) & $\Delta T \times10^{3}=-1.39/-1.27/-1.56$ K, $\ymono\times10^{6}=17.5/25.4/33.4$\\
& & $A_{\rm d}\times10^{-5}=4.02/4.50/4.85$ Jy/sr, $T_{\rm d}=14.6/15.4/16.2$ K,\\
& & $\beta_{\rm d}=1.51/1.52/1.52$, $A_{\rm CO}=2.37/19.2/30.4$ Jy/sr\\
\colrule
5. & Dust + FF  ($\nu_{600}$) & $\Delta T \times 10^{3}= 1.43/1.29/1.59$ K, $\ymono\times10^{6}=23.2/20.8/23.2$\\
& & $A_{\rm d}\times10^{-5}=3.07/3.33/3.52$ Jy/sr, $T_{\rm d}=16.1/17.6/18.7$ K, \\
& & $\beta_{\rm d}=1.51/1.51/1.50$, $A_{\rm FF}\times10^{-3}=8.72/13.3/20.9$ Jy/sr\\
6. & Dust + FF ($\nu_{\rm 800}$) & $\Delta T\times10^{3}=-1.39/-1.27/-1.57$ K, $\ymono\times10^{6}=16.5/14.6/17.2$,\\
& & $A_{\rm d}\times10^{-5}=3.94/4.40/4.56$ Jy/sr, $T_{\rm d}=14.6/15.5/16.5$ K,\\
& & $\beta_{\rm d}=1.51/1.51/1.52$, $A_{\rm FF}\times10^{-3}=2.18/8.51/16.2$ Jy/sr\\
\colrule
7. & Dust + CO + FF ($\nu_{600}$)& $\Delta T\times10^{3}=-1.46/-1.35/-1.66$ K, $\ymono\times10^{6}=40.1/50.2/56.6$,\\
&  & $A_{\rm d}\times10^{-5}=2.18/2.05/2.17$ Jy/sr, $T_{\rm d}=19.1/22.8/24.4$ K,  $\beta_{\rm d}=1.51/1.51/1.51$,\\
& & $A_{\rm CO}=20.1/36.0/41.0$ Jy/sr, $A_{\rm FF}\times10^{-3}=6.30/8.51/15.2$ Jy/sr\\
8. & Dust + CO + FF ($\nu_{\rm 800}$) & $\Delta T\times10^{3}=-1.39/-1.29/-1.59$ K, $\ymono\times10^{6}=16.5/23.2/28.2$,\\
& & $A_{\rm d}\times10^{-5}=3.94/4.35/4.53$ Jy/sr, $T_{\rm d}=14.6/15.5/16.5$ K,  $\beta_{\rm d}=1.51/1.52/1.52$,\\
& & $A_{\rm CO}=-0.05/14.2/18.1$ Jy/sr, $A_{\rm FF}\times10^{-3}=2.19/4.58/11.1$ Jy/sr\\
\colrule
9. & Dust + CIB ($\nu_{600}$) & $\Delta T\times10^{3}=-1.74/-1.60/-1.94$ K, $\ymono\times10^{6}=-45.0/-53.4/-69.6$, \\
& & $A_{\rm d}\times10^{-7}=5.73/6.06/7.41$ Jy/sr, $T_{\rm d}=12.5/12.5/12.5$ K,\\
& & $\beta_{\rm d}=1.51/1.51/1.51$, $A_{\rm CIB}\times10^{-7}=-6.08/-6.42/-7.85$ Jy/sr\\
10. & Dust + CIB ($\nu_{\rm 800}$)& $\Delta T=-1.38/-1.24/-1.53$ K, $\ymono\times10^{6}=15.7/8.67/0.726$, \\
&  & $A_{\rm d}\times10^{-5}=5.84/29.0/83.7$ Jy/sr, $T_{\rm d}=13.9/12.9/12.6$ K,\\
& & $\beta_{\rm d}=1.51/1.51/1.51$, $A_{\rm CIB}\times10^{-5}=-1.95/-25.9/-83.0$ Jy/sr\\
\colrule
11. &Dust + Sync ($\nu_{600}$) & $\Delta T\times10^{3}=-1.38/-1.25/-1.53$ K, $\ymono\times10^{6}=17.3/19.9/22.5$, \\
& & $A_{\rm d}\times10^{-5}=4.20/4.05/4.51$ Jy/sr, $T_{\rm d}=14.2/16.2/16.8$ K, $\beta_{\rm d}=1.51/1.51/1.50$,\\
& & $A_{\rm s}\times10^{-4}=0.00/2.94/4.87$ Jy/sr, $\beta_{\rm s}=0.00/-0.98/-0.93$\\
12. &Dust + Sync ($\nu_{800}$)& $\Delta T\times10^{3}=-1.38/-1.25/-1.52$ K, $\ymono\times10^{-6}=15.8/15.8/19.4$,\\
& & $A_{\rm d}\times10^{-5}=4.06/4.64/4.99$ Jy/sr, $T_{\rm d}=14.5/15.2/16.0$ K, $\beta_{\rm d}=1.51/1.51/1.51$,\\
&& $A_{\rm s}\times10^{-4}=0.00/2.31/4.34$ Jy/sr, $\beta_{\rm s}=0.00/-0.99/-0.97$\\
\colrule
13. &Dust ($\nu_{600}$) & $\Delta T \times 10^{3}=-1.44/-1.29/-1.56$ K, $\ymono \times10^{6}=9.22/2.46/-1.91$,\\
 && $A_{\rm d}\times10^{-4}=3.99/4.99/6.02$ Jy/sr, $T_{\rm d}=100/100/100$ K, $\beta_{\rm d}=0.34/0.41/0.41$\\
14. &Dust ($\nu_{\rm 800}$) & $\Delta T\times10^{3}=-1.34/-1.22/-1.50$ K, $\ymono\times10^{6}=25.9/11.2/5.40$,\\
& &$A_{\rm d}\times10^{-5}=6.70/4.68/3.87$ Jy/sr, $T_{\rm d}=10.1/15.4/20.2$ K, $\beta_{\rm d}=2.34/1.47/1.15$\\
\colrule
\end{tabular}
\caption{\label{tab:models_bestfit_values} Best-fit parameter values for set-ups listed in Table~\ref{tab:models}. Only minimization runs for set-ups 1-2 were initialized at the posterior mean values. Set-ups 3-14 were initialized at the best-fit parameters from set-ups 1-2 with any extra parameter(s) set to zero(s). }
\end{table*}

\begin{table}[t]
\begin{tabular}{|l|c|c|c|c|}
\colrule
Model & $\times 10^{6}$ & P20&P40&P60\\
\colrule
Dust ($\nu_{600}$) & $\ymono$&14.2&12.0&9.8\\
& $\sigma_{\ymono}\ 68\%$ C.L. &+15.6/-11.8&+10.7/-9.7&8.7\\
&$\ymono\leq95\%$ C.L.&40.2&31.2&26.3\\
& Fisher $\sigma_{\langle y \rangle}$&13.4&9.5&8.0\\
\colrule
Dust ($\nu_{\rm 800}$)& $\ymono$ &15.7&11.8&11.9\\
& $\sigma_{\ymono}\ 68\%$ C.L. &9.7&7.0&5.8\\
& $\ymono\leq95\%$ C.L. & 34.5  & 25.3 &23.3\\
& Fisher $\sigma_{\langle y \rangle}$&9.6&7.0&6.1\\
\colrule
Dust ($\nu_{\rm 800}$)& $\ymono$ &13.7&6.4&2.2\\
$\beta_{\rm d}\in[0, 3)$& $\sigma_{\ymono}\ 68\%$ C.L. &14.1&+11.2/-14.6&+8.9/-12.5\\
& $\ymono\leq95\%$ C.L. & 40.2  & 32.5 &24.3\\
& Fisher $\sigma_{\langle y \rangle}$&18.6&13.1&11.0\\
\colrule
\end{tabular}
\caption{$\ymono$ constraints for the different models and sky fractions for the \textit{frequency monopole} method. We list the mean value, the 68$\%$ C.L.~errors, the 95$\%$~C.L. upper limits, and the Fisher forecast error bars. In the first two rows, we list results with a Gaussian prior on $\beta_{\rm d}\in \mathcal{N}(1.51,0.1)$ and a flat prior on $T_{\rm d}\in [0, 100)$~K, and in the bottom row with flat priors for both $T_{\rm d}$ and $\beta_{\rm d}$.\label{tab:y_results}}
\end{table}

\subsection{Baseline Model Results}

\begin{figure*}[h]
\begin{minipage}[c]{0.48\linewidth}
    \centering
    \includegraphics[width=\linewidth]{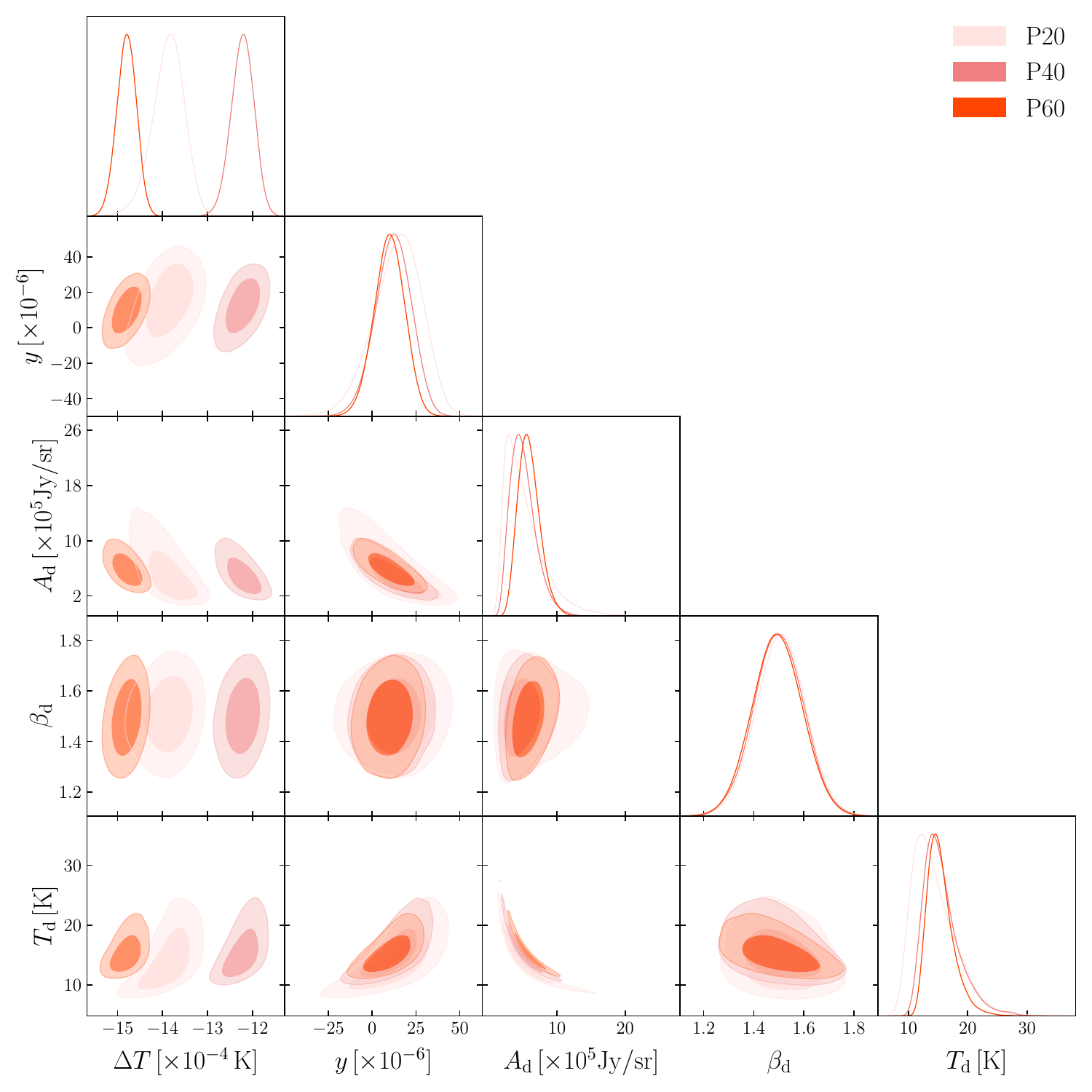}
    \end{minipage}\hfill
\begin{minipage}[c]{0.48\linewidth}
    \centering
    \includegraphics[width=\linewidth]{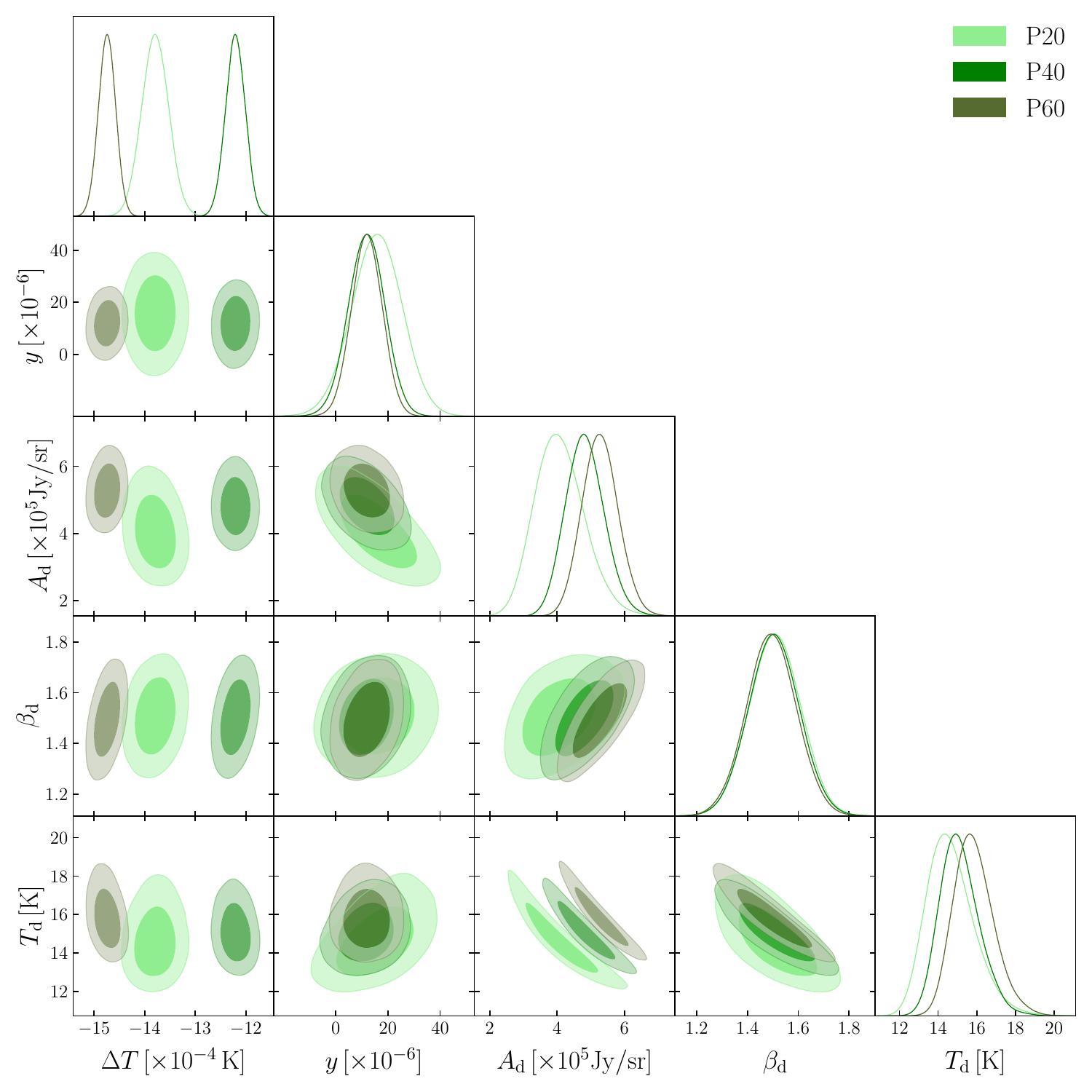}
\end{minipage} 
\caption{\textit{Left:} Marginalized parameter posteriors from fitting the low-frequency data alone ($\nu_{600}$). The sky model consists of the blackbody component, $y$-distortion, and dust (MBB) with a Gaussian prior on $\beta_{\rm d}\in \mathcal{N}(1.51,0.1)$ and a flat prior on $T_{\rm d}\in [0, 100)$~K (set-up 1 in Table \ref{tab:models}). The dust amplitude, $A_{\rm d}$, is consistent with zero at $\approx 2\sigma$ for P20/P40 and shows signs of detection at $3\sigma$ for P60. \textit{Right:} Marginalized posteriors from fitting the data at frequencies up to 800 GHz using the same sky model (set-up 2 in Table \ref{tab:models}). The dust amplitude is detected at $\approx 5/8/10\sigma$ for P20/P40/P60, respectively.}\label{fig:dust_gaussBd_model_data}
\end{figure*}

The original analysis by the \textit{FIRAS} team included data only from the low-frequency instrument because the high-frequency data are more at risk of calibration errors \cite{Fixsen1996_dist}. In the left panel of Fig.~\ref{fig:dust_gaussBd_model_data}, we similarly show the posterior results for the fits where we only include the data from the low-frequency instrument (i.e., $95-626$ GHz --- recall that we remove the two lowest frequency channels due to possible systematics in the lowest bands of the FTS instrument, as described in Sec.~\ref{sec:freq_ranges}).

In this fit, we use the baseline sky model for the \textit{frequency monopole}, which only includes $\Delta B_{\nu}$, $I_{\nu}^{y}$, and the dust MBB as a foreground (set-up 1 in Table~\ref{tab:models}). We find that the low-frequency data do not constrain both $T_{\rm d}$ and $\beta_{\rm d}$, so it is advantageous to use a Gaussian prior on one of the spectral parameters. This is similar to what was found in Ref.~\cite{Fixsen1996_dist}, where the spectral index of dust was fixed to a constant value. In our analysis we use a prior $\beta_{\rm d} \in \mathcal{N}(1.51, 0.1)$ motivated by the results from \planck~\cite{Planck2016FG}, as previously tested on the simulated maps. With this baseline sky model, we find $\ymono=(14.2^{+15.6}_{-11.8})\times10^{-6}/ (12.0^{+10.7}_{-9.7})\times10^{-6}/(9.8\pm8.7)\times10^{-6}$ for P20/P40/P60, corresponding to 95\% CL upper limits of $\ymono < 40.2 \times10^{-6}/31.2\times10^{-6}/26.3\times10^{-6}$. The error bars are summarized in Table~\ref{tab:y_results}. We compute the consistency criterion $C_{\rm mask}$ in Eq.~\eqref{eq:consistency-test} and find that the results using $\nu_{600}$ are consistent across masks.

As seen in Fig.~\ref{fig:dust_gaussBd_model_data}, the dust amplitude is consistent with zero at $\approx 2\sigma$ for P20/P40 and our results only hint at a detection of dust at $3\sigma$ for P60. We further see that we essentially recover the prior for $\beta_{\rm d}$. We find $T_{\rm d}=14.4^{+2.1}_{-4.5}/15.6^{+1.6}_{-3.6}/15.4^{+1.4}_{-2.6}$ K for P20/P40/P60, which is a bit lower than the value expected for Galactic dust (e.g., $T_{\rm d}\approx 21\pm2$~K \cite{Planck2016FG}). This is similar to what we find in the fits from mocks and to the results in Ref.~\cite{Fixsen1996_dist}. Ref.~\cite{Fixsen1996_dist} explained the lower dust temperature with their choice of a higher spectral index ($\beta_{\rm d}=2$) and the fact that the fits were made with high Galactic latitude data. The lower temperature values that we recover are likely due to a combination of the lack of sensitivity of the low frequencies to the exact shape of the dust SED, as described in the tests on mocks, and due to the simplified sky model that we use in these fits. The MBB we use to model dust presumably absorbs other dominant foreground sky emission that we do not account for, such as the CIB, free-free emission, extragalactic CO, and synchrotron.

To improve the constraining power of the \textit{frequency monopole} method, we experiment with including the high-frequency data. We find that including high frequencies results in a worse $\chi^{2}$ even for the lowest sky fractions. Therefore, we choose to limit our main analysis to going up to 789 GHz ($\nu_{800}$). We explore extending our analysis to even higher frequencies in Appendix~\ref{app:high_frequencies}. Ref.~\cite{Fixsen1996_dist} states that calibration of the high-frequency data is more prone to systematic errors. We check for systematic biases in the high-frequency data by fitting an extra amplitude parameter that would shift the high-frequency channels up or down. We find that the extra parameter is consistent with unity and, therefore, does not clearly indicate presence of a systematic bias that would be uniform across all high frequencies. Furthermore, a model with this extra amplitude parameter does not provide a significantly better fit (we find $\Delta \chi^2 \approx 1.6$ for P20 and $\Delta \chi^2<1$ for P40/P60 for one additional free parameter). Since there is not a clear indication of an offset in the data up to $\approx 800$ GHz, we keep the results using the $\nu_{800}$ frequency range in our main analysis. 

The right panel of Fig.~\ref{fig:dust_gaussBd_model_data} shows the full posteriors for the fits using $\nu_{800}$. These results again only include the dust MBB as the foreground component in the sky model (set-up 2 in Table~\ref{tab:models}). As in the case of using the low-frequency data alone, going up to $\sim 800$ GHz still does not constrain both $T_{\rm d}$ and $\beta_{\rm d}$, so we similarly rely on the information from \planck via a Gaussian prior on $\beta_{\rm d}$. Including some of the high-frequency data allows us to detect dust at $\approx 5\sigma/8\sigma/10\sigma$ for P20/P40/P60 and to slightly tighten the constraints on $\ymono$. Using $\nu_{800}$, our $95\%$ C.L. upper limits are: $\ymono < 34.5\times10^{-6}, 25.3\times10^{-6}, 23.3 \times10^{-6}$ for P20, P40, P60, respectively. The P60 mask using $\nu_{800}$ gives the tightest results with the \textit{frequency monopole} approach, but is a poor fit: we find $\chi^{2}=64.9$, corresponding to a low PTE value of $0.0003$. The error bars with these additional frequencies are $\approx 30\%$ tighter than with $\nu_{600}$. This demonstrates how additional high-frequency data can tighten the results from the \textit{frequency monopole}, but the P60 result also illustrates that a more complex sky model is likely needed to provide a good fit to the data.

In Fig.~\ref{fig:baseline_low_vs_800} of Appendix~\ref{app:extra_contours}, we also compare posterior results from fitting $\nu_{600}$ and $\nu_{\rm 800}$ using our baseline sky model. We show the fits for the two different frequency ranges for the same sky masks and find that they are consistent.  Finally, we compute the $C_{\rm mask}$ criterion for our baseline results and find that the results are consistent across all masks.

\begin{figure}[t]
    \centering
    \includegraphics[width=0.6\columnwidth]{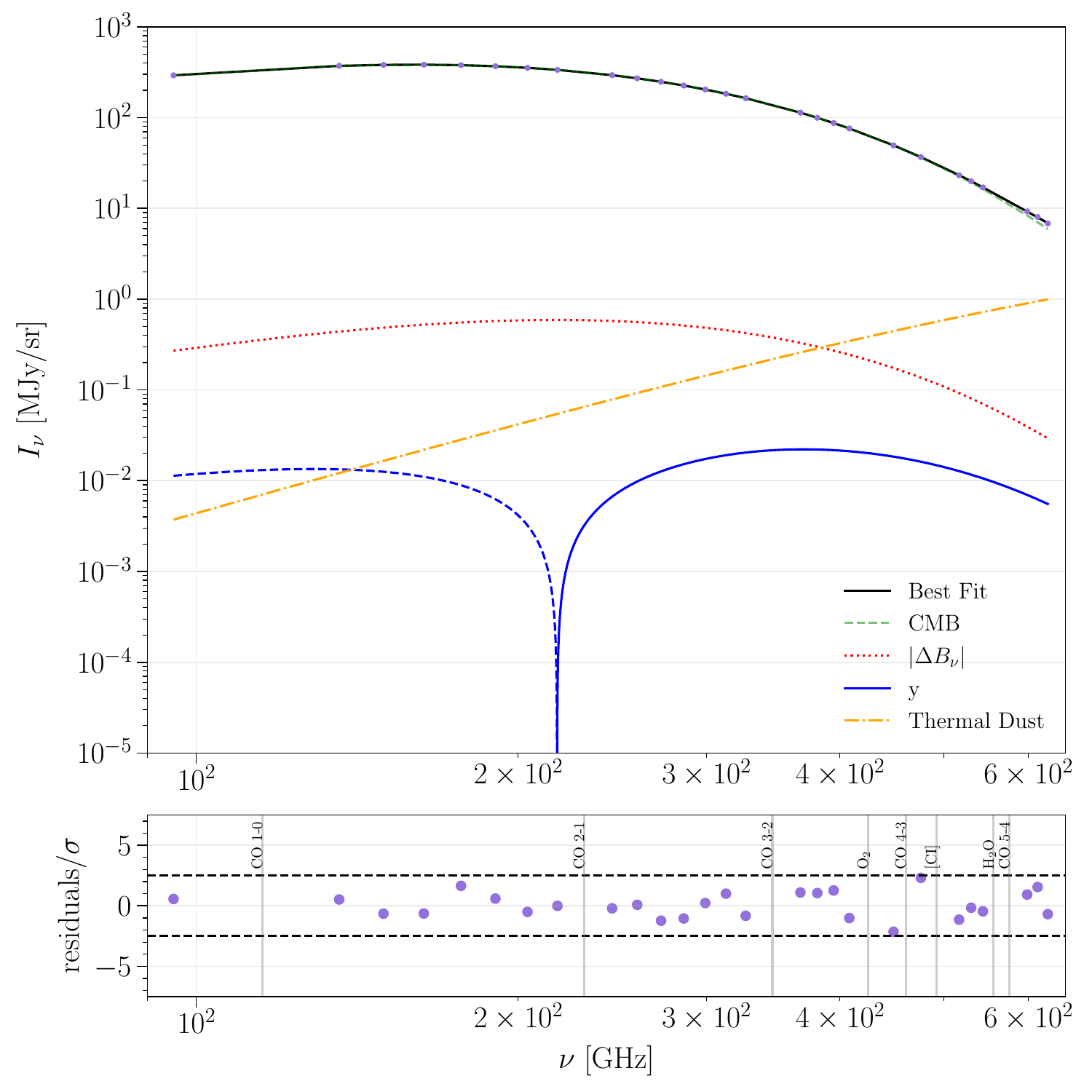}
    \caption{\label{fig:best_fit} Best-fit sky model to the \textit{FIRAS} data, for P40 and $\nu_{600}$, using our baseline model, which includes dust with a Gaussian prior on $\beta_{\rm d}\in \mathcal{N}(1.51,0.1)$ based on \planck. The upper panel shows the best-fit model (solid black) and the low-frequency (light purple markers) data, along with each of the sky model components: the absolute value of the blackbody deviation (red dotted), the thermal dust MBB (orange dot-dashed), and the $y$-distortion with positive values in solid blue and negative values in dashed blue. We also plot a blackbody spectrum at $T_{0}=2.7255$ K (green dashed denoted as `CMB'). The bottom panel shows the residuals of the best-fit model, i.e., (data$-$model)/$\sigma$, which are within 2.5$\sigma$ as indicated by the horizontal dashed lines. We also indicate (in vertical gray lines) the astrophysical emission lines, around which we exclude the data (see Sec.~\ref{sec:methods} and Appendix~\ref{app:emission} for details).}
\end{figure}

\begin{figure}[!h]
    \centering
    \includegraphics[width=0.5\linewidth]{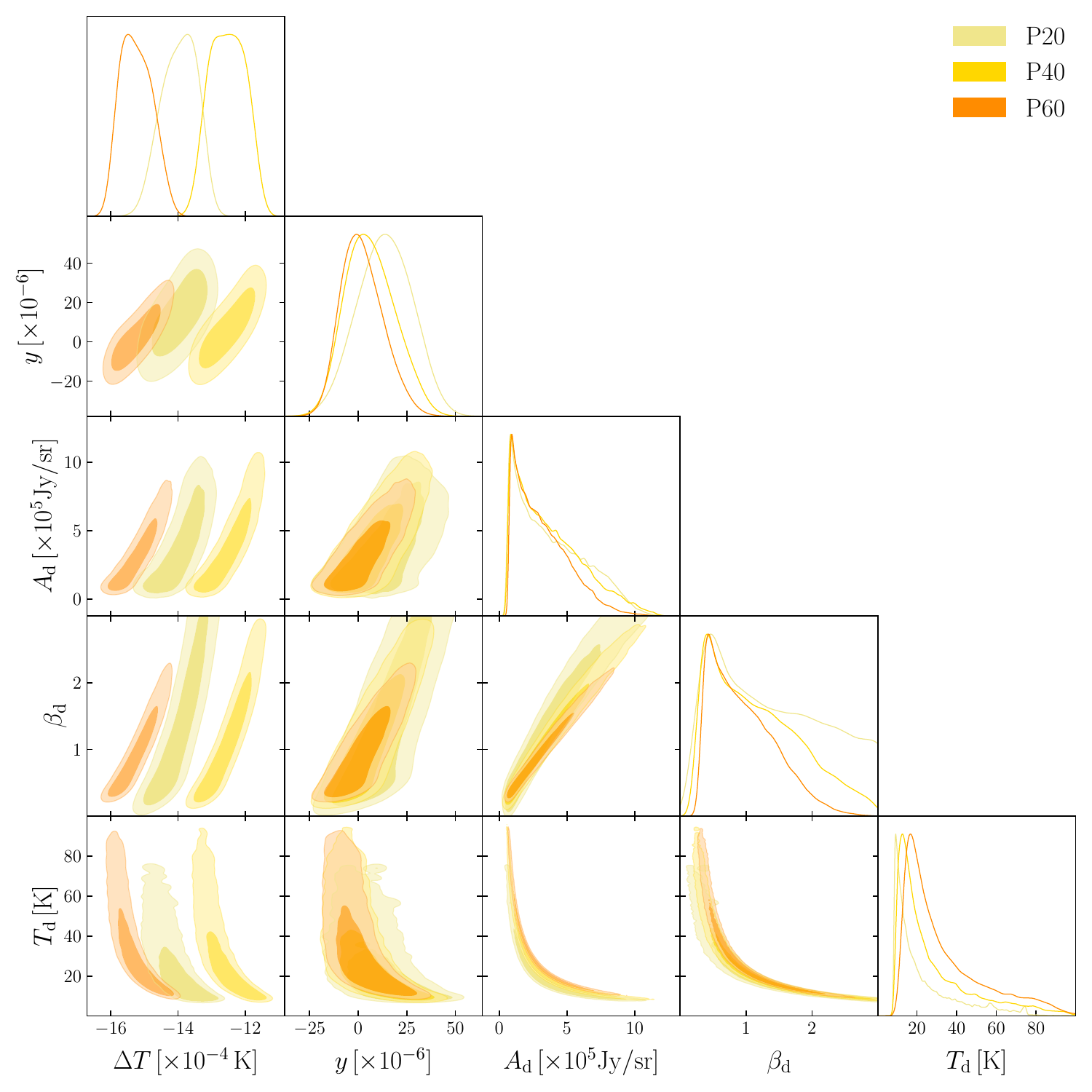}
\caption{Marginalized parameter posteriors from fitting the \textit{frequency monopole} data with a sky model consisting of the blackbody component, $y$-distortion, and dust (MBB) with flat priors on both MBB spectral parameters: $\beta_{\rm d}\in [0, 3)$ and $T_{\rm d}\in [0, 100)$~K using $\nu_{800}$ (set-up 14 in Table \ref{tab:models}).}\label{fig:dust_flat_model_data}
\end{figure}

Fig.~\ref{fig:best_fit} shows the best-fit model for our baseline results: $\nu_{600}$ using P40. The plot includes the fit to the data, as well as the curves for the individual components of the sky model. For clarity, we plot the data and models with the blackbody CMB included (i.e., including a blackbody spectrum at $T_{0}=2.7255$~K, unlike in Fig.~\ref{fig:monopoles} where we plot the data points after subtracting the reference blackbody). In this set-up, we find $\chi^{2}=28.2$ for 22 degrees of freedom (DOF), which corresponds to a PTE of 0.168, indicating a good fit. The residuals after we subtract the model from the data are within 2.5$\sigma$ in all frequency channels, as seen in the bottom panel of the plot.

In order to have a direct comparison with the results from the \textit{pixel-by-pixel} method, we also run MCMCs using analysis set-ups 13-14 in Table~\ref{tab:models}, where we assume flat priors on both $\beta_{\rm d}\in[0, 3)$ and $T_{\rm d}\in[0, 100)$~K. We find that the MCMC chains do not fully converge using $\nu_{600}$ with these flat priors, as previously also found in the tests on mocks. The low-frequency data likely do not have enough sensitivity to constrain both dust parameters, resulting in a very low effective sample size for many of the parameters. On the other hand, the $\nu_{800}$ runs converge and we include those results in Table~\ref{tab:y_results} and in Fig.~\ref{fig:dust_flat_model_data}. As expected, the errors and upper limits slightly worsen with the flat priors in most cases, but are consistent with the results using the Gaussian prior on $\beta_{\rm d}$. We find that the results with flat priors are consistent across different sky masks for both frequency ranges according to the $C_{\rm mask}$ criterion.

Additionally, we consider fitting the relativistic correction to the $y$-distortion to place constraints on the average temperature of the electrons sourcing the tSZ effect. However, the mean temperature is expected to be on the order of 1~keV, resulting in a correction that is about an order of magnitude smaller than the $y$-distortion signal~\cite{Hill2015,Thiele2022}. Therefore, we expect the \textit{FIRAS} data to be even less sensitive to the relativistic correction than to $\ymono$ itself. As a test, we run MCMC fits using $\nu_{600}$ and P40, fitting for the tSZ relativistic correction using the computation implemented in \verb|SZpack| \cite{szpack_1, szpack_2, szpack_3, lee2022, sazanov1998, Challinor1998, Itoh1998, nozawa1998, Nozawa2006}. In this set-up, we again only include dust as the foreground. We find that, indeed, the \textit{FIRAS} data are not sensitive to the relativistic correction: we recover the prior in our MCMCs (i.e., a flat prior on $kT_{\rm e}\in(0, 190]$~keV). 

\subsection{Fisher Forecast Comparison}\label{sec:fisher_comparison}

One of our main motivations for performing the \textit{frequency monopole} analysis on the \textit{FIRAS} data is to check the accuracy of CMB spectral distortion forecasts that have been performed in recent studies, which use the Fisher matrix formalism assuming a sky-averaged intensity spectrum.  For example, the sky model and the Fisher forecast set-up in \citetalias{Abitbol2017} have been used to forecast the detection significance for both $\mu$- and $y$-distortions from future monopole measurements, such as those from the \textit{PIXIE}~\cite{pixie2011, pixie2024} and \textit{BISOU}~\cite{Bisou} experiments and new instrument concepts like \textit{SPECTER}~\cite{specter}. The Fisher forecast relies on the \textit{frequency monopole} modeling of all the sky components, which is different from the \textit{FIRAS} data analyses carried out by the \textit{FIRAS} team and \citetalias{BianchiniFabbian2022}. Therefore, it is important to perform this consistency check to ensure that the spectral distortion forecasts are consistent with actual results from data, when applied to \textit{FIRAS}.

Here, we compare our constraints from the \textit{frequency monopole} to error bars on $\ymono$ computed with the Fisher forecast set-up used in \citetalias{Abitbol2017} \footnote{\url{https://github.com/asabyr/sd_foregrounds/tree/firas} (modified version of \url{https://github.com/mabitbol/sd_foregrounds})}. Assuming a fiducial sky-averaged frequency monopole $I_{\nu}$ with free parameters $\theta$ indexed with $i\,,j$, and a noise covariance matrix $C_{\nu\nu'}$, the Fisher information matrix can be written as 
\begin{equation}
    F_{ij}=\sum_{\nu,\nu'} \frac{\partial I_{\nu}}{\partial \theta_{i}}C_{\nu\nu'}^{-1}\frac{\partial I_{\nu'}}{\partial \theta_{j}}
\end{equation}
Here, $\nu$ are the frequency channels (Sec.~\ref{sec:freq_ranges}) and $C_{\nu\nu'}$ is the \textit{FIRAS} frequency-frequency covariance used in the \textit{frequency monopole} analysis. The parameter covariance matrix can then be computed by inverting $F_{ij}$, i.e., the Fisher error $\sigma_{i}$ is given by $\sigma_{i}=\sqrt{(F^{-1})_{ii}}$. We use the fiducial sky model parameter values from \citetalias{Abitbol2017} (with dust re-parametrized for a fixed $\nu_{0}=353$ GHz) and Gaussian priors on $\beta_{\rm d}$ where applicable:
\begin{itemize}
    \item $\Delta B_{\nu}$: $\Delta T=1.2\times10^{-4}\times T_{0}=3.2706\times10^{-4}$ K
    \item $I_{\nu}^{y}$: $y=1.77\times10^{-6}$
    \item $I_{\nu}^{\rm dust}$: $A_{\rm d}=5.14\times10^{5}$ Jy/sr, $\beta_{\rm d}=1.53$ (with a $6.6\%$ prior on $\beta_{\rm d}$), $T_{\rm d}=21$ K.
\end{itemize}
To account for the integration of the sky signals over the passbands, we compute the sky model averaged over the frequencies within each $13.6$~GHz band of \textit{FIRAS}. We find this to be a very small effect compared to simply using the central frequencies, consistent with what was found in \citetalias{Abitbol2017} and Ref.~\cite{specter}.

Our results using the fiducial set-up from \citetalias{Abitbol2017} are summarized in Table \ref{tab:y_results}. The error bars from our data analysis and the Fisher forecast show very good agreement for the fits with a Gaussian prior on $\beta_{\rm d}$ (column 2 and column 5).  The Fisher and MCMC results agree within $\approx 10\%$ or better across sky masks and frequency ranges.

\begin{table*}[t]
\begin{tabular}{|l|ccc|c|}
\colrule
&&Fisher [$\times 10^{6}$]&&Data [$\times10^{6}$]\\
\colrule
Model & $\sigma_{\ymono} \mathrm{(fid)}$ & $\sigma_{\ymono} \mathrm{(mean)} $& $\sigma_{\ymono}$ (best-fit)&$\sigma_{\ymono}$\\
& P20/P40/P60&P20/P40/P60&P20/P40/P60&P20/P40/P60\\
\colrule
Dust ($\nu_{600}$) &13.4/9.5/8.0&15.4/10.2/8.6&15.6/10.5/8.8&$^{+15.6}_{-11.8}$/$^{+10.7}_{-9.7}$/8.7\\
Dust ($\nu_{\rm 800}$)&9.6/7.0/6.1&10.3/7.1/5.8&10.5/7.1/5.8&9.7/7.0/5.8\\
\colrule
\end{tabular}
\caption{\label{tab:fisher_models} Comparison of error bars from the Fisher forecast using fiducial sky model parameters and parameters found in our fits to the data (the posterior mean and best-fit values).  The error bars from our MCMC runs are listed in the last column.}
\end{table*}

The mean values for the parameters from our MCMC posterior results differ from the fiducial values for the sky model parameters used in \citetalias{Abitbol2017}. Thus, it is promising that the forecast error bars are still consistent using the fiducial values from \citetalias{Abitbol2017}. We explore the effects of using different parameter values in the forecast on the size of the Fisher errors by considering parameter values in the Fisher calculations determined from our data-analysis fits. Table~\ref{tab:fisher_models} summarizes the Fisher error bars for the different sky model parameters: the fiducial set-up in \citetalias{Abitbol2017}, our posterior mean parameters from the MCMC runs, and our best-fit values from the minimization in Table~\ref{tab:models}. The Fisher errors from using the mean or best-fit parameter values are also in good agreement across different sky model assumptions. In the cases of the largest discrepancies (P20, $\approx 10-15\%$ differences), the Fisher error bars tend to be more conservative. 

For flat priors, we see larger discrepancies between Fisher error bars and the results obtained from data. Using $\nu_{800}$ and fiducial \citetalias{Abitbol2017} parameters, the Fisher error bars are $\approx30\%$ larger for P20 but only a $\sim$ few percent larger for P40/P60. When using posterior mean and best-fit parameter values, we find that the Fisher error bars can be $>10\%$ larger across all sky fractions compared to the constraints from data. Using $\nu_{600}$, we find that the Fisher errors are not consistent with results from data. Since we find that these runs are not fully converged and we do not expect to be able to constrain both dust parameters with these low frequencies, such discrepancies are not unexpected and unlikely to be meaningful.

\section{Results from Data: \textit{pixel-by-pixel} Method}\label{sec:data_results_pixpix}
\subsection{Constraints}

Similarly to the \textit{frequency monopole} method, we apply the \textit{pixel-by-pixel} method using our baseline foreground model, which includes $A_{\rm d}, T_{\rm d}$, and $\beta_{\rm d}$, to the \textit{FIRAS} data. In Table~\ref{tab:pix-pix-data} we report a summary of our results together with some variations and consistency tests. Our reference set-up adopting flat priors on both $T_{\rm d}$ and $\beta_{\rm d}$, and using the low-frequency data ($\nu_{600}$), recovers values of $\ymono$ that are consistent with zero.

Fluctuations between the constraints obtained from different sky fractions are consistent at the $\approx 2.3\sigma$ level with the hypothesis of statistical fluctuations following our consistency criterion, $C_{\rm mask}$, discussed in Sec.~\ref{sec:mock_results} and Eq.~\eqref{eq:consistency-test}. 

The measurements on the P20/P40/P60 masks have $\approx 4$ times smaller errors than those obtained with the \textit{frequency monopole} method for the same sky fractions. The corresponding 1-sided upper limits $\ymono < 4.3/7.7/8.3\times 10^{-6}$ at 95\% C.L. are $\approx 9/4/3$ times tighter than what is achieved with the \textit{frequency monopole} method. We note that the low value of the P20 upper limit is due to the fluctuation of $\ymono$ to negative values, and 2-sided upper limits are all consistent with $\approx 3-4$ times improvements, consistent with the error bar improvements.  Similarly, we find an improvement of a factor of $\approx3-4$ on the 68\% C.L.~error bars, consistent with the expectations from the mock data.  We also note that the \textit{pixel-by-pixel} estimate, taking P60 as the baseline, improves upon the original \textit{FIRAS} results ($\ymono^{\rm FIRAS} = -1\pm 6\ (\rm stat.) \pm 4\ (\rm syst.) \times 10^{-6}$ and $\ymono  \leq 15 \times 10^{-6}$ at 95\% C.L.) by a factor of $\approx2$ in terms of the upper limit and by a factor of $\approx3$ in terms of the statistical error bar. In this paper, we focus on the comparison of the component separation methods and we refer the reader to Ref.~\cite{Fabbian2023} for discussion on the implications and robustness of this measurement, as well as ways to improve it including additional frequency bands.

Since in the \textit{frequency monopole} case we had to resort to sub-optimal inverse-variance weighting of the sky pixels due to the uncertainty in the modeling of the \textit{FIRAS} pixel-pixel covariance matrix, we also compute an estimate of $\ymono$ using a simple inverse-variance weighted average of all the individual pixel $y$ values. In this case, we obtain $\ymono^{\texttt{ivar}} = (5.7\pm 2.8) \times 10^{-6}$ on the P60 mask and an upper limit $\ymono < 11.3 \times 10^{-6}$ at 95\% C.L. The uncertainty on the estimate and the upper limit are about $\sim 30\%$ higher than our optimal baseline analysis, leading to an upper limit closer to (but still tighter than) the \textit{frequency monopole} results and with an improvement on the statistical error by a factor of 2 with respect to the original \textit{FIRAS} analysis.  This is similar to the results of \citetalias{BianchiniFabbian2022}, where no optimal inverse-covariance weighting obtained from the MCMC samples was employed.

Similar degradations (by $\approx 20-50\%$) due to suboptimal weighting are seen for all the set-ups that we investigate and discuss later in the text, pointing to the importance of accurate noise modeling in achieving the tightest possible constraint on the quantities, which are averaged over many sky pixels as in the case of the monopole signal.

\begin{table}[t]
\begin{tabular}{|l|c|c|c|c|}
\colrule
Model & $\times 10^{6}$ & P20&P40&P60\\
\colrule
Dust ($\nu_{600}$)& $\ymono$ &-3.3&2.8&4.3\\
&$\sigma_{\ymono}\ 68\%$ C.L.& $\pm 3.8$ &$\pm 2.4$&$\pm 2.0$\\
&$\ymono\leq 95\%$ C.L.&4.3 &7.7 &8.3\\
\colrule
Dust ($\nu_{\rm 800}$) & $\ymono$&10.2&7.2& 4.3\\
&$\sigma_{\ymono}\ 68\%$ C.L.&$\pm 2.7$&$\pm 1.9$&$\pm 1.6$\\
&$\ymono\leq 95\%$ C.L.&--& --& 7.5\\
\colrule
Dust ($\nu_{600}$)& $\ymono$ &-0.7&11.8&16.1\\
$\beta_{\rm d}\in\mathcal{N}(1.5,0.1)$& $\sigma_{\ymono}\ 68\%$ C.L.& $\pm 4.1$ &$\pm 2.5$&$\pm 2.0$\\
&$\ymono\leq 95\%$ C.L.&7.5 &-- &--\\
\colrule
Dust ($\nu_{\rm 800}$) & $\ymono$&34.0&37.1& 37.2\\
$\beta_{\rm d}\in\mathcal{N}(1.5,0.1)$&$\sigma_{\ymono}\ 68\%$ C.L.&$\pm 2.6$&$\pm 1.8$&$\pm 1.5$\\
&$\ymono\leq 95\%$ C.L.&--&--&--\\
\colrule
\end{tabular}
\caption{Summary of $\ymono$ constraints obtained with the \textit{pixel-by-pixel} method. We do not quote the upper limits for the cases where non-zero $\ymono$ is detected due to biases as discussed in Sec.~\ref{sec:data_results_pixpix} and Sec.~\ref{sec:mock-resuls-pixpix}.}
\label{tab:pix-pix-data}
\end{table}

We check the robustness of our results using a $\beta_{\rm d}$ prior that matches the one adopted in the \textit{frequency monopole} case, i.e., $\beta_{\rm d}\in \mathcal{N} (1.51, 0.1)$. In this set-up, we find that the results are not consistent across the sky masks and that we obtain a non-zero value for $\ymono$ at high significance using P40 and P60. For P20, the constraints are consistent with zero and with the results obtained with a flat prior on $\beta_{\rm d}$. Such findings are consistent with our results based on mocks, where we found that imposing a tight prior on $\beta_{\rm d}$ seems to prevent the possibility to re-absorb the superposition of Galactic dust and CIB emission in regions of the sky where they are both important. At high Galactic latitudes where the Galactic dust is less bright, the degeneracy between $T_{\rm d}$ and $\beta_{\rm d}$ in this frequency range allows us to fit the MBB spectrum without biases since the emission is presumably dominated by the CIB. We also note that the uncertainties are consistent with our findings on mock data. Comparing the P20 results with a Gaussian prior to the \textit{frequency monopole} results, we find that \textit{pixel-by-pixel} still outperforms by about a factor of $\approx3$ in the statistical error bar.

Next, we use the $\nu_{800}$ frequency range, adopting the same Gaussian prior on $\beta_d$.  In this case, we find that we detect large values of $\ymono \approx 30 \times 10^{-6}$ on all masks, consistent with our findings on mock data. Relaxing the prior on $\beta_{\rm d}$ to our baseline flat prior leads to a detection of lower values of $\ymono$, except on the P60 mask. All these findings are consistent with the expectations from mock data, including in terms of the errors. Such findings indicate that more complex foreground models with more flexibility must be used if higher frequencies are included in the \textit{pixel-by-pixel} analysis, and that component separation becomes more challenging, leading to a more complex behavior of the foreground residuals that is potentially highly dependent on the sky fraction assumed. Given the noise level of \textit{FIRAS} and the large calibration uncertainties in the high-frequency instrument of \textit{FIRAS}, we conclude that adopting $\nu_{600}$ yields more stable results across all the different masks and is less affected by the parameter degeneracies and the discrepancies between the simple parametric sky model and the true sky signal complexity.

\subsubsection{Consistency tests}
As a consistency check, we estimate the value of the CMB monopole temperature, similar to what was done in \citetalias{BianchiniFabbian2022}. We compute the temperature via the inverse-variance weighted average of the $\Delta T$ map extracted from the data on the P60 mask. We find a CMB monopole temperature of $T_0\approx 2.724$~K, which is slightly lower than the commonly used $T_0=2.7255$~K, but compatible with the value found in \citetalias{BianchiniFabbian2022}. We note that the latter value has been obtained estimating the frequency spectrum of the CMB dipole and used the\textit{WMAP} measurement of the CMB dipole anisotropy to recalibrate \textit{FIRAS} data and not from a direct measurement of the CMB monopole temperature. Shifts as high as $5$~mK from the value of $T_0$ estimated from the CMB monopole spectrum are expected in the \textit{FIRAS} data release that we employed in this work. As explained in the data release documentation \footnote{\url{https://lambda.gsfc.nasa.gov/product/cobe/firas_tpp_info.html}}, the procedure used to correct the temperature bias in the high-current readout of the instrument, which is presumably due to self-heating of the thermometers themselves, should produce shifts in the direct measurement of $T_0$ from the monopole compared to the original analysis results ($T_0=2.728\pm 0.004$~K, dominated by systematic uncertainties \cite{Fixsen1996_dist}). As such, systematic uncertainties dominate on the statistical error of our $T_0\approx 2.724$~K.

We also find the direction and the amplitude of the CMB dipole to be\footnote{We obtain the estimates of the direction and amplitude of the CMB dipole as point estimates using the \texttt{remove\_dipole} routine of the \texttt{healpy} package. As an accurate estimate of these quantities is beyond the scope of this work, we assess the consistency with other data sets using their specific error bars.}  $A_{\rm dipole} \sim 3376\,\mu$K  pointing in the $(l, b) \sim (263.04^\circ, 48.16^\circ)$ direction in Galactic coordinates. The dipole amplitude is consistent with the original estimate from the \textit{FIRAS} team $A_{\rm dipole}^{\rm FIRAS} = 3369 \pm 40 \,\mu $K  \cite{Fixsen1996_dist}, while the direction is only slightly discrepant with the original estimate ($(l, b)_{\rm FIRAS} = (264.14^\circ \pm 0.15^\circ, 48.26^\circ \pm 0.15^\circ )$) at $\approx 3\sigma$ level, driven by the differences in the $l$ direction. As noted in \citetalias{BianchiniFabbian2022}, similar weak discrepancies are removed if no spectral distortions are fit from the data. We note that using mock data, we recover a dipole amplitude and direction that are consistent with the values measured on \textit{Planck} data and used to generate the mocks.

Finally, we compute the monopole of the $A_{\rm d}$ map. For our reference set-up using $\nu_{600}$ with no priors on $\beta_{\rm d}$, we find $\langle A_{\rm d} \rangle = 0.26 \pm 0.01$ MJy/sr. To interpret this measurement, we assume that the foreground brightness can be described by our Galactic foreground model based on the Commander maps and a CIB component, neglecting low-frequency foregrounds, which are negligible at the pivot frequency where $A_{\rm d}$ is fit. For the CIB component, in addition to the SED used in the generation of the mock data, we consider two other SEDs proposed in the literature: Ref.~\cite{fixsen1998} (hereafter Fixsen98) and Ref.~\cite{gispert2000} (hereafter Gispert00). Our $\langle A_{\rm d}\rangle$ value is consistent with the expectation of our Galactic foreground model plus a Fixsen98 CIB component. Results of $\langle A_{\rm d} \rangle$ obtained when keeping the same maximum frequency but adding a Gaussian prior on $\beta_{\rm d}$ retrieve results more consistent with a CIB component based on the SED of Gispert00. All set-ups with $\nu_{800}$ produce measurements of $\langle A_{\rm d} \rangle$ that cannot be described either as the sum of a Galactic foreground monopole plus a CIB component with any of the major SEDs previously mentioned at $\gtrsim 3\sigma$ significance, or with the assumption of a CIB component alone. We interpret this as a further indication of the robustness of our baseline set-up. While repeating the analysis on the P40 mask gives consistent results, repeating the measurement on the P20 mask to minimize the Galactic foreground contamination gives $\langle A_{\rm d}\rangle = 0.20 \pm 0.01$ MJy/sr. This value is consistent with a Galactic component plus a CIB component based on any of the CIB SEDs, preventing us from clearly distinguishing which one of the CIB SEDs better describes the data on all regions of the sky. Although the Commander maps on which our Galactic foreground models are based might contain a CIB residual (and as such overestimate the average foreground brightness temperature), the hypothesis of a CIB component alone is always inconsistent with our $\langle A_{\rm d} \rangle$ estimates, irrespective of the Galactic mask, suggesting the dust emission cannot be neglected even on the cleanest regions of the sky.

\section{Conclusions}\label{sec:conclusion}
\begin{figure}
    \centering
    \includegraphics[width=0.4\linewidth]{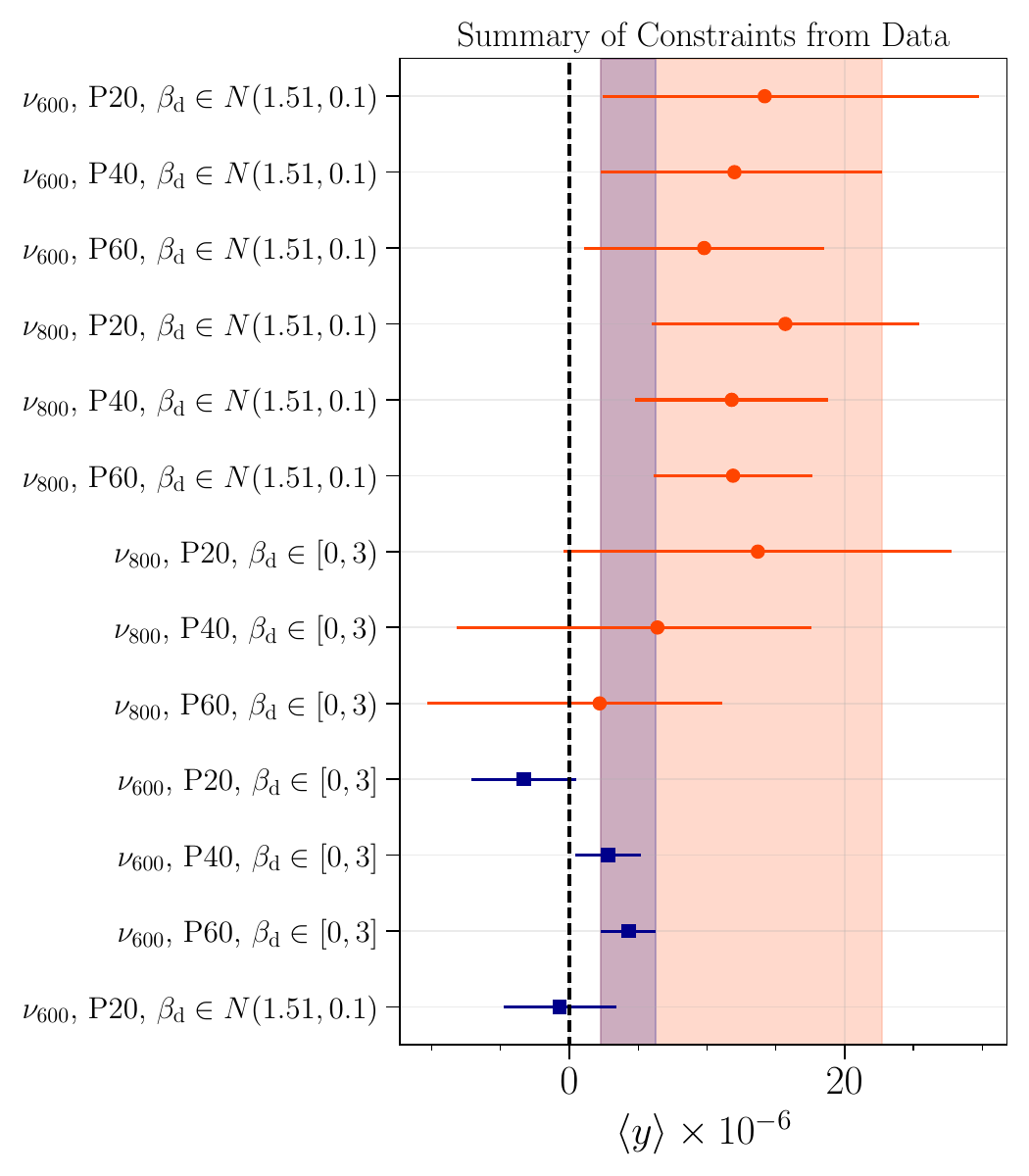}
    \caption{Summary of the results from data using both the \textit{frequency monopole} (orange circle) and \textit{pixel-by-pixel} (blue square) methods. All results are shown for a sky model that includes $\Delta B_{\nu}$, $y$-distortion, and dust.  All error bars here are 68\% C.L.}
    \label{fig:summary_results}
\end{figure}

In this work, we perform a detailed re-analysis of the \textit{COBE/FIRAS} data to constrain the monopole of the $y$-distortion. In particular, we investigate two complementary analysis approaches to foreground removal, one based on the modeling of the sky-averaged signal (\textit{frequency monopole}) and the other on the intensity spectra in each pixel (\textit{pixel-by-pixel}). The summary of the $68\%$ C.L.~errors from the two methods is shown in Fig.~\ref{fig:summary_results}. In the plot, we show the error bars obtained using the baseline sky model (blackbody temperature deviation + $\ymono$ + dust) for both methods across different sky fractions and frequency ranges. Our main results and conclusions can be summarized as follows:

\begin{itemize}
    \item [(1)] Using the \textit{frequency monopole}, we obtain the baseline upper limit $\ymono\lesssim31\times10^{-6}$ ($95\%$ C.L.), using $\nu_{600}$ and P40 (the $1\sigma$ error bar is $\approx10\times10^{-6}$). The upper limit and the statistical error are about a factor of two larger than the results obtained in the original \textit{FIRAS} analysis \cite{Fixsen1996_dist}. We find that with the \textit{FIRAS} data, a simple sky model that includes an MBB SED to model dust as the only foreground component is sufficient to capture any emission in the sky without biasing the final results.
    \item [(2)] Our \textit{pixel-by-pixel} method delivers superior performance compared to the \textit{frequency monopole} method.  We find that it improves the upper limits from the original \textit{FIRAS} analysis by a factor of $\approx2$, giving $\ymono \lesssim 8.3 \times 10^{-6}$ ($95\%$ C.L.). A more complete analysis of the data based on the \textit{pixel-by-pixel} method and the implications for galaxy formation models is presented in our companion paper, Ref.~\cite{Fabbian2023}.
    \item [(3)] For a common set-up where both methods give unbiased and converged constraints (i.e., $\nu_{600}$, P20, and a dust model with a Gaussian prior on $\beta_{\rm d}$ --- see row 1 in Table~\ref{tab:y_results} and row 3 in Table~\ref{tab:pix-pix-data}), we find that the \textit{pixel-by-pixel} method outperforms the \textit{frequency monopole} by a factor of $\approx3$ in statistical error (the 1$\sigma$ errors from \textit{pixel-by-pixel} are $\approx70\%$ smaller). For this set-up, \textit{pixel-by-pixel} provides an upper limit on $\ymono$ that is $\approx$5 times tighter than \textit{frequency monopole}. Comparing the baseline results from both methods (i.e., the $\nu_{600}$, P40, and dust with a Gaussian prior on $\beta_{\rm d}$ results from the \textit{frequency monopole}, and the $\nu_{600}$, P60, and dust with a flat prior on $\beta_{\rm d}$ results from the \textit{pixel-by-pixel}), we find that the \textit{pixel-by-pixel} method gives $\approx5$ times smaller error bar and $\approx4$ times tighter upper limit. This stems from the complexity and variation of the sky signal that is not easily captured by a single parametric model. Taking into account the spatial variation of the sky emission via the \textit{pixel-by-pixel} approach leads to significant improvement in the constraints. In the high noise regime of \textit{FIRAS}, we also find the \textit{frequency monopole} method to be more sensitive to imperfections in the pixel-pixel correlations of the noise modeling, while the \textit{pixel-by-pixel} method is naturally more sensitive to mismodeling of the correlated noise across frequencies. No realistic noise simulations of the \textit{FIRAS} data are publicly available, which prevents us from quantifying or investigating these aspects further, but we find the \textit{pixel-by-pixel} method to be better characterized in the case of the \textit{FIRAS} data. 
    \item [(4)] We validate, for the first time in a spectral distortion analysis, our findings on realistic mocks that include astrophysical foregrounds and \textit{FIRAS} noise.  We find very good agreement with the conclusions derived from data. We stress that at the \textit{FIRAS} noise levels, the \textit{frequency monopole} approach is able to successfully fit more complex foreground models, while the \textit{pixel-by-pixel} method is limited by degeneracies and lack of convergence due to the large noise in each sky pixel compared to the sky-averaged signal. 
    \item [(5)] We show that the results obtained from the \textit{frequency monopole} method are consistent with error bars from Fisher forecasts, which have been used to assess future spectral distortion missions and assume a sky-averaged monopole data vector. Since we find that the \textit{pixel-by-pixel} method substantially outperforms the \textit{frequency monopole} for \textit{FIRAS}, we conclude that Fisher-based spectral distortion forecasts in recent studies are likely conservative. Our results directly show that tighter constraints can be achieved by taking into account the spatial variation of the foregrounds, which thus suggests promise for tighter spectral distortion constraints from upcoming experiments.
\end{itemize}

Our findings show that both the \textit{pixel-by-pixel} and the \textit{frequency monopole} analysis methods have unique advantages and shortcomings. While the results in this work indicate that the \textit{pixel-by-pixel} approach delivers both robust and more constraining results, the details of some of our findings depend on the noise properties and the frequency range of the \textit{FIRAS} instrument. Our work suggests that in order to take full advantage of the data from future spectral distortion experiments (e.g., \textit{BISOU}~\cite{Bisou, Bisou2022}, \textit{Voyage2050}~\cite{Voyage2050}, \textit{PIXIE}~\cite{pixie2011, pixie2024}, \textit{COSMO}~\cite{Cosmo}, or \textit{SPECTER}~\cite{specter}), it will be highly useful to perform similar analyses and assess the most optimal strategy to remove foregrounds.

\section{Acknowledgements}

We are very grateful to Oliver Philcox for the assistance with \verb|NUTS| implementation in \verb|numpyro|. We also thank Chirag Modi, Dan Foreman-Mackey, and Soichiro Hattori for recommendations regarding sampling, and Jens Chluba and Fiona McCarthy for moment expansion discussions. We are also very grateful to Jens Chluba for useful discussions about fitting the tSZ relativistic correction and assistance with the \verb|SZpack| software.  AS and JCH acknowledge support from NASA grant 80NSSC23K0463 (ADAP).  JCH also acknowledges support from the Sloan Foundation and the Simons Foundation.  GF acknowledges the support of the European Research Council under the Marie Sk\l{}odowska Curie actions through the Individual Global Fellowship No.~892401 PiCOGAMBAS, of the European Union’s Horizon 2020 research and innovation program (Grant agreement No. 851274) and of the STFC Ernest Rutherford fellowship during the final stages of this work. GF also acknowledges the hospitality of the Center for Computational Astrophysics at the Flatiron Institute during the initial stages of this work. We acknowledge computing resources from the Flatiron Institute and Columbia University's Shared Research Computing Facility project, which is supported by NIH Research Facility Improvement Grant 1G20RR030893-01, and associated funds from the New York State Empire State Development, Division of Science Technology and Innovation (NYSTAR) Contract C090171, both awarded April 15, 2010. We acknowledge the use of the NSF XSEDE facility for the data analysis in this study. The authors acknowledge the Texas Advanced Computing Center (TACC) at The University of Texas at Austin for providing computational resources that have contributed to the research results reported within this paper. We acknowledge the use of the following open-source software throughout this work: \verb|GetDist|~\cite{getdist}, \verb|numpy|~\cite{numpy}, \verb|matplotlib|~\cite{matplotlib}, and \verb|scipy|~\cite{scipy}.

\appendix

\section{Additional Contour Plots}\label{app:extra_contours}

\begin{figure*}[t]
\begin{minipage}[c]{0.5\linewidth}
    \centering
    \includegraphics[width=\linewidth]{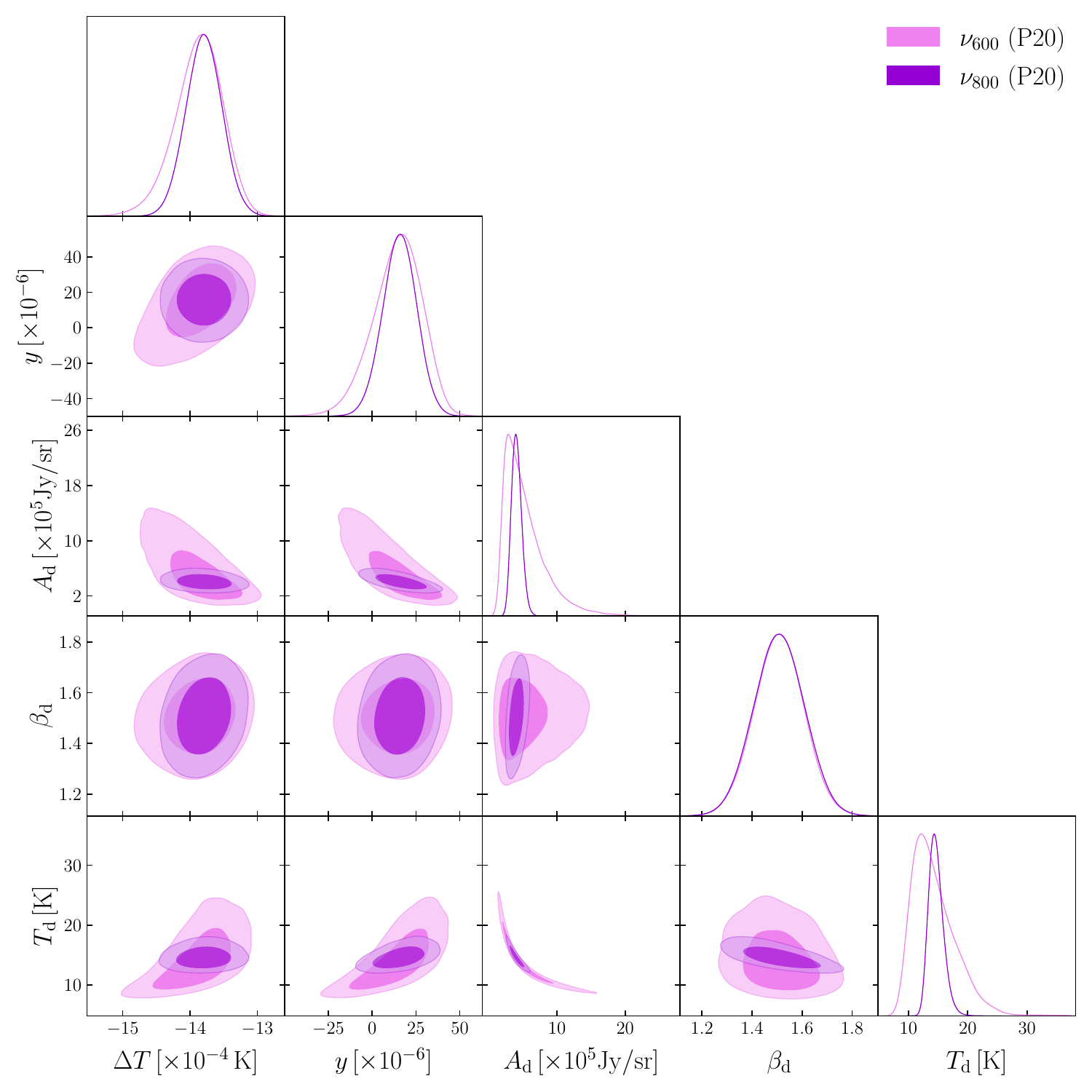}
\end{minipage}\hfill
\begin{minipage}[c]{0.5\linewidth}
    \centering
    \includegraphics[width=\linewidth]{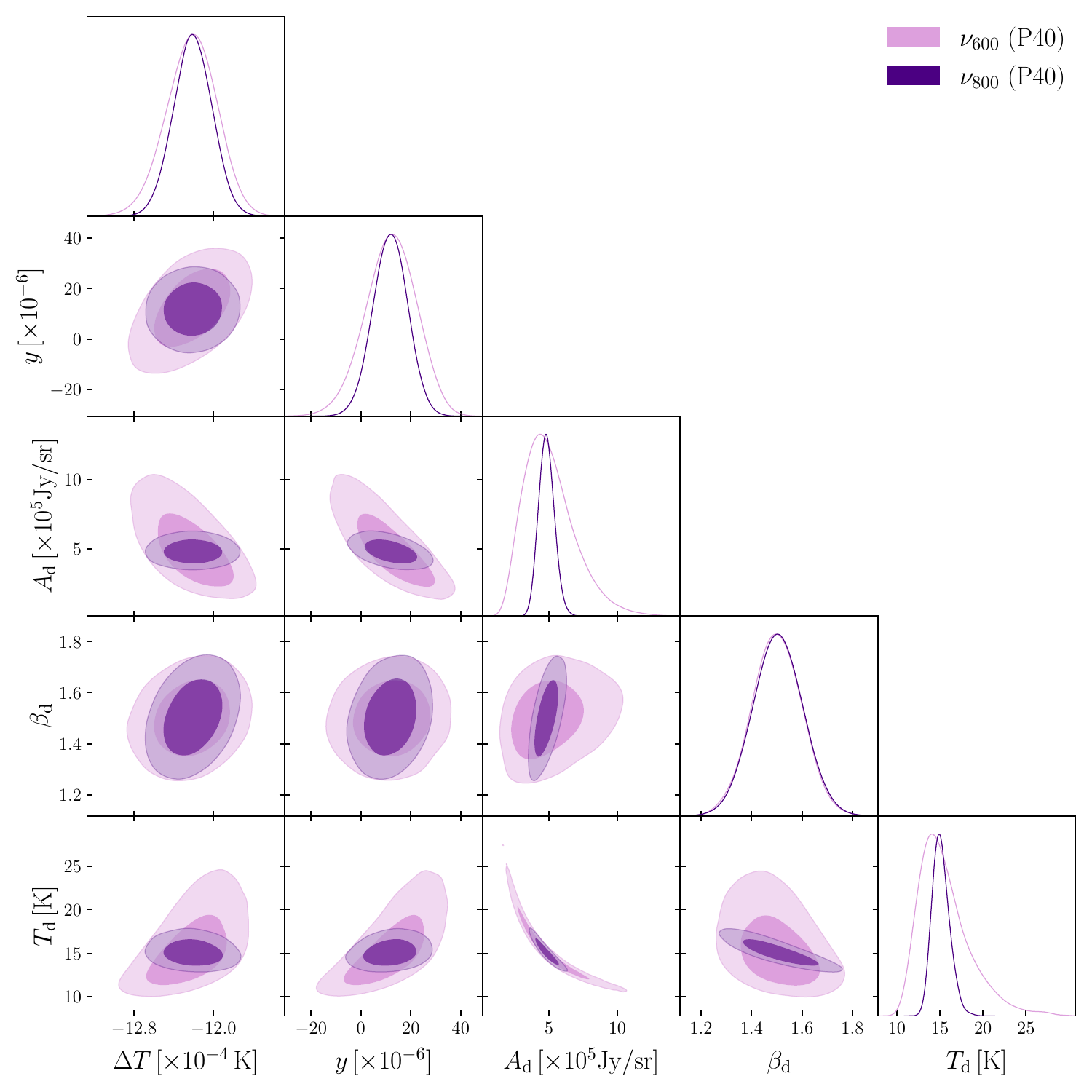} 
\end{minipage} 
\caption{Comparison of marginalized posteriors from fitting low-frequency data alone versus including some channels from the high-frequency instrument. The plots show the results from fitting the same sky model (including only dust as the foreground) and sky fraction (P20/P40) but for different frequency ranges ($\nu_{600}/\nu_{800}$). Both CMB and foreground components are consistent within errors between the two frequency ranges.}
\label{fig:baseline_low_vs_800}
\end{figure*}

In this appendix, we include additional contour plots from our main analysis. In Fig.~\ref{fig:baseline_low_vs_800}, we show the posterior results from fitting the baseline sky model to $\nu_{600}$ and $\nu_{800}$ (set-ups 1-2 in Table~\ref{tab:models}) for P20/P40. The plots show that the results are consistent between the two frequency ranges.

\section{Additional Foreground Components}\label{sec:more_fgs}
We find very low PTE values for the P60 \textit{frequency monopole} fits with our baseline model that only includes dust. We check if the fits are improved with additional foreground components in the sky model. Table \ref{tab:models} summarizes some of the set-ups that we try, which include adding extragalactic CO, free-free, synchrotron, and a fixed CIB SED. At the frequencies used in this analysis, extragalactic CO and Galactic free-free are most likely the strongest sources of emission after dust (see Fig.~2 in \citetalias{Abitbol2017} for contributions from different foregrounds using 70$\%$ of the sky).

\begin{figure*}[t]
\begin{minipage}[c]{0.5\linewidth}
    \centering
    \includegraphics[width=\linewidth]{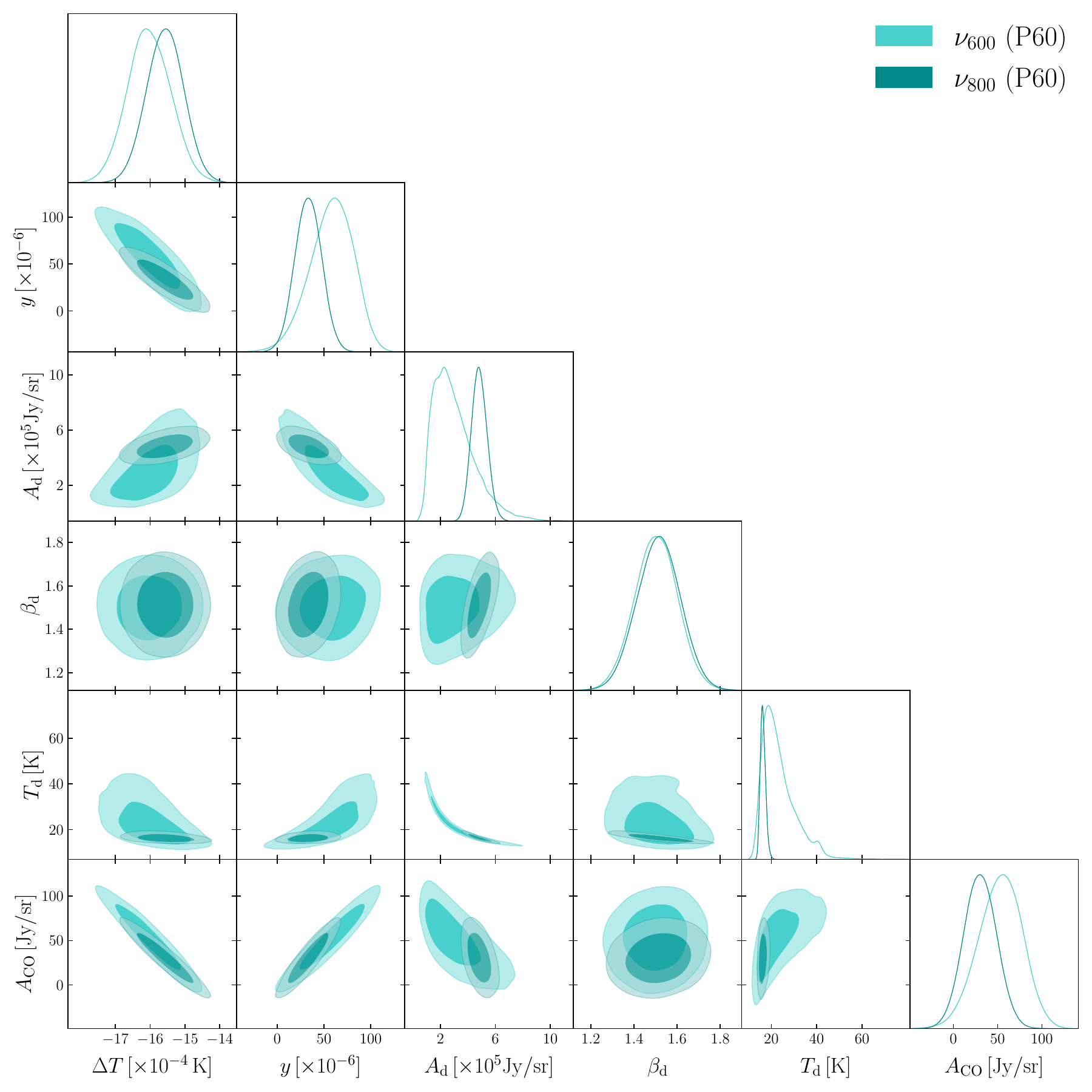}
\end{minipage}\hfill
\begin{minipage}[c]{0.5\linewidth}
    \centering
    \includegraphics[width=\linewidth]{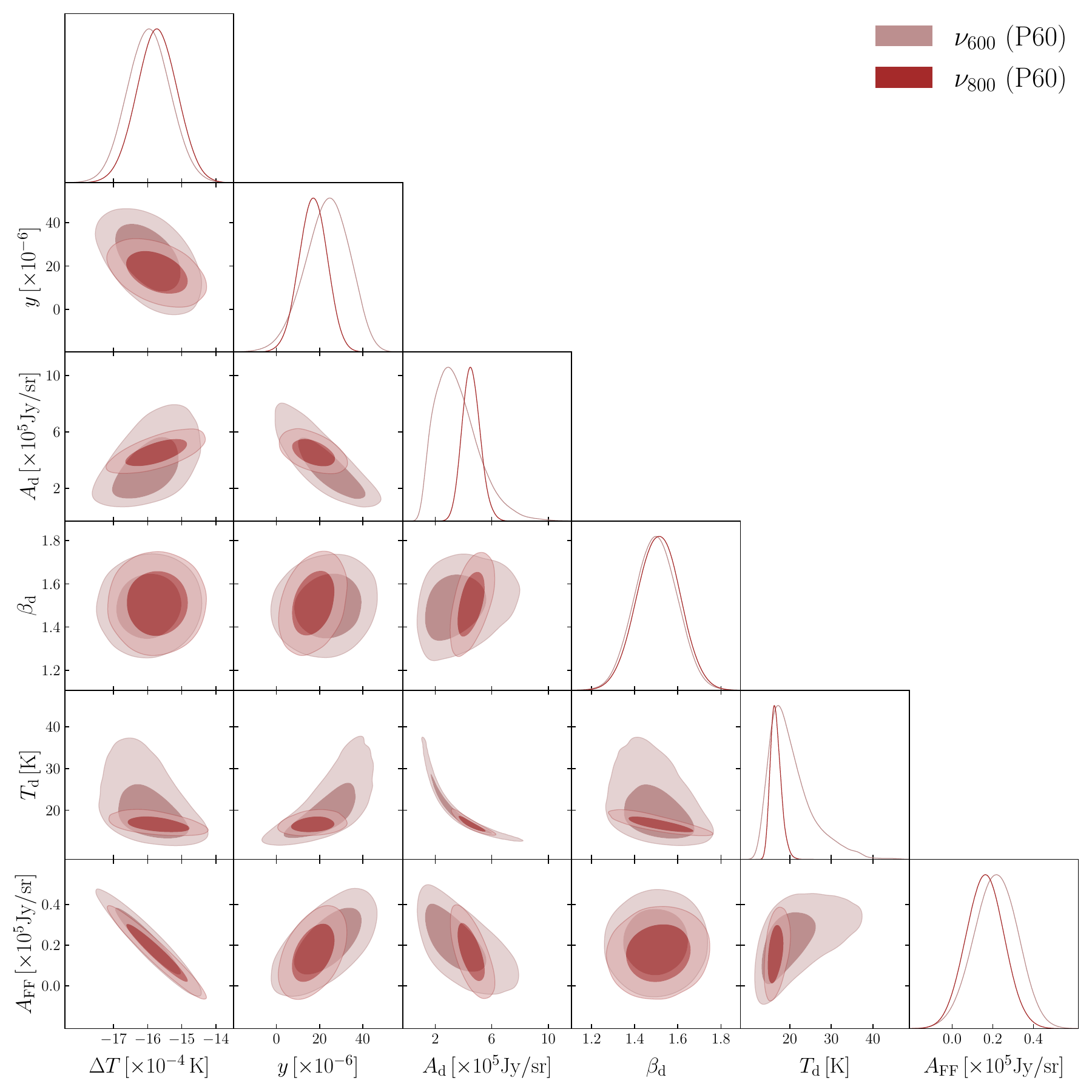}
\end{minipage} 
\caption{Marginalized posterior results from fitting sky models that include extragalactic CO (left) and free-free (right) in addition to dust (set-ups 3-4 and 5-6 in Table~\ref{tab:models}, respectively). The plots show results for the two frequency ranges considered in this paper using the P60 sky mask.}\label{fig:CO_FF}
\end{figure*}

\begin{figure}[!h]
    \begin{minipage}[c]{0.5\linewidth}
    \centering
    \includegraphics[width=\linewidth]{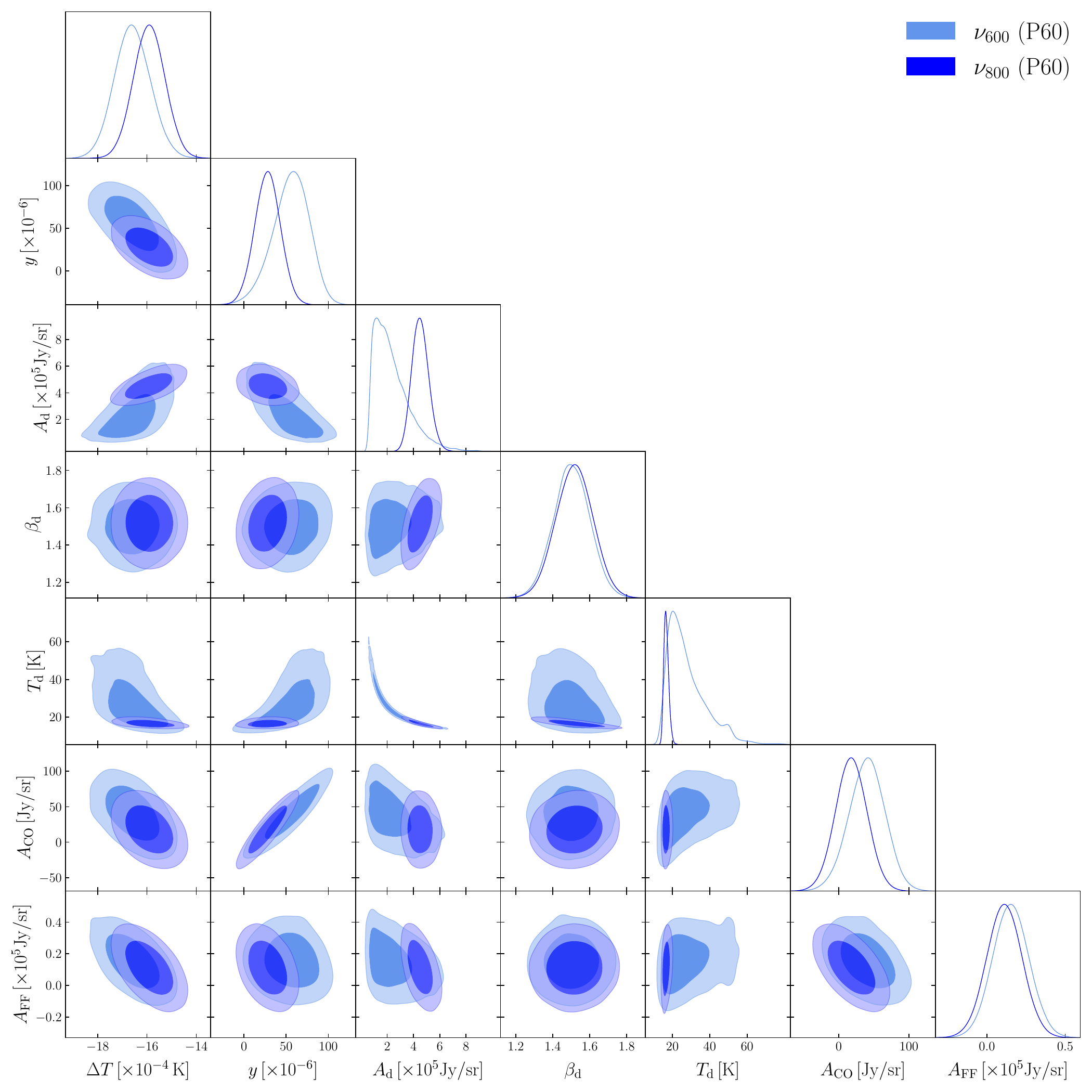}
\end{minipage}\hfill
\begin{minipage}[c]{0.5\linewidth}
    \centering
    \includegraphics[width=\linewidth]{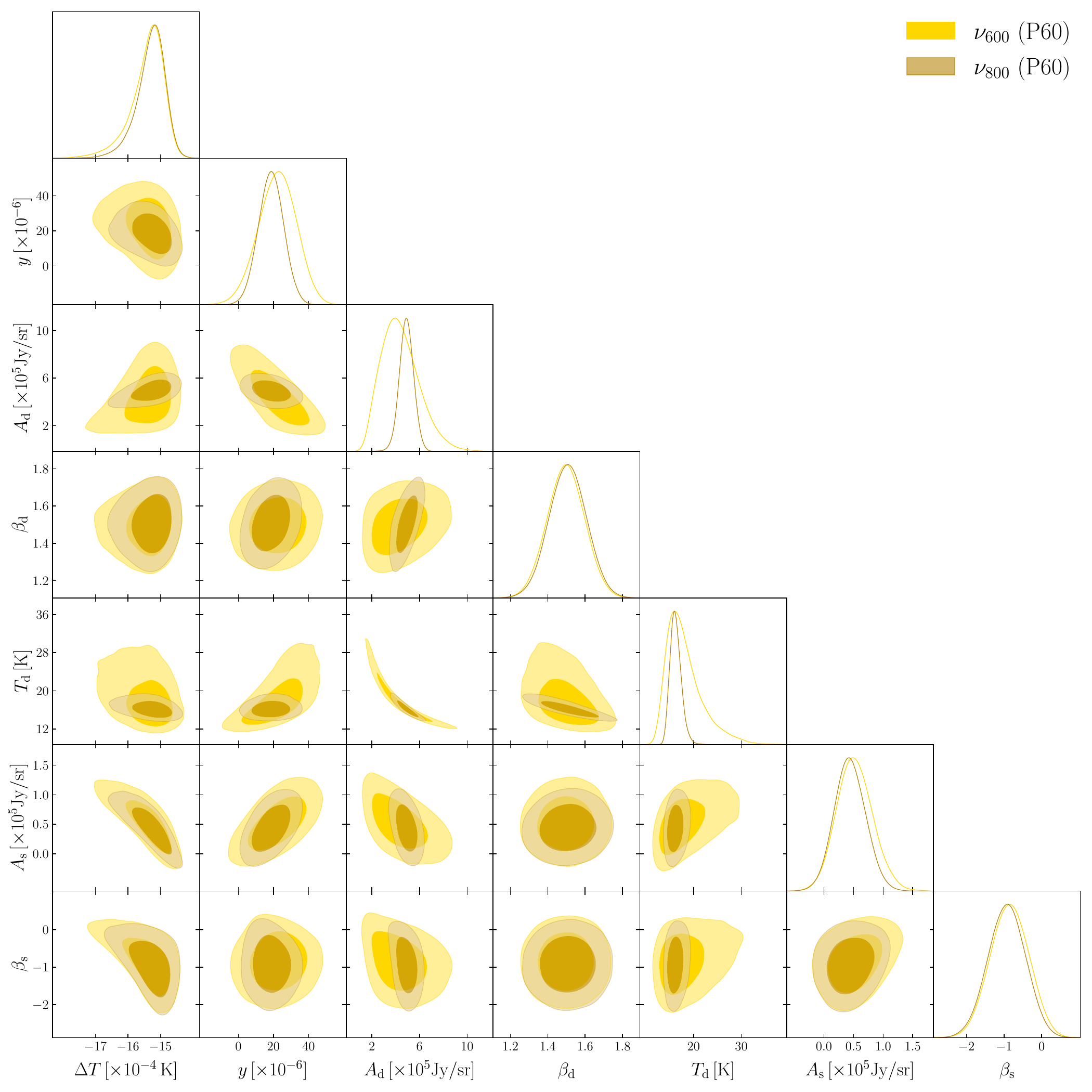}
\end{minipage} 
\caption{\textit{Left:} Marginalized posterior results for fitting sky models, which include extragalactic CO and FF (set-ups 7-8). \textit{Right:} Results for fits that include synchrotron (set-ups 11-12). The results are shown for two frequency ranges and for P60 sky mask. \label{fig:CO+FF_sync}}
\end{figure}

We find that the PTE for P60 is indeed improved with addition of these two components and becomes marginally acceptable for set-ups that only fit the low frequencies ($\nu_{600})$. Adding a fixed CIB SED also helps fit the P60 spectrum better. Although at very low frequencies the foreground emission is dominated by synchrotron radiation, when we check adding synchrotron to our sky model, we find that the fits are not significantly improved ($\Delta \chi^{2}\approx -3$ for P60). This is likely because our data only includes frequencies $> 95$ GHz. In this fit, we include a Gaussian prior on the synchrotron spectral index, $\beta_{\rm s} \in  \mathcal{N}(-1.0,0.5)$, motivated by \planck (note that $\beta_{\rm s} \sim -1$ in intensity corresponds to $\sim-3$ in temperature units). The data do not have enough sensitivity to detect any of the aforementioned additional foregrounds with either the $\nu_{600}$ or $\nu_{\rm 800}$ frequency ranges. Contrary to what we expect from the improved $\chi^{2}$/PTEs, the value of $\ymono$ shifts away from zero and is biased with the addition of these foreground components, which suggests that systematics may be at play (see Table~\ref{tab:y_results_other_FGs}). We include posterior results from some of these fits in Figs.~\ref{fig:CO_FF} and~\ref{fig:CO+FF_sync}. We find that the MCMC run that includes CIB does not converge (low effective sample size), so we do not include the posteriors here. We also compute the Fisher errors using the models listed in Table~\ref{tab:y_results_other_FGs}. For these, we use the following fiducial model parameters: $A_{\rm CO} = 1$ Jy/sr, $A_{\rm FF} = 300$ Jy/sr, $A_{\rm s} = 288$ Jy/sr, $\beta_{\rm s} = -0.82$ (with a $50\%$ prior on $\beta_{\rm s}$).

\begin{table}
\begin{tabular}{|l|c|c|}
\colrule
Model & $\times 10^{6}$ & P60\\
\colrule
Dust + CO ($\nu_{600}$) & $\ymono$&57.8\\
& $\sigma_{\ymono}\ 68\%$ C.L.&+25.9/-20.0\\
&$\ymono\leq95\%$ C.L.&101\\
& Fisher $\sigma_{\langle y \rangle}$&24.1\\
\colrule
Dust + FF ($\nu_{600}$) & $\ymono$&23.4\\
& $\sigma_{\ymono}\ 68\%$ C.L. &+11.3/-9.2\\
&$\ymono\leq95\%$ C.L.&42.8\\
& Fisher $\sigma_{\langle y \rangle}$&10.2\\
\colrule
Dust + CO + FF ($\nu_{600}$) & $\ymono$&55.3\\
& $\sigma_{\ymono}\ 68\%$ C.L.&+23.8/-18.7\\
&$\ymono\leq95\%$ C.L.&96.2\\
& Fisher $\sigma_{\langle y \rangle}$&24.2\\
\colrule
Dust + Sync ($\nu_{600}$) & $\ymono$&22.1\\
& $\sigma_{\ymono}\ 68\%$ C.L.&+11.3/-10.2\\
&$\ymono\leq95\%$ C.L.&42.6\\
& Fisher $\sigma_{\langle y \rangle}$&10.5\\
\colrule
\end{tabular}
\caption{\label{tab:y_results_other_FGs}$\ymono$ constraints for additional sky models, which include foreground parameters besides dust (see set-ups 3-12 in Table \ref{tab:models}). We list the mean, $68\%$ C.L. errors, $95\%$ C.L. upper limits, and Fisher forecast error bars for models applied to P60 data, which have best-fit PTE $>1\%$, only. $\ymono$ is more biased when additional foregrounds are included except in the case of synchrotron despite improved PTEs, which suggests the presence of systematic errors.}
\end{table}

\section{High-Frequency Data}\label{app:high_frequencies}

\begin{table*}[t]
\begin{tabular}{|l|l|l|l|}
\colrule
Foregrounds& Free parameters &$\chi^{2}$ or $\Delta\chi^{2}$ [DOF]& PTE\\
(frequency range)& + extra priors&P20/P40/P60& P20/P40/P60\\
\colrule
1. Dust  & $\Delta_{\rm T}$, $y$, $A_{\rm d}$, $T_{\rm d}$, $\beta_{\rm d}$ & 129.9/185.2/286.8 [99] & 0.020/$3.50\times10^{-7}$/0.000\\
2. Dust ($\beta_{\rm d}$ prior)  & $\Delta_{\rm T}$, $y$, $A_{\rm d}$, $T_{\rm d}$, $\beta_{\rm d} \in \mathcal{N}(1.51,0.1)$ & +7.0/+6.3/+2.9 [99] & 0.007/$7.37\times10^{-8}$/0.000\\
\colrule
3. Dust+Sync  & $\Delta_{\rm T}$, $y$, $A_{\rm d}$, $T_{\rm d}$, $\beta_{\rm d}$, $A_{\rm s}$, $\beta_{\rm s}$ &-6.6/-13.1/-26.6 [97]& 0.037/$4.20\times10^{-6}$/$1.11\times10^{-16}$\\
4. Dust+Sync ($\beta_{\rm s}$ prior)  & $\Delta_{\rm T}$, $y$, $A_{\rm d}$, $T_{\rm d}$, $\beta_{\rm d}$, $A_{\rm s}$, $\beta_{\rm s}\in \mathcal{N}(-1, 0.5)$ & -1.0/-3.8/-23.4 [97]&0.017/$4.62\times10^{-7}$/0.000\\
5. Dust+CO  & $\Delta_{\rm T}$, $y$, $A_{\rm d}$, $T_{\rm d}$, $\beta_{\rm d}$, $A_{\rm CO}$&-1.5/-4.2/-9.2 [98]&0.021/$6.94\times10^{-7}$/0.000\\
6. Dust+FF  & $\Delta_{\rm T}$, $y$, $A_{\rm d}$, $T_{\rm d}$, $\beta_{\rm d}$, $A_{\rm FF}$ & -2.1/-5.7/-14.3 [98]&0.023/$1.00\times10^{-6}$/0.000\\
7. Dust+CIB (fixed SED) & $\Delta_{\rm T}$, $y$, $A_{\rm d}$, $T_{\rm d}$, $\beta_{\rm d}$, $A_{\rm CIB}$&-6.9/-17.4/-33.2 [98]& 0.044/$1.48\times10^{-5}$/$9.99\times10^{-16}$\\
\colrule
8. Dust+Sync+FF & $\Delta_{\rm T}$, $y$, $A_{\rm d}$, $T_{\rm d}$, $\beta_{\rm d}$, $A_{\rm s}$, $\beta_{\rm s}$, A$_{\rm FF}$ & -7.3/-17.8/-40.5 [96]&$0.035/8.92\times10^{-6}/3.66\times10^{-15}$\\
9. Dust+Sync ($\beta_{\rm s}$ prior) & $\Delta_{\rm T}$, $y$, $A_{\rm d}$, $T_{\rm d}$, $\beta_{\rm d}$, $A_{\rm s}$, & -5.3/-12.8/-33.7 [96]&0.026/$2.86\times10^{-6}$/$4.44\times10^{-16}$\\
+FF & $\beta_{\rm s}\in \mathcal{N}(-1, 0.5)$, $A_{\rm FF}$ &&\\
\colrule
10. Dust+Sync+curv.  &$\Delta_{\rm T}$, $y$, $A_{\rm d}$, $T_{\rm d}$, $\beta_{\rm d}$, A$_{\rm s}$, $\beta_{\rm s}$, $\omega_{\rm s}$ & -7.7/-17.7/-40.0 [96]&0.037/$8.80\times10^{-6}$/$3.11\times10^{-15}$\\
11. Dust+Sync ($\beta_{\rm s}$ prior) & $\Delta_{\rm T}$, $y$, $A_{\rm d}$, $T_{\rm d}$, $\beta_{\rm d}$, $A_{\rm s}$ & -7.2/-16.1/-30.8 [96]&0.034/$6.17\times10^{-6}$/$2.22\times10^{-16}$\\
+curv. & $\beta_{\rm s}\in \mathcal{N}(-1, 0.5)$, $\omega_{\rm s}$ &&\\
\colrule
12. Dust+Sync& $\Delta_{\rm T}$, $y$, $A_{\rm d}$, $T_{\rm d}$, $\beta_{\rm d}$, $A_{\rm s}$, $\beta_{\rm s}$, $A_{\rm plaw}$, $\beta_{\rm plaw}$ & -7.8/-18.2/-40.2 [95]&0.032/$7.4\times10^{-6}$/$2.11\times10^{-15}$\\
+ power law & & &\\
13. Dust+Sync ($\beta_{\rm s}$ prior) & $\Delta_{\rm T}$, $y$, $A_{\rm d}$, $T_{\rm d}$, $\beta_{\rm d}$, $A_{\rm s}$, $\beta_{\rm s}\in \mathcal{N}(-1, 0.5)$, & -7.2/-17.8/-40.7 [95]&0.029/$6.62\times10^{-6}$/$2.44\times10^{-15}$\\
+ power law & $A_{\rm plaw}$, $\beta_{\rm plaw}$ & &\\
\colrule
14. Moments 1st order &  $\Delta_{\rm T}$, $y$, $A_{\rm d}$, $\omega_{2},\omega_{3}\in[-10^{3}, 10^{3})$ &+1.8/+4.1/+3.3 [99]&0.015/$1.28\times10^{-7}$/0.000\\
15. Moments 1st order &  $\Delta_{\rm T}$, $y$, $A_{\rm d}$, $\omega_{2},\omega_{3}\in[-10^{3}, 10^{3})$ &0.0/-2.1/-5.9 [99]&0.020/$5.78\times10^{-7}$/0.000\\
(best-fit)\footnote{Moments fit with $\beta_{\rm d}$ and $T_{\rm d}$ fixed to best-fit values from the \textit{Dust} fit (Model 1) to P20.}&&&\\
16. Moments 2nd order in $\beta_{\rm d}$  &  $\Delta_{\rm T}$, $y$, $A_{\rm d}$, $\omega_{2},\omega_{3},\omega_{22}\in[-10^{3}, 10^{3})$ &-6.2/-18.1/-35.5 [98]&0.040/$1.73\times10^{-5}$/$2.00\times10^{-15}$\\
\colrule
\end{tabular}
\caption{\label{tab:high_frequencies}Models we try to fit to the frequency monopole data going up to $\approx 1.9$ THz along with the list of free parameters, best-fit $\chi^2$, degrees of freedom (DOF $=$ number of data points $-$ free parameters), and PTE.  We use a flat prior on $T_{\rm d}\in[0,100)$~K in all these cases. For the Gaussian priors on $\beta_{\rm d}$ and $\beta_{\rm s}$, we adopt the bounds based on the posterior results from \planck as in the main analysis \cite{Planck2016FG}. Unless otherwise noted, we adopt a flat prior on $\beta_{\rm d}\in[0,3)$ and $\beta_{\rm s}\in[-5,5)$. The best-fit values are summarized in Table~\ref{tab:high_frequencies_bestfit}.}
\end{table*}

\begin{table*}[t]
\begin{tabular}{|l|l|}
\colrule
Foregrounds& Best-fit parameter values \\
(frequency range)& P20/P40/P60\\
\colrule
1. Dust  & $\Delta T\times10^{3}=-1.40/-1.25/-1.51$ K, $\ymono\times10^{6}=22.0/13.9/14.2$,\\
& $A_{\rm d}\times10^{-5}=1.94/2.08/2.44$ Jy/sr, $T_{\rm d}=23.7/27.6/27.6$ K, $\beta_{\rm d}=0.99/0.87/0.93$\\
2. Dust ($\beta_{\rm d}$ prior)  & $\Delta T\times10^{3}=-1.37/-1.24/-1.50$ K, $\ymono\times10^{6}=35.9/22.8/19.1$,\\
& $A_{\rm d}\times10^{-5}=2.03/2.15/2.48$ Jy/sr, $T_{\rm d}=20.8/25.6/26.7$ K, $\beta_{\rm d}=1.25/1.00/0.98$\\
\colrule
3. Dust+Sync  & $\Delta T\times10^{3}=-1.63/-1.49/-1.78$ K, $\ymono\times10^{6}=-2.60/-8.08/-10.3$\\
& $A_{\rm d}\times10^{-4}=7.98/12.3/16.1$ Jy/sr, $T_{\rm d}=12.9/18.7/20.2$ K, $\beta_{\rm d}=3.00/1.91/1.76$, \\
& $A_{\rm s}\times10^{-4}=6.32/7.07/8.25$ Jy/sr, $\beta_{\rm s}=1.13/1.04/0.97$\\
\colrule
4. Dust+Sync ($\beta_{\rm s}$ prior)  & $\Delta T\times10^{3}=-1.44/-1.34/-1.78$ K, $\ymono=26.9/19.3/2.56$, $A_{\rm d}\times10^{-5}=1.93/2.07/2.03$ Jy/sr,\\
& $T_{\rm d}=23.2/26.6/23.2$ K, $\beta_{\rm d}=1.04/0.94/1.37$,\\
&$A_{\rm s}\times10^{-4}=4.11/5.87/10.5$ Jy/sr, $\beta_{\rm s}=-0.90/-0.55/0.5$\\
\colrule
5. Dust+CO  & $\Delta T\times10^{3}=-1.48/-1.35/-1.63$ K, $\ymono\times10^{6}=40.0/34.1/38.8$,\\
& $A_{\rm d}\times10^{-5}=1.95/2.10/2.46$ Jy/sr, $T_{\rm d}=23.0/26.8/26.9$ K\\
& $\beta_{\rm d}=1.05/0.92/0.98$, $A_{\rm CO}=31.4/36.0/44.4$ Jy/sr\\
\colrule
6. Dust+FF  & $\Delta T\times10^{3}=-1.51/-1.39/-1.68$ K, $\ymono\times10^{6}=25.6/17.5/18.2$, \\
& $A_{\rm d}\times10^{-5}=1.85/2.02/2.36$ Jy/sr, $T_{\rm d}=22.2/25.9/25.8$ K, $\beta_{\rm d}=1.16/1.01/1.08$,\\ &$A_{\rm FF}\times10^{-4}=2.01/2.28/2.97$ Jy/sr\\
7. Dust+CIB (fixed SED) & $\Delta T\times10^{3}=-1.47/-1.34/-1.61$ K, $\ymono\times10^{6}=-2.55/-13.3/-17.5$,\\
&$A_{\rm d}\times10^{-5}=6.10/4.70/5.20$ Jy/sr, $T_{\rm d}=19.6/24.3/25.1$, $\beta_{\rm d}=0.86/0.70/0.71$,\\
&$A_{\rm CIB}\times10^{-5}=-6.31/-4.90/-5.55$ Jy/sr\\
\colrule
8. Dust+Sync+FF & $\Delta T\times10^{3}=-1.60/-1.33/-1.60$ K, $\ymono\times10^{6}=-12.0/-2.18/-4.41$, \\
& $A_{\rm d}\times10^{-4}=7.51/6.49/14.3$ Jy/sr, $T_{\rm d}=13.2/84.1/46.4$ K, $\beta_{\rm d}=3.00/0.44/0.56$,\\
&$A_{\rm s}=9.85\times10^{4}/-1.28\times10^{2}/-1.13$ Jy/sr, $\beta_{\rm s}=0.95/3.65/5.00$,\\
& $A_{\rm FF}\times10^{-3}=-21.3/3.68/6.93$ Jy/sr \\
9. Dust+Sync ($\beta_{\rm s}$ prior)+FF & $\Delta T\times10^{3}=-1.71/-1.58/-1.45$ K, $\ymono\times10^{6}=4.33/-5.12/-44.3$,\\
 & $A_{\rm d}\times10^{-5}=1.50/1.74/1.15$ Jy/sr, $T_{\rm d}=18.8/22.5/18.3$ K, $\beta_{\rm d}=1.68/1.39/2.24$, \\
 & $A_{\rm s}\times10^{-5}=-2.87/-3.29/636$, $\beta_{\rm s}=-0.90/-0.80/-0.16$, \\ & $A_{\rm FF}\times10^{-5}=1.11/1.26/-197$ Jy/sr\\
\colrule
10. Dust+Sync+curv.  &$\Delta T\times10^{3}=-1.49/-1.32/-1.57$ K, $\ymono\times10^{6}=-17.3/-4.37/-7.86$, \\
& $A_{\rm d}\times10^{-4}=6.78/3.62/9.56$ Jy/sr, $T_{\rm d}=13.6/77.1/68.2$ K, $\beta_{\rm d}=3.00/0.00/0.44$,\\
&$A_{\rm s}=5.65/4.56\times10^{3}/8.42\times10^{-3}$ Jy/sr, $\beta_{\rm s}=-0.010/2.86/3.71$,
\\& $\omega_{\rm s}=5.59\times10^{4}/-0.203/-2.28\times10^{3}$\\
11. Dust+Sync ($\beta_{\rm s}$ prior)+curv. & $\Delta T\times10^{3}=-1.66/-1.50/-1.76$ K, $\ymono\times10^{6}=-25.7/-33.1/-34.5$, \\
& $A_{\rm d}\times10^{-5}=1.26/1.51/1.81$ Jy/sr, $T_{\rm d}=17.2/20.8/21.4$ K, $\beta_{\rm d}=2.04/1.66/1.64$, \\
& $A_{\rm s}\times10^{-4}=2.81/2.84/4.07$ Jy/sr, $\beta_{\rm s}=-0.99/-0.86/-0.75$, $\omega_{\rm s}=40.1/34.2/21.9$ \\
\colrule
12. Dust+Sync& $\Delta T\times10^{3}=-1.67/-1.33/-1.56$ K, $\ymono\times10^{6}=-2.28/-8.51/-6.26$,\\
+power law&  $A_{\rm d}\times10^{-4}=8.48/1.40/-1.39$ Jy/sr, $T_{\rm d}=13.0/16.5/11.9$ K, $\beta_{\rm d}=3.00/3.00/3.00$, \\
&$A_{\rm s}\times10^{-4}=7.86/6.14/41.1$ Jy/sr, $\beta_{\rm s}=0.99/2.53/2.88$, \\
& $A_{\rm plaw}=1.50\times10^{-21}/-4.31\times10^{4}/-3.97\times10^{5}$, $\beta_{\rm plaw}=20.7/2.64/2.89$\\ 
13. Dust+Sync ($\beta_{\rm s}$ prior) & $\Delta T\times10^{3}=-1.62/-1.32/-1.59$ K, $\ymono\times10^{6}=-10.4/-2.12/-3.21$,\\
+power law&$A_{\rm d}\times10^{-4}=7.70/5.95/14.7$ Jy/sr, $T_{\rm d}=13.1/92.2/45.9$ K, $\beta_{\rm d}=3.00/0.41/0.54$, \\
&$A_{\rm s}\times10^{-4}=-3.97/0.95/2.21$ Jy/sr, $\beta_{\rm s}=-0.98/-1.00/-1.02$,\\
&$A_{\rm plaw}=7.75\times10^{4}/-117/-0.524$ Jy/sr, $\beta_{\rm plaw}=1.03/3.68/5.23$\\
\colrule
14. Moments 1st order &  $\Delta T\times10^{3}=-1.38/-1.23/-1.49$ K, $\ymono\times10^{6}=26.7/19.3/18.5$,\\
& $A_{\rm d}\times10^{-5}=1.89/2.17/2.55$ Jy/sr, $\omega_{2}=-0.44/-0.61/-0.59$, $\omega_{3}=2.15/4.67/5.00$\\
15. Moments 1st order (best-fit) &  $\Delta T\times10^{3}=-1.40/-1.26/-1.52$ K, $\ymono\times10^{6}=22.0/12.0/10.9$,\\
& $A_{\rm d}\times10^{-5}=1.94/2.03/2.31$ Jy/sr, $\omega_{2}=-0.003/-0.189/-0.158$, $\omega_{3}=0.02/4.74/5.68$\\
16. Moments 2nd order in $\beta_{\rm d}$  &  $\Delta T\times10^{3}=-1.46/-1.32/-1.59$ K, $\ymono\times10^{6}=0.20/-11.7/-15.8$,\\
&$A_{\rm d}\times10^{-5}=4.66/5.34/6.06$ Jy/sr, $\omega_{2}=-0.45/-0.52/-0.52$,\\
&$\omega_{3}=-5.33/-4.28/-3.92$, $\omega_{22}=0.81/0.81/0.79$\\
\colrule
\end{tabular}
\caption{\label{tab:high_frequencies_bestfit} Best-fit parameter values for models listed in Table~\ref{tab:high_frequencies}. Note that the minimization runs were not initialized using posterior mean values.}
\end{table*}

We attempt to extend our \textit{frequency monopole} analysis to include higher frequencies, which would improve the constraining power from this method, as well as allow us to fit the foreground parameters without relying on external knowledge from \planck. Table~\ref{tab:high_frequencies} summarizes our results from fitting different models to frequencies up to $\approx 1.9$ THz --- the data beyond this range are very noisy, and the best-fit parameter values are included in Table \ref{tab:high_frequencies_bestfit}. The exact frequency range includes 27 channels from the low-frequency instrument ($\sim95-626$ GHz) and 77 channels from the high-frequency instrument ($\sim653-1864$ GHz). We find that including higher frequencies results in a high $\chi^{2}$ for P40 and P60 even when we have several components in our sky model and in only marginally acceptable PTEs for P20 for some of the models (see Table \ref{tab:high_frequencies}). We therefore limited our main analysis to the maximum frequency of $\approx800$ GHz.

To test the fits to high frequencies, we start with the simplest sky model, which contains only dust as the foreground and then add more foreground components to the model. We check imposing the same prior on $\beta_{\rm d}$ as in the main analysis but find that the dust model with flat priors on both parameters fits better (lower value of $\beta_{\rm d}$ is preferred). Adding synchrotron to the sky model improves PTE for P20. Since we do not force the power law index to be negative in set-up 3, this set-up essentially fits for the sum of the rest of the foregrounds in addition to dust. Indeed, the data prefers a low and positive $\beta_{\rm s}$ possibly due to the presence of other foregrounds that are not modeled and therefore imposing a synchrotron prior (set-up 4) with a negative spectral index does not help reduce the $\chi^{2}$. We try fitting CO, FF and CIB as in the main analysis but the fits for P40/P60 still have PTE=0. We also try adding curvature parameter in synchrotron parametrization (i.e., higher order term in the synchrotron spectrum based on the moment expansion approach, see \citetalias{Abitbol2017} for more details) or another simple power-law SED\footnote{We test a power-law initialized with amplitude of the free-free emission and a spectral index equal to 1.}, but including such components does not improve our fits. 

Finally, we try a moment-based approach to modeling our foregrounds, which was introduced in Ref.~\cite{Chluba2017} (set-ups 14-16 in Table \ref{tab:high_frequencies}). The foreground sky emission represents the composite signal of individually emitting sources with differently parameterized SEDs and is subject to averaging both across the sky and along the line of sight. With broad frequency and sky coverage, a simple single SED cannot accurately describe the cumulative and averaged sky emission. Ref.~\cite{Chluba2017} presented a general method to capture the effects of beam and line-of-sight averaging on a given spectrum, by Taylor expanding in free spectral parameters, evaluated at some fixed average parameters. In this prescription, if the sky emission is exactly described by some chosen SED, only first-order moments are needed and the other terms go to zero. The first-order moments capture any deviations of the true mean parameters from the chosen values that were fixed in the Taylor expansion (i.e., if the average parameters of a given SED are known exactly, the first order moments would also be zero). If, however, the sky emission is not fully described by a simple SED (e.g. power law, MBB etc.), higher-order terms capture the complexity beyond the simple spectrum. Using this moment approach, we try fitting the dust spectrum up to second-order moment in $\beta_{\rm d}$:
\begin{equation}
\begin{split}
    I^{\rm dust+moments}_{\nu}=I^{\rm dust}_{\nu}+\omega_{\beta_{\rm d}}\partial_{\beta_{\rm d}}I^{\rm dust}_{\nu}
    +\omega_{T_{\rm d}}\partial_{T_{\rm d}}I^{\rm dust}_{\nu}+\frac{1}{2}\omega_{\beta_{\rm d}\beta_{\rm d}}\partial_{\beta_{\rm d}}\partial_{\beta_{\rm d}}I^{\rm dust}_{\nu}\\
    = A_{\rm d}\left(\frac{\nu}{\nu_{0}}\right)^{\bar{\beta}_{\rm d}+3}\frac{1}{e^{\bar{x}_{\rm d}}-1}\times
    \left[1+\omega_{\beta_{\rm d}}\ln\left(\frac{\nu}{\nu_{0}}\right)+\omega_{T_{\rm d}}\frac{\bar{x}_{\rm d}e^{\bar{x}_{\rm d}}}{e^{\bar{x}_{\rm d}}-1}\frac{1}{\bar{T}_{\rm d}}+\frac{1}{2}\omega_{\beta_{\rm d}\beta_{\rm d}}\ln^{2}\left(\frac{\nu}{\nu_{0}}\right)\right]\\
\end{split}
\end{equation}

Here, free parameters are the overall amplitude $A_{\rm d}$ and moments ($\omega_{T_{\rm d}}$, $\omega_{\beta_{\rm d}}$ and $\omega_{\beta_{\rm d}\beta_{\rm d}}$). We do not include the moment associated with the amplitude since it is not a spectral parameter and its derivative can be written as $\partial_{A_{\rm d}}I^{\rm dust}_{\nu}=I^{\rm dust}_{\nu}/A_{\rm d}$. Therefore, $\bar{\beta}, \bar{T}_{\rm d}$ are the parameters with respect to which we take the derivatives and are fixed, while $A_{\rm d}$ is fit for together with the moments. We test three set-ups, which include moments: moments to 1st order  where $\bar{T}_{\rm d}=21$ K and $\bar{\beta}_{\rm d}=1.51$ are set to values based on $\planck$ constraints (set-up 14), moments to 1st order where $\bar{T}_{\rm d}=23.7$ K and $\bar{\beta}_{\rm d}=0.99$ are fixed based on best-fit results from fitting set-up 1 to P20 (set-up 15), and moments to 2nd order in $\beta_{\rm d}$ (set-up 16).

\begin{figure}[t]
    \centering
    \includegraphics[width=0.5\columnwidth]{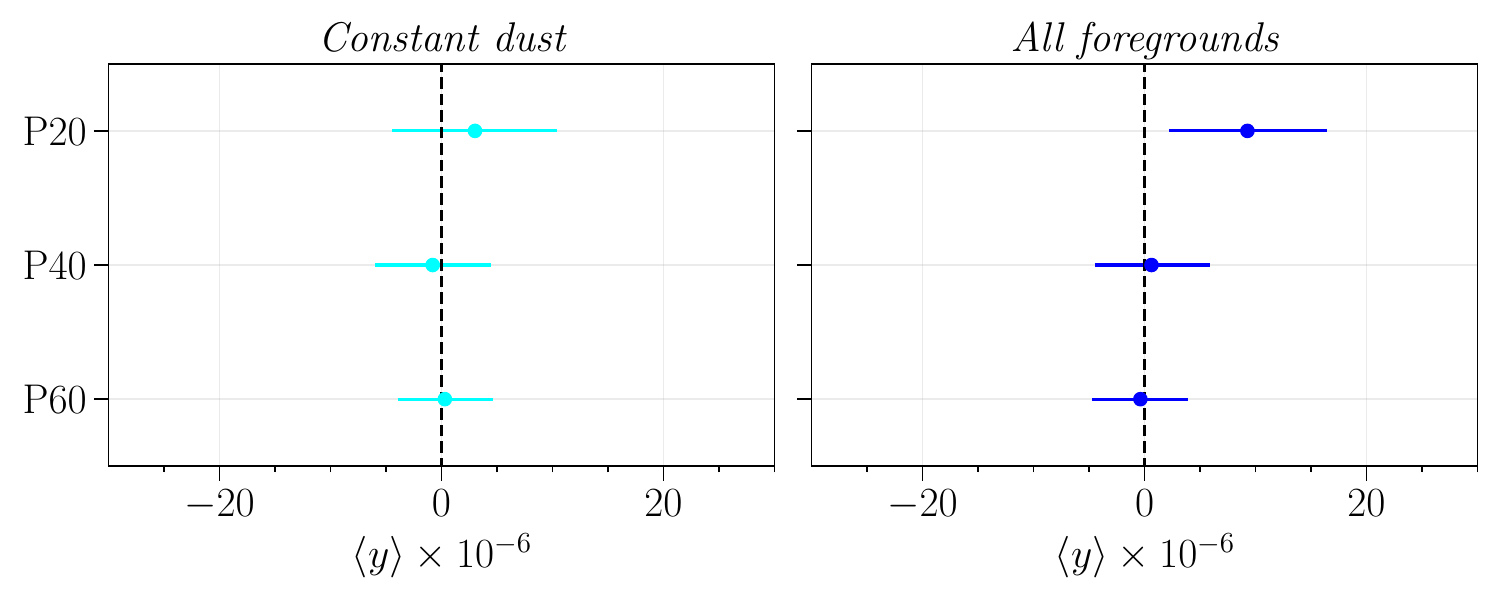}
    \caption{Results from mocks using high frequencies up to 1.9 THz on the \textit{constant dust} and the most complex mock. Here we fit a simple sky model, which only includes dust as the foreground, and has only uniform priors for both $T_{\rm d}$ and $\beta_{\rm d}$ (set-up 1 in Table~\ref{tab:high_frequencies}). Results are stable across all sky masks. }\label{fig:1pt89_dust_mock_tests}
\end{figure}

Finally, in Fig.~\ref{fig:1pt89_dust_mock_tests} we show fits to the simple \textit{constant dust} mock and the most complex mock using the simple sky model with dust as the only foreground (set-up 1 in Table \ref{tab:high_frequencies}) when we extend the frequency range. We find stable results for all sky fractions and about $\times\approx 2-2.5$ tighter error bars than when using $\nu_{600}$ (see Fig.~\ref{fig:summary_mock_monopole}). In the case of the simplest mock, only using P20, we find $T_{\rm d}$ and $\beta_{\rm d}$ that are consistent with what is in the mock. This suggests that either the varying amplitudes across pixels contribute to the shifts in $\beta_{\rm d}$ and $T_{\rm d}$ recovered in the MCMC runs or that the high frequencies suffer from systematic effects and therefore we do not find consistent values for dust parameters with higher sky fractions.

To further demonstrate the constraining power that we gain from including higher frequencies, we run MCMC using some of the models listed in Table \ref{tab:high_frequencies} on the data (example constraints are listed in Table \ref{tab:y_results_1pt89}). With high frequencies, we are able to fit the simple dust model to P20 and constrain both $\beta_{\rm d}$ and $T_{\rm d}$. Even with no priors, the $1\sigma$ error bar on $\ymono$ is $8\times10^{-6}$, which is $\approx1.7-2.2\times$ tighter than the errors from $\nu_{600}$. However, the upper limits are similar due to highly biased constraint from including these high frequencies. 

We also fit the first-order dust moment, which is similar to fitting the dust alone, except all free parameters have linear dependence. We include the results from fitting moments where $\bar{T}_{\rm d}$ and $\bar{\beta}_{\rm d}$ are set to best-fit parameter values when fitting the data (set-up 15). With this set-up, $\ymono$ is biased at $\approx 3\sigma$, similar to the level of bias found in the dust MBB fit. We find that $\ymono$ is more biased when we fit moments setting the $\bar{T}_{\rm d}$ and $\bar{\beta}_{\rm d}$ to \planck values rather than the best-fit determined from data. The constraints are also very similar to fitting the dust MBB. 

\begin{figure*}[h]
\begin{minipage}[c]{0.5\linewidth}
    \centering
    \includegraphics[width=\linewidth]{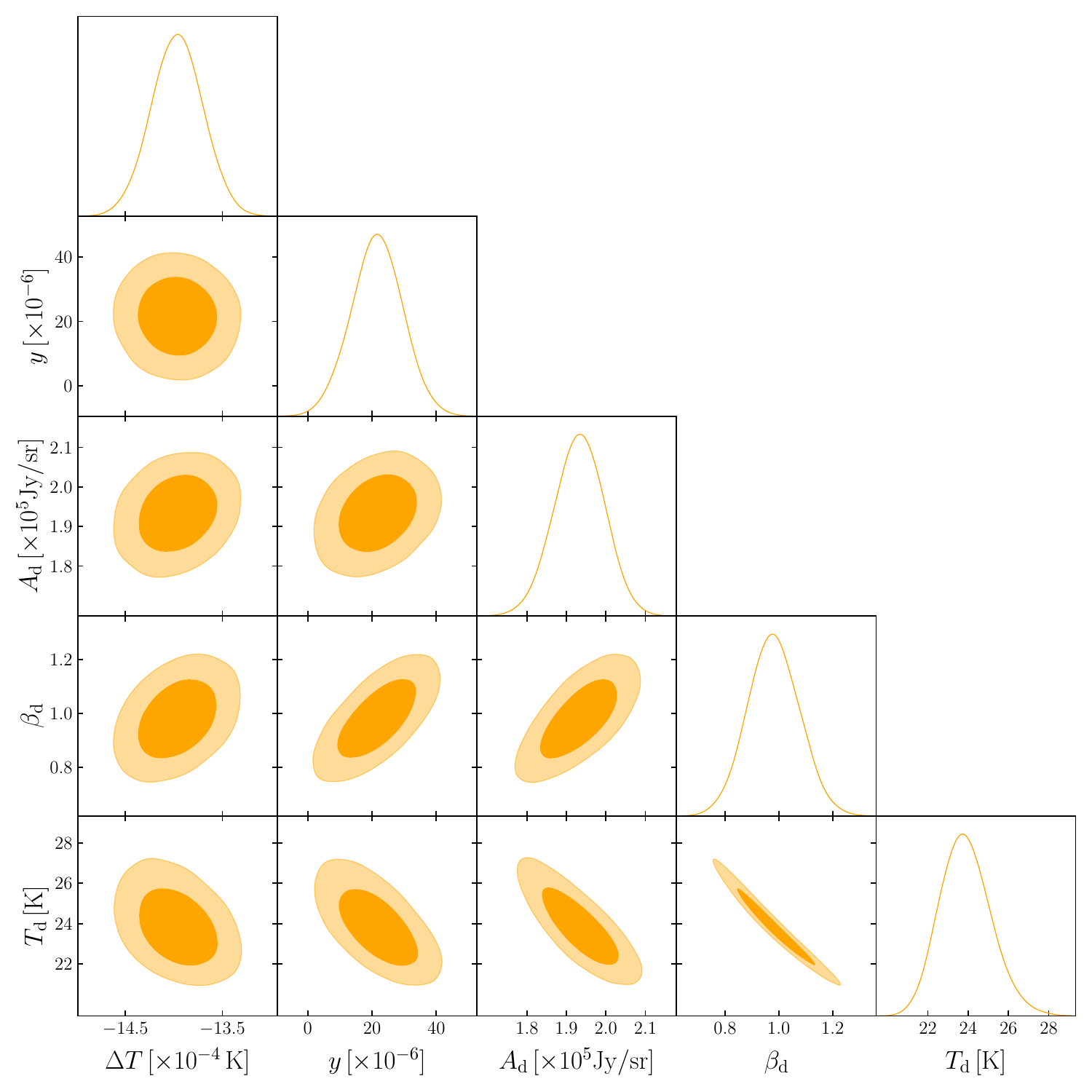}
\end{minipage}\hfill
\begin{minipage}[c]{0.5\linewidth}
    \centering
    \includegraphics[width=\linewidth]{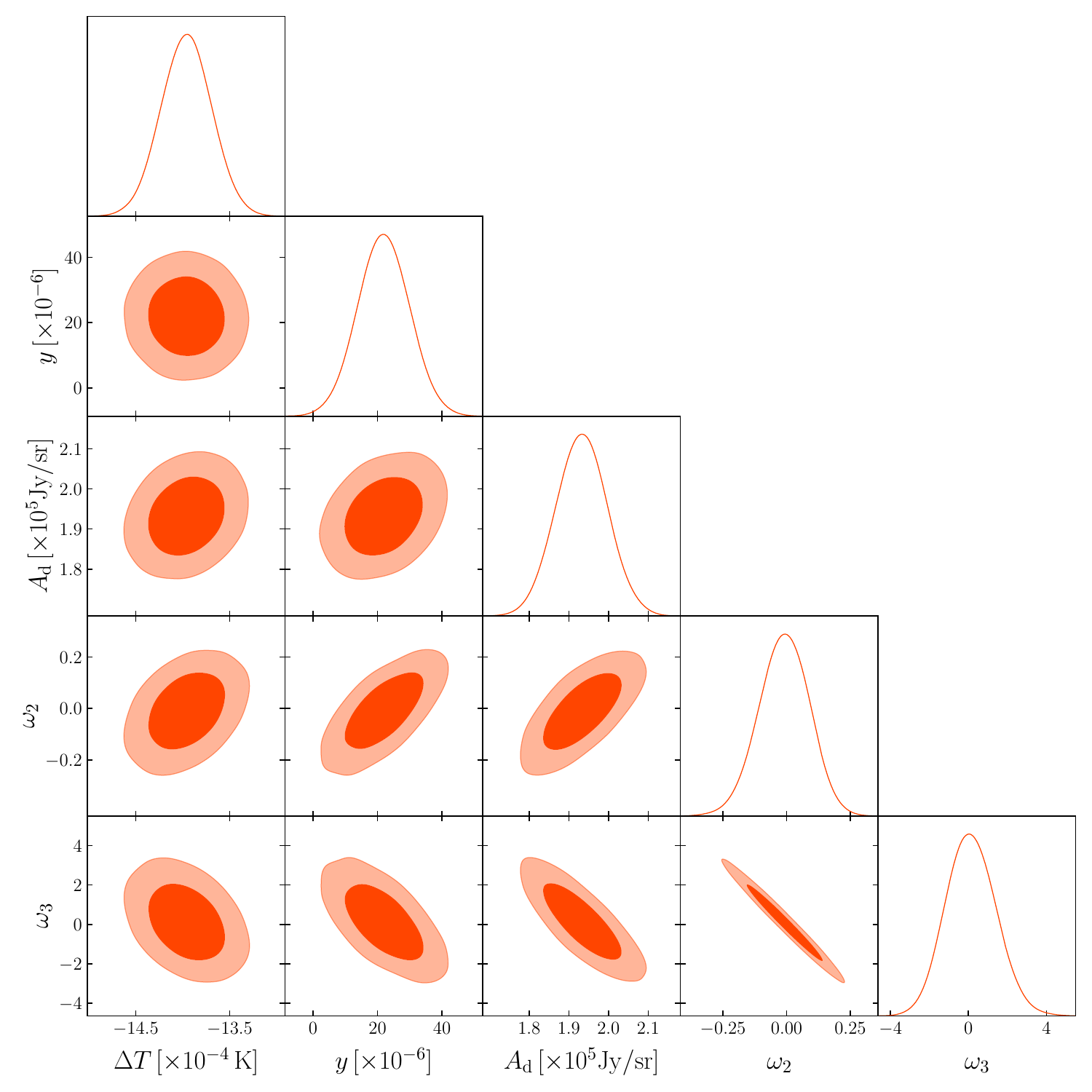}
\end{minipage} 
\caption{\textit{Left:} Marginalized posteriors fitting the sky model that includes the dust SED with flat priors on $T_{\rm d}$ and $\beta_{\rm d}$ using frequencies up to $\approx 1.9$ THz (set-up 1 in Table \ref{tab:high_frequencies}). \textit{Right:} Marginalized posteriors fitting the sky model that includes the moment expansion for the dust SED to first order (set-up 15 in Table~\ref{tab:high_frequencies}). 
}
\end{figure*}

\begin{table}[t]
\begin{tabular}{|l|c|c|}
\colrule
Model & $\times 10^{6}$ & P20\\
\colrule
Dust & $\ymono$&21.6\\
&$\sigma_{\ymono}\ 68\%$ C.L.&8.0\\
&$\ymono\leq95\%$ C.L. &37.5\\
\colrule
Moments 1st order (best-fit)& $\ymono$ &21.9\\
&$\sigma_{\ymono}\ 68\%$ C.L.&+8.0/-7.9\\
&$\ymono\leq95\%$ C.L. & 37.7\\
\colrule
\end{tabular}
\caption{\label{tab:y_results_1pt89} $\ymono$ constraints on a few models fit to high-frequency data up to $\approx 1.9$ THz. We only include P20 due to low PTE values for P40/P60.}
\end{table}

\section{Frequency Channels}\label{app:emission}

\begin{figure}
    \centering
    \includegraphics[width=0.8\linewidth]{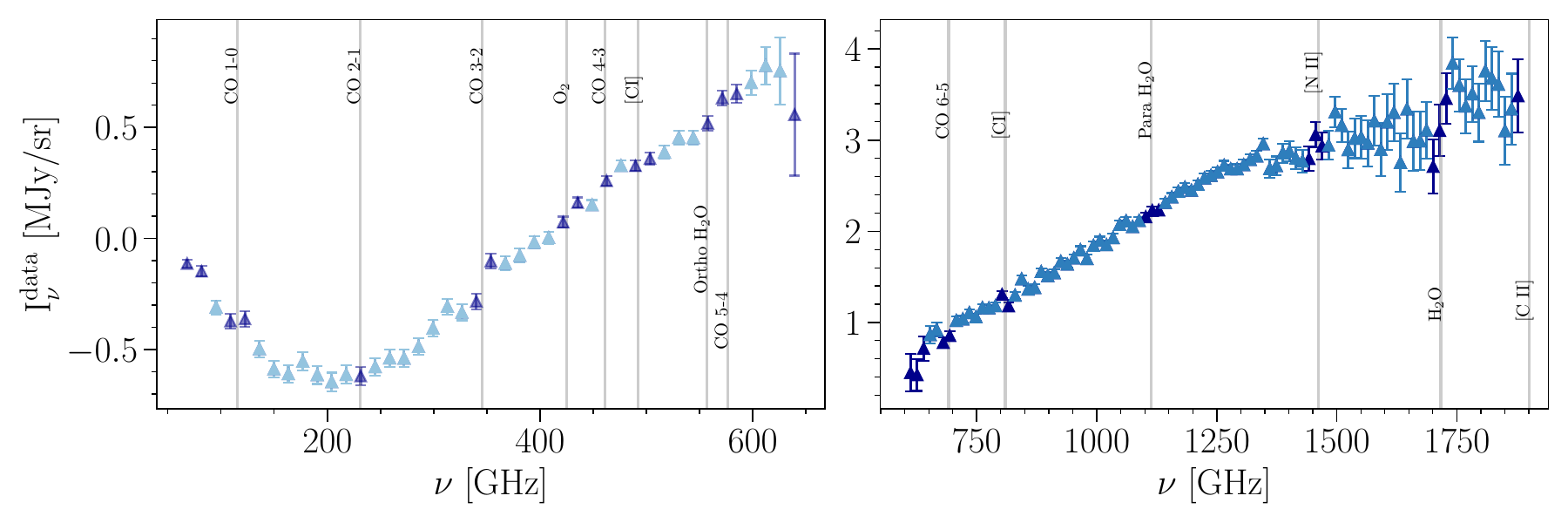}
    \caption{Frequency channels used in the \textit{frequency monopole} analysis presented in this paper. The left/right panels show the low/high-frequency data, respectively.  Frequency channels that are not used are plotted in a dark shade of blue; grey lines correspond to emission features.}\label{fig:plot_channels}
\end{figure}

In Table~\ref{tab:emission_lines} we list the spectral lines that we used to determine which frequency channels to remove in the \textit{frequency monopole} analysis as detailed in Sec.~\ref{sec:freq_ranges}. In Fig.~\ref{fig:plot_channels} we plot all the \textit{FIRAS} frequency channels within the frequency ranges considered in this paper from the \textit{frequency monopole} for the low-frequency (left) and the high-frequency (right) instruments. The frequencies that are excluded from the \textit{frequency monopole} results are plotted in darker shades of blue.

\begin{table}[t]
\begin{tabular}{|c|c|}
\colrule
Spectral Line & Frequency [GHz] \\
\colrule
CO 1-0 & 115.27\\
CO 2-1 & 230.54\\
CO 3-2 & 345.8\\
$O_{2}^{*}$ & 424.75\\
CO 4-3 & 461.04\\
$[\rm C I]$ & 492.23\\
Ortho H$_{2}\rm O^{*}$ & 556.89\\
CO 5-4 & 576.27\\
CO 6-5 & 691.47\\\relax
[C I] & 809.44\\
Para H$_{2}$O & 1113.3\\\relax
[N II] & 1461.1\\
H$_{2}O^{*}$ & 1716.6\\\relax
[C II] & 1900.5\\\relax
[O I]& 2060.1\\\relax
[Si II] & 2311.7\\\relax
[N II] & 2459.4\\
CH 2-1 & 2589.6\\
\colrule
\end{tabular}
\caption{Galactic emission lines considered in this work. We remove these lines from the sky model fitting in the \textit{frequency monopole} approach, while we find that the lines are not problematic for the \textit{pixel-by-pixel} approach outside the Galactic plane.}
 \label{tab:emission_lines}
\end{table} 

\section{$\Delta$T as a Function of Sky Fraction}\label{app:dT}
\begin{figure}[h]
    \centering
    \includegraphics[width=0.3\linewidth]{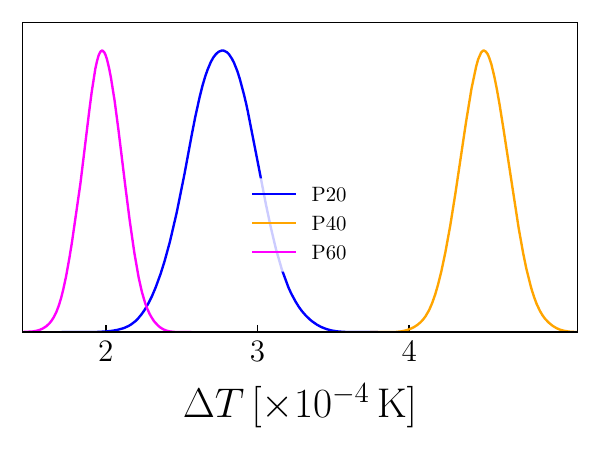}
    \caption{Posterior results from fitting a sky model consisting of $\Delta B_{\nu}$ alone to simulated data, which only includes the CMB. $\Delta T$ shifts for different sky fractions --- likely due to varying contributions of the CMB dipole signal.}\label{fig:CMB_dT}
\end{figure}

As discussed in Sec.~\ref{sec:sky_models}, the blackbody deviation component in our sky model, $\Delta B_{\nu}$, fits any residual CMB contribution to our spectrum including the CMB dipole. Since we use sky masks in all of our fits, the dipole contribution is not symmetric and does not fully cancel out. Therefore, we can expect that $\Delta B_{\nu}$ will contain contributions from the CMB dipole and is different across the sky masks used in our analysis. To test this, we use a mock that only contains the CMB signal and fit for $\Delta B_{\nu}$. Our results are shown in Fig.~\ref{fig:CMB_dT}. We find that the direction of the shift is consistent with what we obtain from fits to the data, which include additional sky components.  

\section{Robustness of mock data and pixel-by-pixel analysis pipeline}\label{sec:mock-robustness}

\begin{figure}[!htbp]
\centering
\includegraphics[width=0.7\columnwidth]{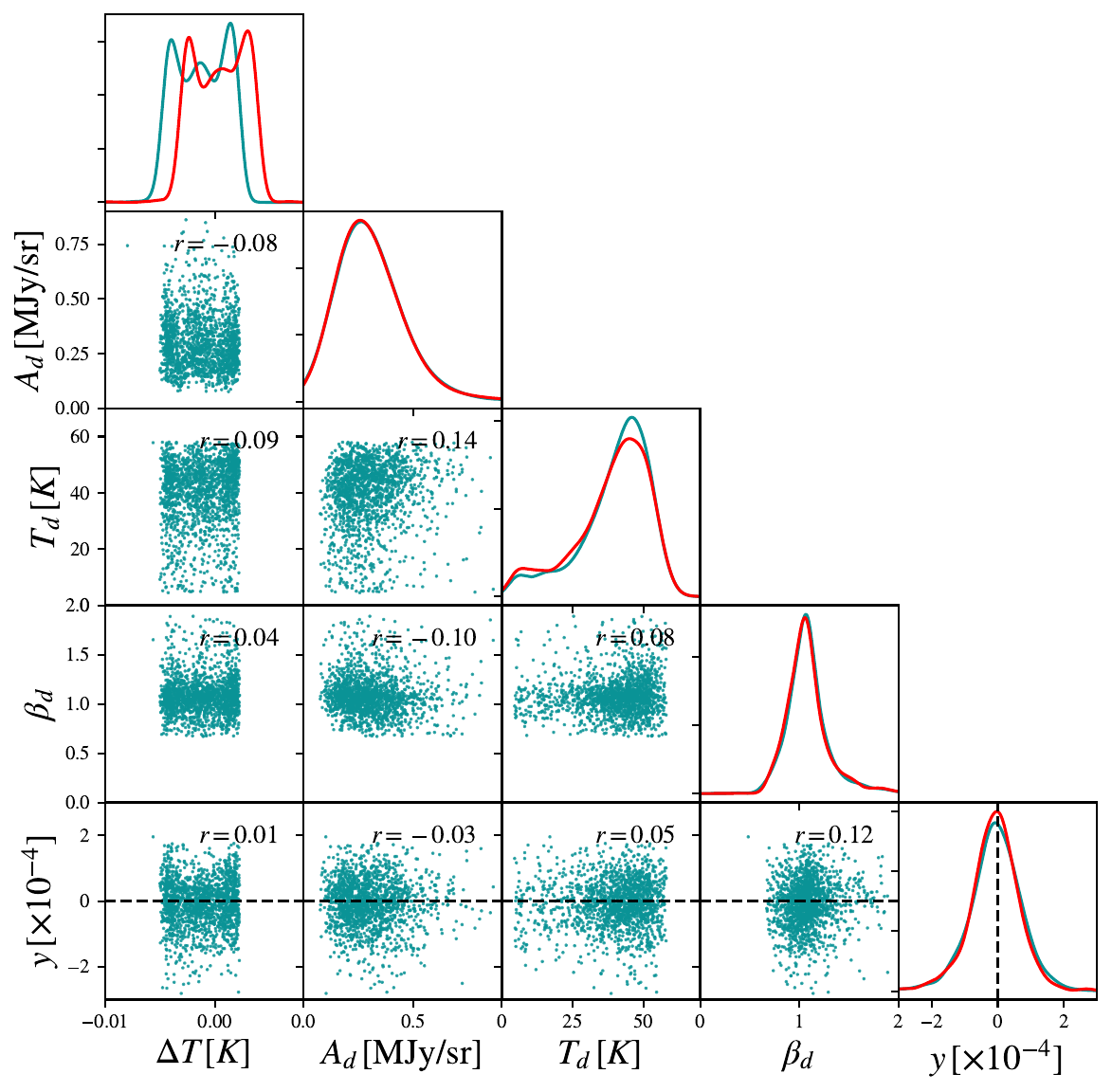}
\caption{Scatter plot matrix of the $y$ pixel values on a P60 Galactic mask. The marginalized posterior shows a Gaussian kernel density estimate of the pixel value distributions in each map (blue) and the same quantity on mock data (red). The numbers in the boxes show the Spearman’s rank correlation coefficient between the two variables under consideration.}
\label{fig:y_scatter}
\end{figure}

\begin{figure}[!htbp]
\centering
\includegraphics[width=0.5\columnwidth]{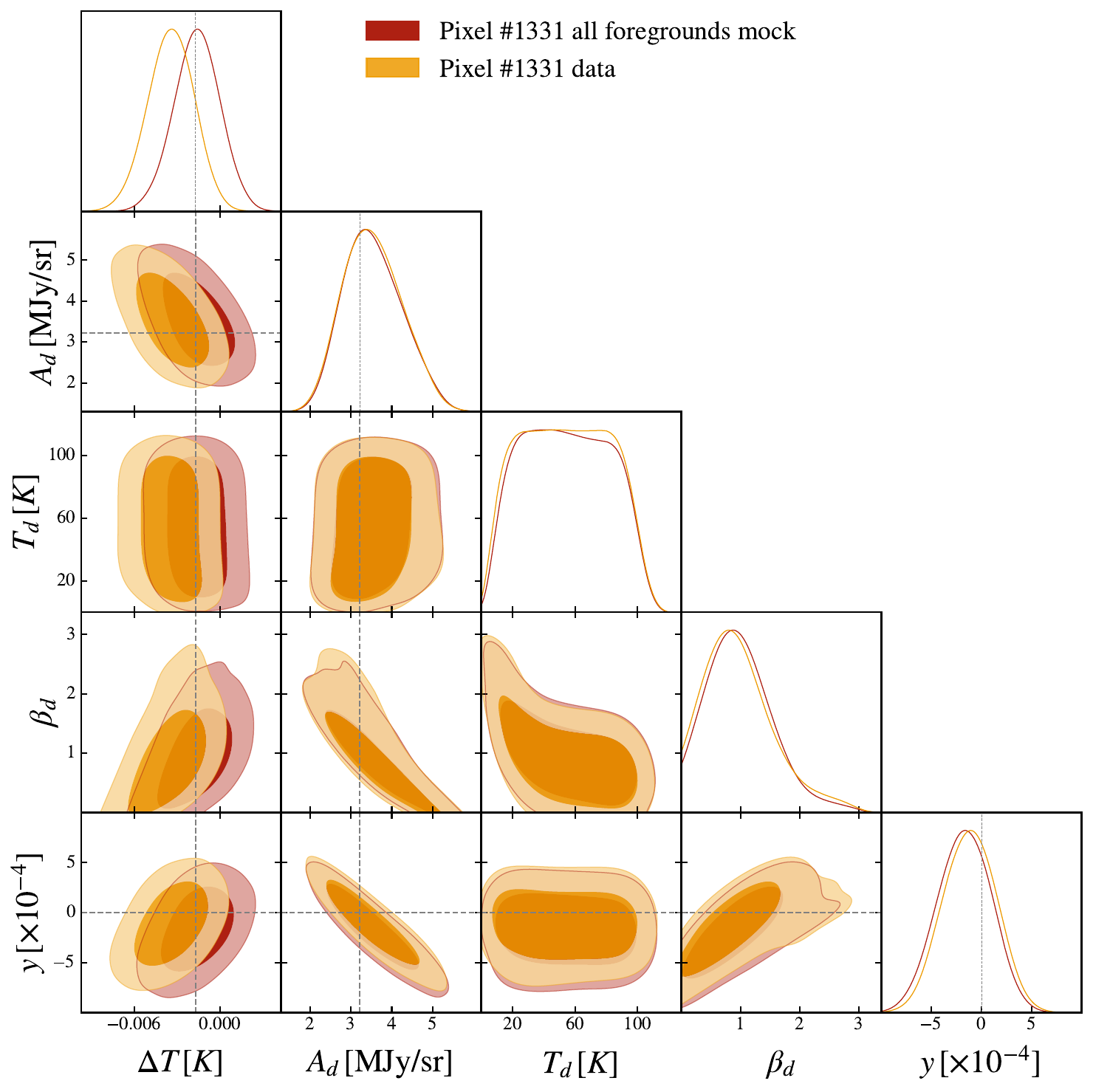}\\
\includegraphics[width=0.5\columnwidth]{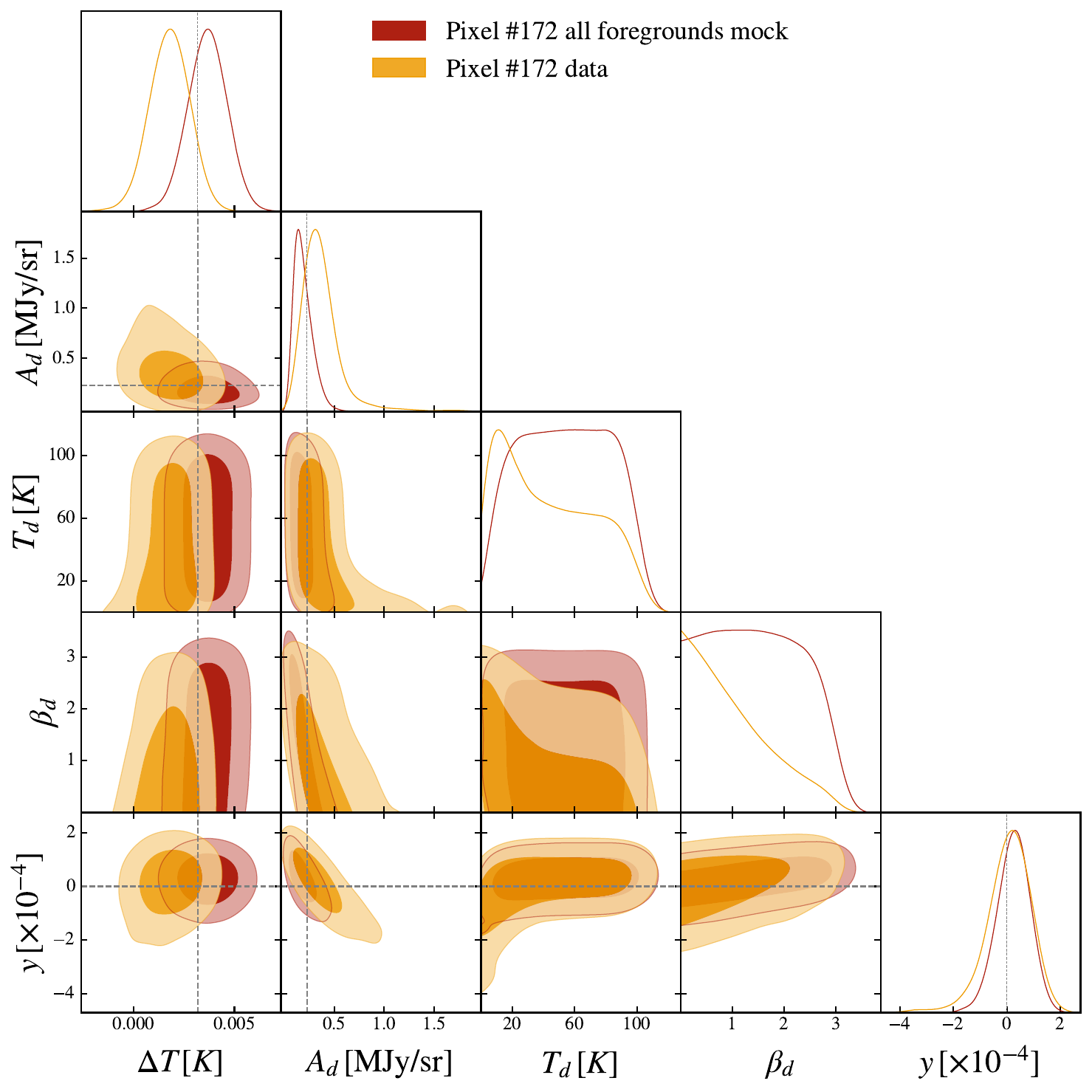}
\caption{Comparison of the sky model constraints obtained on mock data including extragalactic and Galactic sky components. The top panel show results for a pixel in the Galactic plane while the bottom for a pixel close to the Galactic north polar cap. Dashed grey line shows the expectation for the total CMB anisotropy, $y$, and the foreground amplitude from the mock data in that specific pixel without including the fluctuations due to noise.}
\label{fig:fg_mock_pixel_posteriors}
\end{figure}

Our analysis relies heavily on the use of mock data for the identification of optimal set-ups of the analysis and in the interpretation of our results on real data. In Fig.~\ref{fig:fg_mock_pixel_posteriors} we show a comparison between the MCMC chains in specific sky pixels for both mock data including all foreground components and real data, and in Fig.~\ref{fig:y_scatter} a scatter plot showing the correlation between the estimated maps of the parameters. For signal-dominated pixels, i.e., in the Galactic plane, the results obtained on mock data agree with the findings on real data including the recovered posteriors and the correlations between parameters. 

As we move towards the Galactic poles, we still find a very good level of consistency with foreground parameters displaying differences in the posteriors that become more important when moving more away from the Galactic plane where the foreground intensity gets weaker.
This is in particular the case in regions of the sky where the CIB dominates (and hence the uncertainties in the modeling) and the foreground amplitude we recover is slightly higher for the data compared to what is assumed in the mock data. Nonetheless, the estimates of $y$ display a very good level of consistency. A more quantitative comparison would require a more detailed analysis involving multiple noise realizations. We repeat the full analysis adding a different noise realization and find consistent trends and minimal scatter in the final $\ymono$ measurements, concluding that the final uncertainty is dominated by the scatter induced by the imperfect component separation. The overall estimates of the parameters shown in Fig.~\ref{fig:y_scatter} show that $y$, CMB, and foreground amplitudes are largely uncorrelated. In particular, the reduction in the correlations between the estimated $y$ and the foreground parameters improves on the original results of Ref.~\cite{Fixsen1996_dist} where the correlation between the spectral residuals describing astrophysical foreground was a non-negligible effect. The overall distribution of the estimates is consistent with the expectations from mock data. The only relevant difference is observed for the CMB distribution, which is due to the difference in the $T_{0}$ temperature recovered from the data (2.740 K) and the one assumed in the mock data.

\begin{figure}[!htbp]
\includegraphics[width=0.5\textwidth]{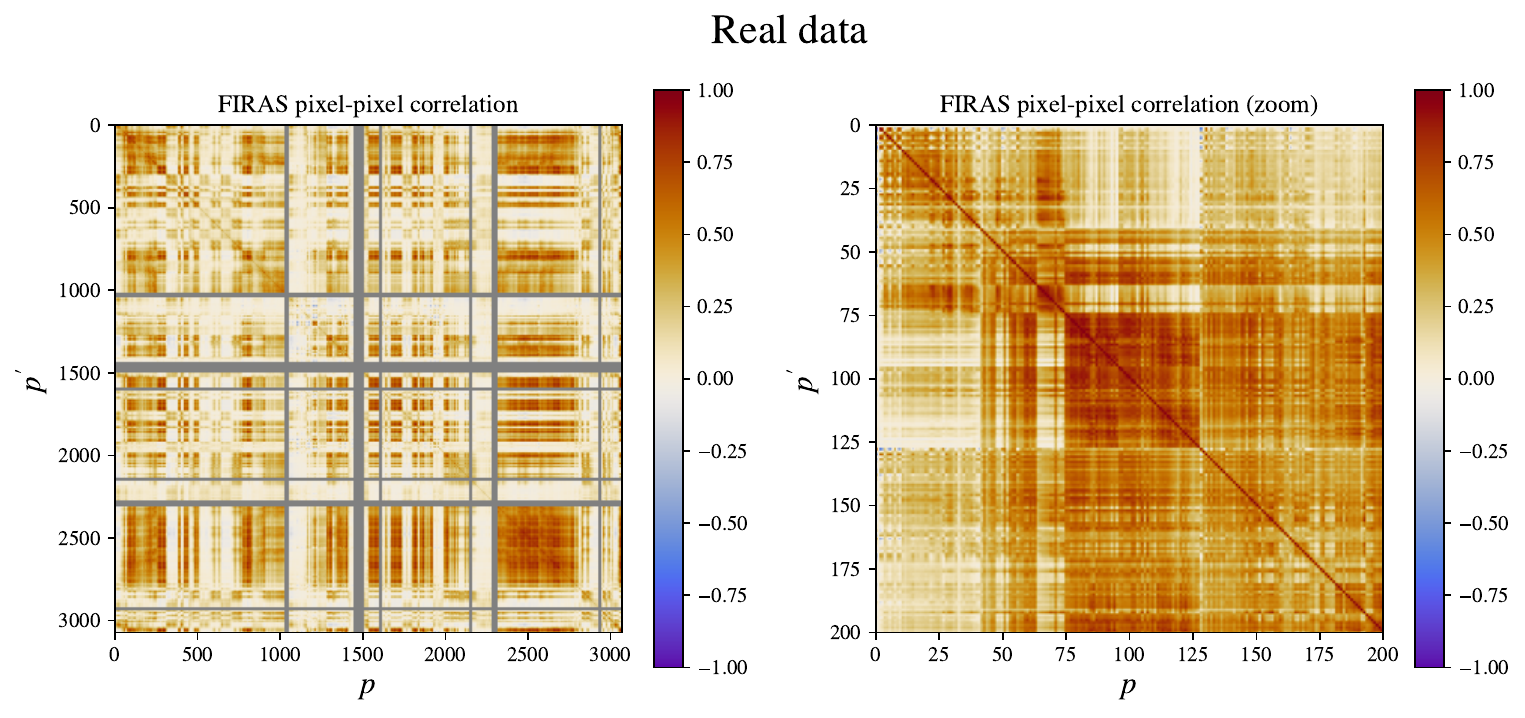}\\
\includegraphics[width=0.5\textwidth]{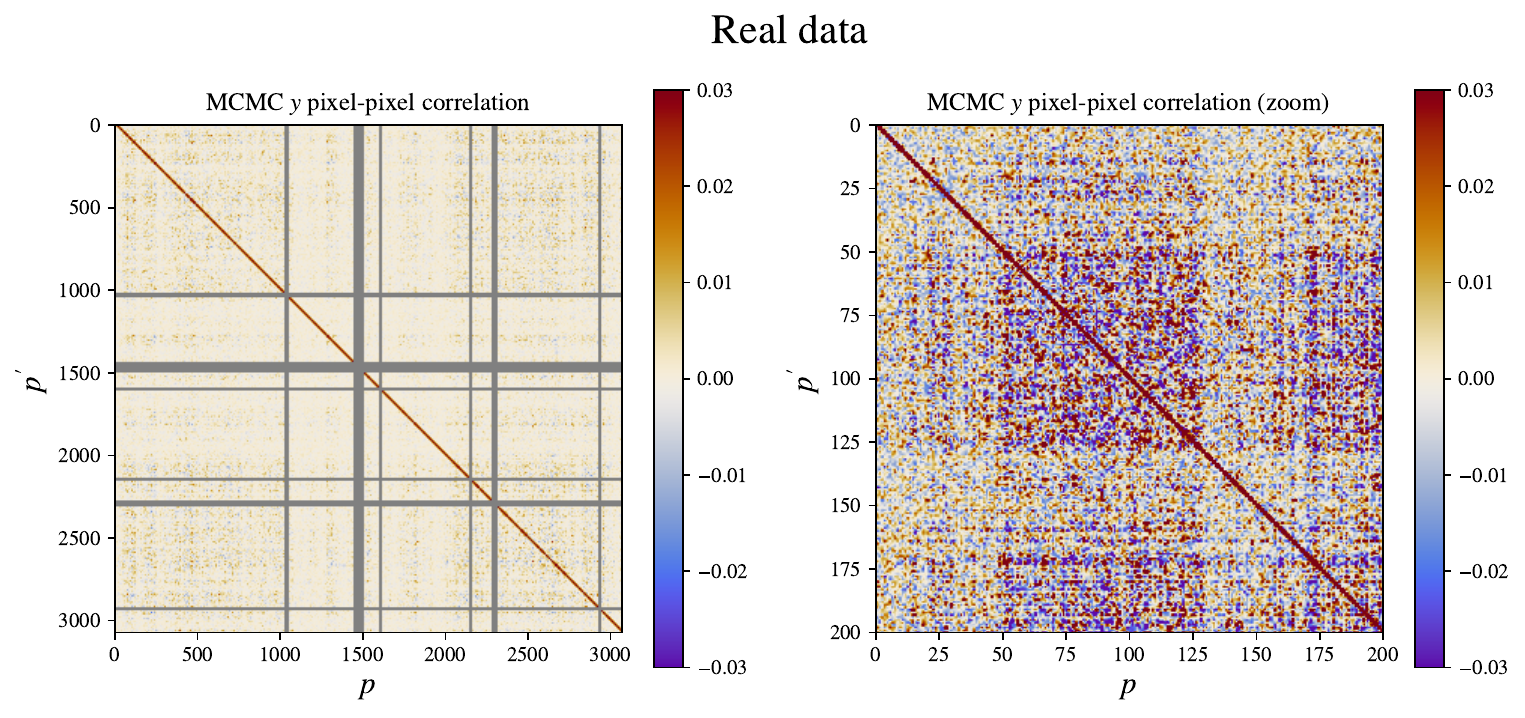}\\
\includegraphics[width=0.5\textwidth]{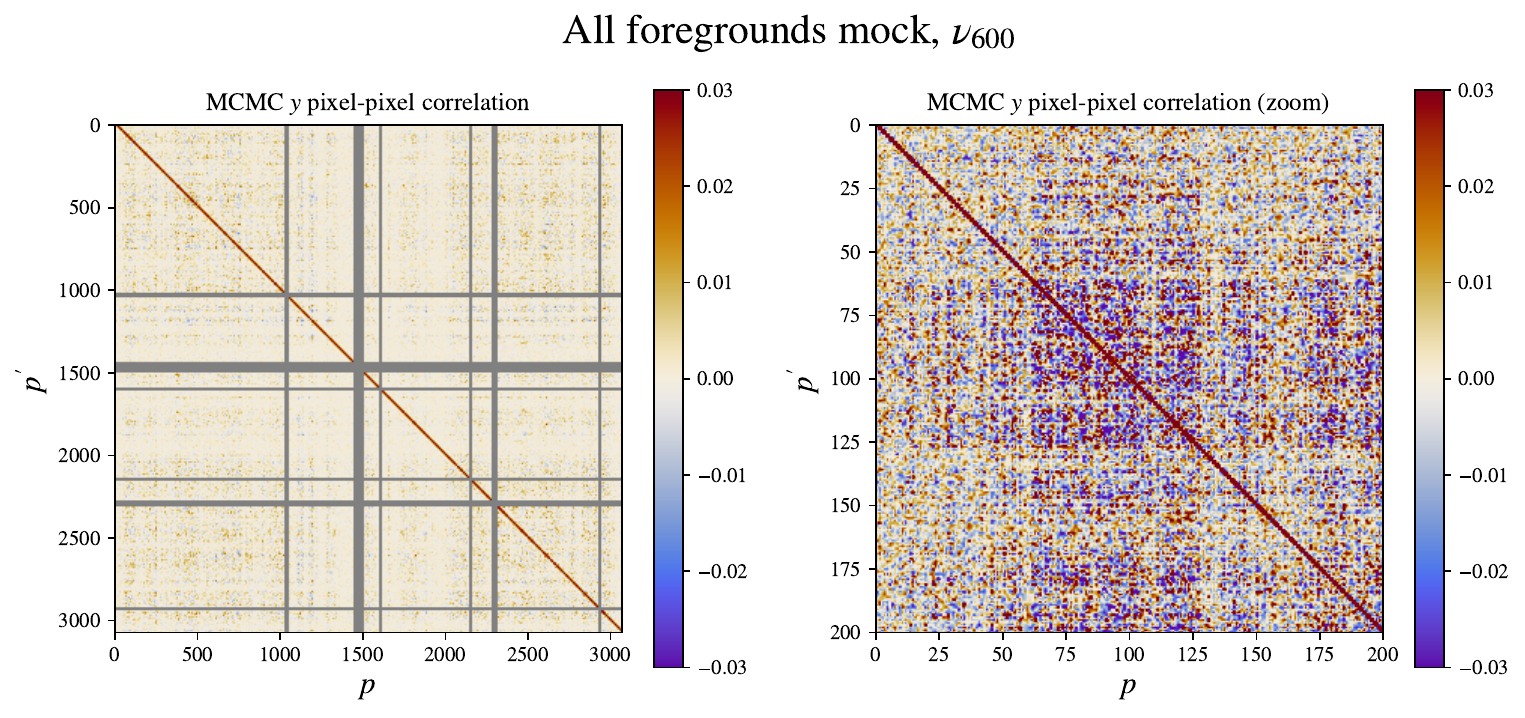}\\
\caption{Example of pixel-pixel correlation of the recovered $y$ map from \textit{FIRAS} data for our baseline \textit{pixel-by-pixel} analysis set-up (middle) compared to the one predicted by the original \textit{FIRAS} model (top). Such correlation was estimated from the correlation between MCMC chains ran on different pixels. The bottom panel shows the same correlation recovered from mock data. The left panels show the full matrix while the right panel show a zoom in of a subset. The row and column indices follow a HEALPix nested ordering scheme. Pixels not observed by \textit{FIRAS} are shown in grey.}
\label{fig:pixpix-corr-recap}
\end{figure}

The second point we investigate is the reliability of our pixel averaging method. As discussed throughout the text, we find multiple sources of evidence that the pixel-pixel correlation modeling in the original \textit{FIRAS} covariance is inaccurate. In Fig.~\ref{fig:pixpix-corr-recap} we show the pixel-pixel correlation predicted by the original \textit{FIRAS} model and the one recovered from the MCMC chains both on real data and mock data. While the \textit{FIRAS} model predicts a very high degree of correlation between pixels, our estimates on data result in amplitudes that are somewhat lower despite showing a similar enhanced level of correlation between pixels in similar regions (see, e.g., pixels with index between 50 and 125). As we recover similar structures in the analysis of mock data, we conclude that these correlations are mainly induced by noise and foreground residuals and, as such, using this covariance in the averaging process would also allow us to reduce their impact in the final $\ymono$ measurements. If an end-to-end reprocessing of \textit{FIRAS} data were to be done, characterizing this pixel-pixel correlation more reliably could greatly improve the accuracy of the $\ymono$ constraints. However, in our final analysis, we show that the choice of the correlation model has only a marginal impact and variations in the final constraints are well below the factor of $\approx 2$-$3$ improvement over the original \textit{FIRAS} constraints that we observe on real data.

 \section{Robustness of the previous $\mu$-distortion constraints}\label{sec:mu}
\begin{figure}[htbp]
\includegraphics[width=0.5\columnwidth]{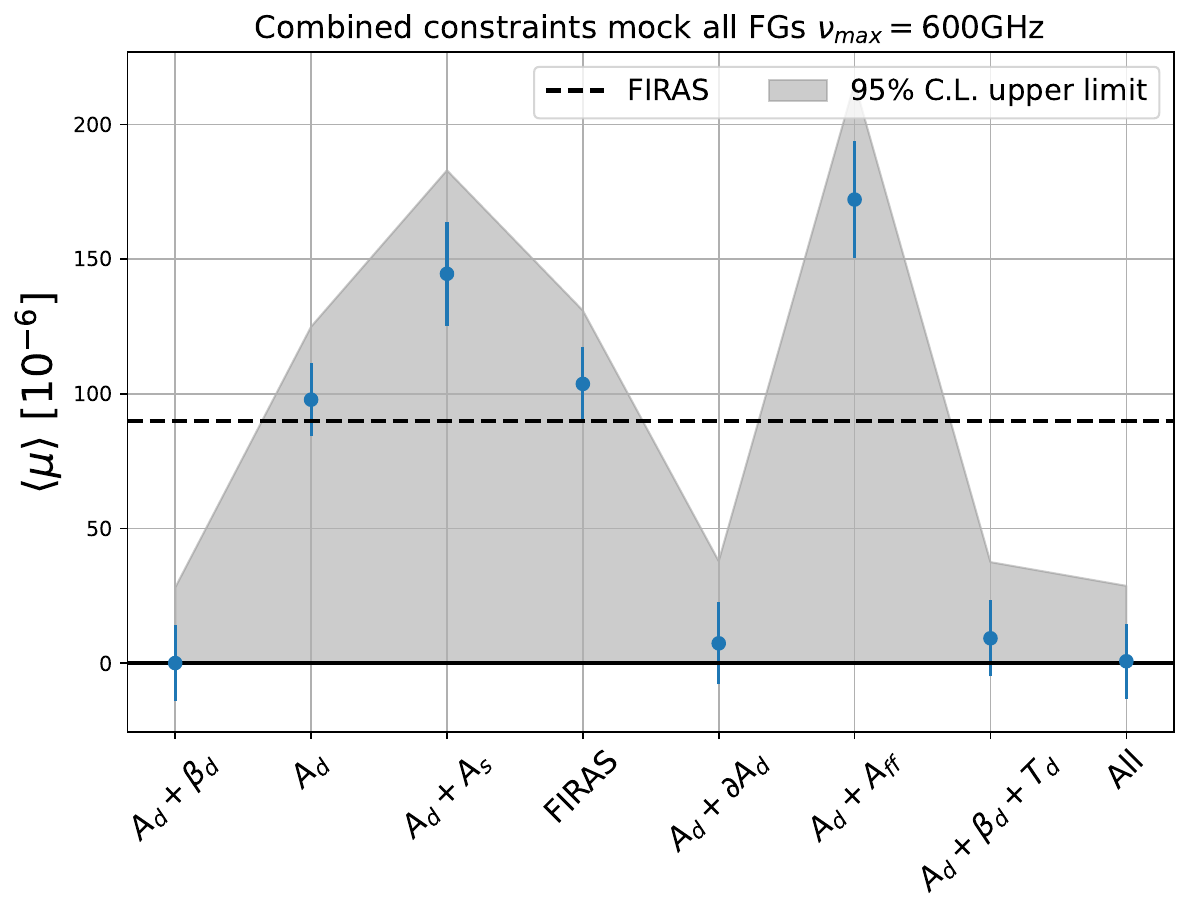}
\caption{Upper limit on $\langle\mu\rangle$ obtained on our \textit{all foregrounds} mocks for different component separation methods employed in \citetalias{BianchiniFabbian2022}. Each data point was computed averaging the $\mu$ map on a P60, P40, P20 footprint and combining those measurements together accounting for the correlation induced by overlapping sky area. The point labeled as ``All'' is the average between the methods consistent with 0 accounting for the correlation between the estimate given by the covariance between the three methods averaged over all pixels.}
\label{fig:mumu-summary}
\end{figure}

As a final consistency check of our analysis pipeline we use our simulated data to constrain the $\mu$ distortion monopole $\langle\mu\rangle$ instead of the $y$ distortion that is the focus of this work. The purpose of this is repeating the analysis of \citetalias{BianchiniFabbian2022} to assess the robustness of their constraints to foregrounds residuals and test if the modified procedure to account for the pixel-pixel correlation in the $\mu$ map could affect the original result. The original baseline foreground model included a free dust amplitude $A_{\rm d}$ and spectral index $\beta_{\rm d}$ with a fixed $T_{\rm d}=19.6$ K. Additional models tested in the paper include:  $\beta_{\rm d}=1.6$ fixed and only $A_{\rm d}$  free to vary, $A_{\rm d}+A_{\rm synch}$, $A_{\rm d}+A_{\rm free-free}$, as well as one fitting only the amplitude of the reference spectral foreground template released by the \textit{FIRAS} team (hereafter \textit{FIRAS} model). Our results are summarized in Fig.~\ref{fig:mumu-summary}. We find that once the inverse covariance weighting is applied, the only robust method among the original ones employed in \citetalias{BianchiniFabbian2022} is the reference one, which can better account for the variability of the emissivity across the sky outside of the Galactic plane. We tested in this context two additional foreground models for this analysis that can account for spatial variability of the foregrounds: the $A_{\rm d}, \beta_{\rm d}, T_{\rm d}$ model used throughout this work and the one where we fit $A_{\rm d}$ and the amplitude of the first derivative of the dust SED with respect to the temperature  $\partial A_{\rm d}$. These models recover similar results to the reference one. The final obtained upper limit is $\mu\lesssim 40\times 10^{-5}$, similar to what is obtained on real data in \citetalias{BianchiniFabbian2022}.  
\bibliography{main}
\end{document}